\documentclass[twocolumn]{aa520} % for a paper on 2 columns  

 \pdfoutput=1
                                   
\usepackage{graphicx}
\usepackage{natbib}

\usepackage{txfonts}
\usepackage{longtable}
\usepackage{supertabular}
\newcommand{\ms}{\,m\,s$^{-1}$}
\newcommand{\cms}{\,cm\,s$^{-1}$}
\newcommand{\rhk}{$\log(R^{\prime}_{\rm HK})$\,}
\begin{document}
   \title{The HARPS search for Earth-like planets in the habitable zone}
   \subtitle{I -- Very low-mass planets around  \object{HD20794}, \object{HD85512} and \object{HD192310}\thanks{Based on observations made with the HARPS instrument on ESO's 3.6\,m telescope at the La Silla Observatory in the frame of the HARPS-Upgrade GTO program ID 69.A-0123(A)}}
   \author{F.~Pepe\inst{1}
          \and 
          C.~Lovis\inst{1}
           \and
	 D.~S\'egransan\inst{1}
	 \and
          W.~Benz\inst{2}
	 \and
          F.~Bouchy\inst{3,4}
	 \and
	 X.~Dumusque\inst{1}
	 \and
	 M.~Mayor\inst{1} 
	  \and
	 D.~Queloz\inst{1}
	 \and
          N.~C.~Santos\inst{5,6} 
	 \and
	 S.~Udry\inst{1}	  	  
          }

   \offprints{\newline F. Pepe, \email{Francesco.Pepe@unige.ch}}

   \institute{
          Observatoire de Gen\`eve, Universit\'e de Gen\`eve, 51 ch. des Maillettes, CH--1290 Versoix, Switzerland
          \and
	 Physikalisches Institut Universit\"at Bern, Sidlerstrasse 5, CH--3012 Bern, Switzerland
	 \and
	 Institut d'Astrophysique de Paris, UMR7095 CNRS, Universit\'e Pierre \& Marie Curie, 98bis Bd Arago, F--75014 Paris, France
	 \and
	 Observatoire de Haute-Provence/CNRS, F--04870 St.Michel l'Observatoire, France
	 \and
	 Centro de Astrof\'isica da Universidade do Porto, Rua das Estrelas, P--4150-762 Porto, Portugal
	 \and
	 Departamento de F\'isica e Astronomia, Faculdade de Ci\^encias, Universidade do Porto, Portugal}

   \date{received; accepted}

    \abstract
     {In 2009 we started an intense radial-velocity monitoring of a few nearby, slowly-rotating and quiet solar-type stars within the dedicated HARPS-Upgrade GTO program. The goal of this campaign is to gather very-precise radial-velocity data with high cadence and continuity to detect tiny signatures of very-low-mass stars that are potentially present in the habitable zone of their parent stars. Ten stars were selected among the most stable stars of the original HARPS high-precision program that are uniformly spread in hour angle, such that three to four of them are observable at any time of the year. For each star we recorded 50 data points spread over the observing season. The data points consist of three nightly observations with a total integration time of 10 minutes each and are separated by two hours. This is an observational strategy adopted to minimize stellar pulsation and granulation noise. We present the first results of this ambitious program. The radial-velocity data and the orbital parameters of five new and one confirmed low-mass planets around the stars  \object{HD\,20794}, \object{HD\,85512}, and \object{HD\,192310} are reported and discussed, among which is a system of three super-Earths and one that harbors a 3.6\,M$_{\oplus}$-planet at the inner edge of the habitable zone. This result already confirms previous indications that low-mass planets seem to be very frequent around solar-type stars and that this may occur with a frequency higher than 30\%.}

 \maketitle

%________________________________________________________________
%
\section{Introduction}
%______________________________________________________________
During the past years the field of extra-solar planets evolved toward the exploration of very low-mass planets down to the regime of super-Earths, i.e., to objects of only few times the Earth mass. Although finding Earth-like planets is probably the main trigger for these searches, one has to consider that their characterization contributes in a significant way to building up a general picture of how exoplanets form and evolve. Knowing the frequency and nature of these planets may enable us to distinguish between various theories and models, deliver new inputs and constraints to them, and this knowledge may contribute to refining their parameters. On the other hand, the models provide us with predictions which can be verified by observations. For instance, the presently discovered low-mass planets are predicted to be only the tip of the iceberg of a huge, still undiscovered planetary population \citep{Mordasini:2009}. If this is confirmed, hundreds of new planets will be discovered in a near future as the radial-velocity (RV) precision improves and the various programs increase their effectiveness and their time basis.

\par
ESO's HARPS instrument \citep{Mayor:2003} has certainly played a key role by delivering more than 100 new candidates in its first eight years of operation. The most important and impressive contribution of this instrument lies more specifically in the field of super-Earths and Neptune-mass planets. Indeed, about 2/3 of the planets with mass below 18\,M$_{\oplus}$ known to date have been discovered by HARPS. This new era started in 2004 with the discovery of several Neptune-mass planets such as  $\mu$\,Ara\,c (\citet{Santos:2004-b}, see also \citet{Pepe:2007} for an update of the parameters), 55\,Cnc \citep{McArthur:2004}, and GJ\,436 \citep{Butler:2004-b}. Many more followed, but the detection of the planetary system HD\,69830 containing three Neptune-mass planets \citep{Lovis:2006}, and that of HD\,40307 with its three super-Earths \citep{Mayor:2009-a} best illustrate the huge potential of HARPS. HARPS also revealed to us the system Gl\,581, with two possibly rocky planets c and d with masses of 5 and 8\,M-Earth, respectively, both lying at the edge of the habitable zone (HZ) of their parent star \citep{Udry:2007, Selsis:2007b}, and the Gl\,581\,e of only 1.9\,M$_{\oplus}$  \citep{Mayor:2009-b}. Last but not least, we should mention the system around HD\,10180, with seven extra-solar planets of low mass, among which is the lightest ever detected exoplanet HD\,10180\,b, with $m*\sin i$ of only 1.5\,M$_{\oplus}$  \citep{Lovis:2011a}.
\par
\begin{figure}
\includegraphics[width=\columnwidth]{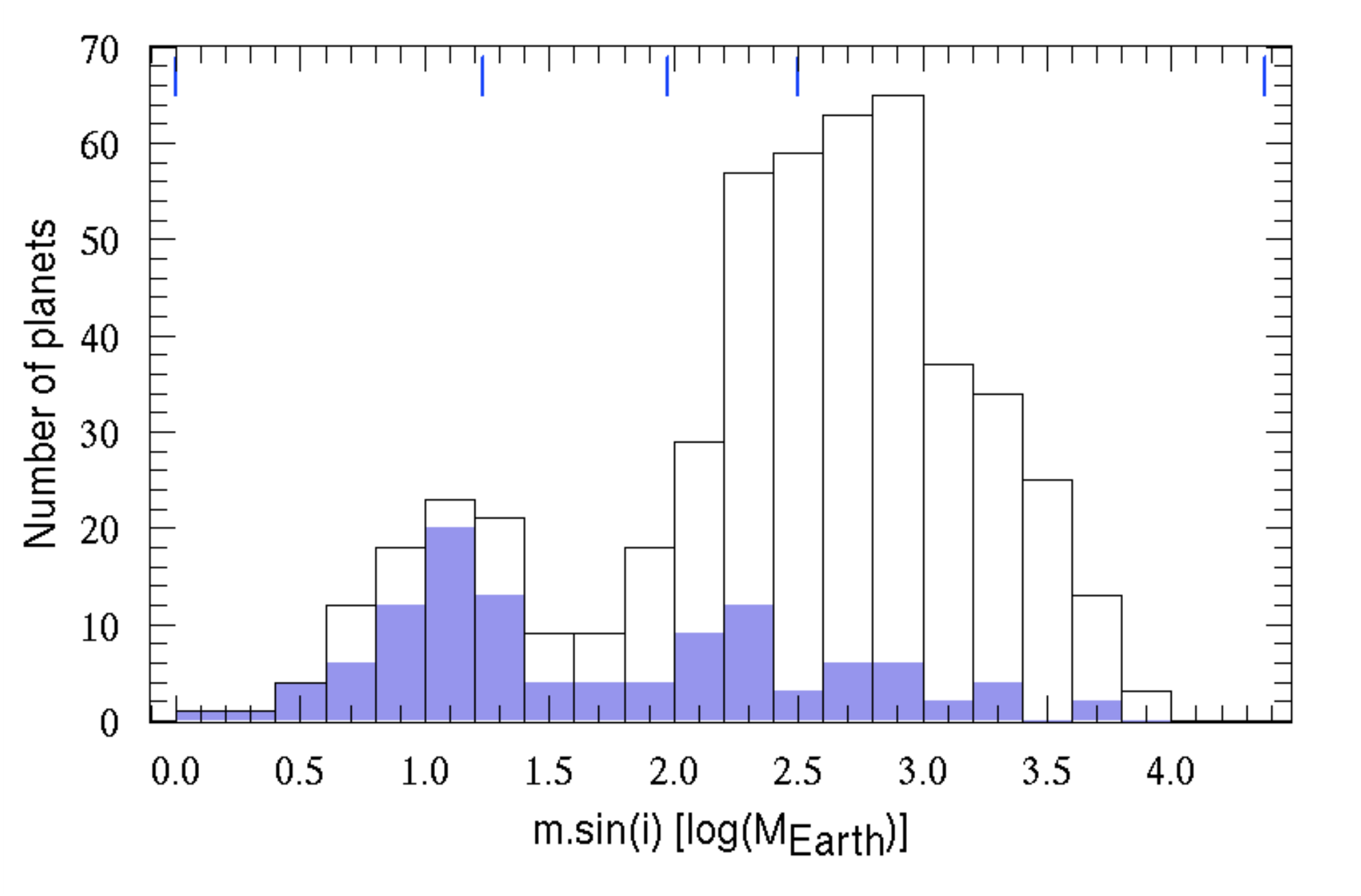}
\centering
\caption{Histogram of companion masses. The dark area shows the companions discovered with HARPS and includes the candidates which are in preparation for publication. }
\label{fi:Mass_histogram}
\end{figure}

Figure\,\ref{fi:Mass_histogram} shows the mass distribution of companions around solar-type stars of all published planets, including those of HARPS that are in preparation. The HARPS detections (dark area) probe a mass range that suffered from strong detection biases before. It must be noticed that the planet frequency increases again below 20\,M$_{\oplus}$. This increase is highly significant, because higher-mass planets induce higher RV variations on their parent star and are therefore detected more easily. Moreover, a recent investigation of the HARPS high-precision sample  has shown that about 1/3 of all sample stars exhibit RV variations, which indicates the presence of super-Earths or ice giants \citep{Lovis:2009}. Indeed, planet formation models \citep{Ida:2008,Mordasini:2009,Alibert:2011} show that only a small fraction (on the order of 10\%) of all existing embryos will be able to grow and become giant  planets. \emph{Hence, we expect that the majority of solar-type stars will be surrounded by low-mass planets.} The inclusion of telluric planets can only increase the planet frequency  further and consequently the probability of detection. This implies that even a small sample of target stars is likely to reveal, if followed with high enough accuracy, several Earth-like planets. Pushing the measurement precision even more farther should therefore naturally increase the probability of finding new planets, especially of very low-mass and rocky planets that are possibly in the habitable zone of their host star.

In the following section we will describe our program for the specific search for low-mass and rocky extra-solar planets that possibly lie in the habitable zone of their parent star. The observations of these targets is presented in Section\,\ref{se:observations}.  The three host stars of the program around which we have detected new planets will be described in greater detail in Section\,\ref{se:hosts}. Section\,\ref{se:results} presents the three newly discovered planets as well as the confirmation of a recently announced planet. Finally, a short discussion and an outlook on the occurrence frequency of low-mass planets is given in Section\,\ref{se:discussion}.

%__________________________________________________________________
%
\section{Searching for Earth analogs around nearby stars}                                
%__________________________________________________________________

The recent HARPS discoveries made us aware that discovering Earth-like exoplanets is already within the reach of HARPS, although important questions still remain open: How frequent are low-mass planets? At what level is the detection bias? Where is the precision limit set by stellar noise? Can we detect low-mass planets in the habitable zone of their parent star if sufficient observations on a high-precision instrument are invested? Driven by the very encouraging results of the HARPS high-precision program \citep{Udry:2008} and the hints for high planet occurrence announced by \citet{Lovis:2009}, we have decided to investigate these questions in more detail. For this purpose we have defined a specific program on HARPS for the search of rocky planets in the habitable zone HZ based on Guaranteed Time Observations (GTO). 

\par
We decided to observe ten nearby stars with the highest possible radial-velocity precision and sampling during a period of four years. Our goal is to probe the presence of extra-solar planets as similar as possible to our Earth. Therefore, we have selected targets with the following characteristics: 
\begin{itemize}
\item The stellar magnitude must be such that we are not limited by photon noise within a 15-minute exposure, i.e., ensure an RV-precision of better than 30\cms.
\item Stars should lie within 10\,pc from the Sun.
\item The targets must have characteristics that guarantee the best RV precision, i.e, they should be of late-G to early-K spectral type and have low rotational velocity (a rotational velocity below what can be detected with HARPS, i. e., $v \sin i<1$\,km\,s$^{-1}$) and low activity indicators ($log(R$'$_{HK}) < -4.85$).
\item The RV scatter measured with HARPS over the first years of operation must be lower than 2\,m/s rms over several years and based on a minimum of ten recorded data points.
\item The ten targets must be evenly distributed in terms of right-ascension to allow the observation of at least three to four targets at any period of the year.
\end{itemize}

Combining all these criteria resulted in a strong down-selection of the 400 possible candidates of the original HARPS high-precision program. In particular the requirement on the distance from the Sun turned out to be too limiting considering all the other requirements, so that we had to relax this requirement and allow targets to enter the list with distances up to 16\,pc.
\par
The ten selected targets are given in Table\ref{ta:targets}. The most prominent member is Alpha Cen\,B (\object{HD\,128621}). This star is of particular interest because of its brightness and its short distance to the Sun. If planets were detected around Alpha Cen\,B, they would be ideal candidates for additional follow-up by spectroscopy, astrometry, photometry, etc. Furthermore, this star is part of a triple system, which makes the question even more interesting. However, what appears to be the greatest advantage generates some observational challenge: On the one hand, the bright magnitude of Alpha Cen B is a limiting factor for the telescope guiding system and may result in poorer RV precision due to incomplete light scrambling across the spectrograph's entrance slit. On the other hand, because Alpha Cen B is in a triple system, the RVs must be analyzed by considering a complete - and precise - orbital model, especially when looking at long-period planets.
\par
Another prominent candidate is $\tau$\,Ceti (\object{HD\,10700}). This star was already known to be an 'RV-standard' star in the sense that it is followed-up by several groups and shows a very low RV dispersion.  Indeed, the RV dispersion measured with HARPS over a time span of more than seven years and involving 157\, data points is 0.93\,ms! Because of its magnitude, the spectral type, the RV stability, and its low chromospheric activity level,  $\tau$\,Ceti  is close to the perfect candidate for our program.
\par
The remaining eight candidates show similar characteristics. They are all part of the original HARPS high-precision program. As such, they have all been observed since the very beginning of the HARPS operations, which allowed us to make a preliminary selection also on the basis of preliminary RV-dispersion and chromospheric activity indicator values. Since the start of our specific program on April 1st, 2009, the number of measurements has increased and the mentioned values refined.
\par
An important aspect to be pointed out is that the presented program carries some 'risks': By construction this program preselected only stars very stable in terms of radial velocities. In other terms, we were sure that they would not harbor any \emph{known} planet. Even worse, their raw RV dispersion does not leave much room for 'strong' RV signals. Of course, we were aware of this and that we might end up with no positive result.

\begin{table}
\centering
\caption{Targets of the HARPS  search for Earth analogs around nearby stars program}
\label{ta:targets}

\begin{tabular}{l l l l l l l l}
\hline
 Target		& R.A.	& DEC		& Sp. type	& $V$Mag	& Dist.	\\
			& [h:m:s]	&  [$^\circ$:':"]	& 		& 			& [pc]	\\
 \hline
 \hline
 HD\,1581		 & 00:20:04	 & -64:52:29	& F9V	& 4.23	& 8.59 	\\
 HD\,10700	 & 01:44:04	 & -15:56:15	& G8V	& 3.49	& 3.65	\\
 HD\,20794	 & 03:19:55	 & -43:04:11	& G8V	& 4.26	& 6.06	\\
 HD\,65907A	 & 07:57:46	 & -60:18:11	& G2V	& 5.59	& 16.19	\\
 HD\,85512	 & 09:51:07	 & -43:30:10	& K5V	& 7.67	& 11.15	\\
 HD\,109200	 & 12:33:31	 & -68:45:20	& K0V	& 7.13	& 16.17	\\
 HD\,128621	 & 14:39:35	 & -60:50:14	& K1V	& 1.35	& 1.35	\\
 HD\,154577	 & 17:10:10	 & -60:43:43	& K0V	& 7.38	& 13.69	\\
 HD\,190248	 & 20:08:44	 & -66:10:55	& G5IV-V	& 3.55	& 6.11	\\
 HD\,192310	 & 20:15:17	 & -27:01:58	& K3V	& 5.73	& 8.82	\\
\hline
\end{tabular}
\end{table}

There is no doubt that successful detections depend on the observational precision, in particular when searching for signals with semi-amplitudes below 1-2\ms, as is the case for the presented program. Five years of observations have proved that, on quiet and bright dwarf stars, a radial-velocity precision well below 1\ms can be achieved on both short-term (night) \emph{and} long-term  (years) timescales. A direct confirmation is offered by the raw RV dispersions measured for the candidates of our program and given in Table\,\ref{ta:measurements}. 
\par
To make a rough estimation of the detection limits we may expect, let us assume a precision of 0.7\ms on each data point all included, i.e. stellar noise, instrumental errors, atmospheric effects, etc. The photon noise is 'by design' well below this level and can be neglected. With 50 well-sampled data points we should then be able to detect a 50\cms RV semi-amplitude at better than a 3-sigma level. This signal corresponds to a 2.2\,M$_\oplus$ on a one-month orbit or a 4\,M$_\oplus$ on a six-month orbit. When extended to three years, the same observational strategy will allow us to increase the number of data points and consequently reduce the detectable semi-amplitude to about 30\,\cms or the equivalent of a 1.3\,M$_\oplus$ planet on a one-month orbit, or a 2.3\,M$_\oplus$ on a six-month orbit.  On the other hand, it is interesting to note that such a planet would lie at the inner edge of the habitable zone of Alpha Cen\,B or even inside the habitable zone of \object{HD\,85512} or  \object{HD\,192310}, which are later-type stars.
\par
In the previous discussion it was assumed of course that no RV signal induced by stellar activity is present at the orbital period of investigation. In the opposite case, the detection limit will not decrease proportionally to the square-root of the number of data points. Even worse, if the stellar activity signal has a coherence time similar to or longer than the observational period, it becomes indistinguishable from a possible radial-velocity signal of similar periodicity. However, other indicators such as precise photometry of the star, line bisector or stellar activity indicators can help to identify possible correlations with the radial-velocity signal and thus provide a tool to distinguish stellar noise from a planetary signal. Another powerful tool is to verify that the RV signal is persistent and coherent over time.

%__________________________________________________________________
%
\section{Observations} 
%__________________________________________________________________
\label{se:observations}

\subsection{HARPS-Upgrade GTO}
All observations presented here have been carried out using HARPS on ESO's 3.6-m telescope at the La Silla Observatory, Chile \citep{Mayor:2003}. A total of 30\,nights per year over four years are allocated to this program on HARPS within our Guaranteed-Time Observations (HARPS-Upgrade GTO). Our goal was to observe each target about 50 times per season. These constraints, together with the need of observing each target several times per night to average stellar noise, set the actual number of targets we are able to follow-up. In the following we shall discuss the observational strategy. For this purpose we have to distinguish different types of measurements and adopt the following definitions: An \emph{exposure} represents the single integration (one open-close shutter cycle). The \emph{observation} is the average of all subsequent exposures. The \emph{data point}, finally, denominates the average of all the exposures in a given night. 

\subsection{Observational strategy}
Thanks to the experience issued from the HARPS high-precision program we have been able to define an optimum strategy of observation to reduce the effects of stellar noise and its impact on the RV measurements. From the study by \citet{Dumusque:2011a} we have learned that stellar pulsations are best averaged out by extending the measurement over a time longer than the typical stellar oscillation period. Because our targets are mainly bright G and K dwarfs, we ended up with exposure times of typically 15 minutes. To avoid saturation on bright objects, the exposures where split into several shorter exposures covering a total of 15\,minutes, including overheads, leading to typical 'open-shutter' time of about 10 minutes. From the same study we can also deduce that super-granulation noise, which is of a longer time scale, is not averaged out by the 15-minutes exposure. For this purpose, the daily time scale must be covered. This led us to observe each target at least twice, if possible three times during the same night, but in any case with the single observations separated as far as possible in time. A total observation time per target of 35-45\, minutes per night had to be considered. If we assume ten hours per night in average, we end up with about 40 to 50 possible observations per target and per year.
\par
\citet{Dumusque:2011a} show that longer-period granulation noise and RV-jitter caused by spots and activity are best averaged out by evenly sampling over the observational season. If we fix the amount of available observation time, an equidistant time grid would be the best solution. However, the visitor-observing scheme applied to ESO's 3.6-m telescope does not allow us to obtain optimum sampling. Nevertheless, it has been possible to join our GTO time with observations of other large programs. The result is that, during nights allocated to this joint program, about two to three targets were observed on average every night, while no data are obtained during other nights, leaving some observational gap with a typical period of one to a few weeks.

\subsection{Obtained measurements}
Table\,\ref{ta:measurements} summarizes the measurements made with HARPS on each of the program's targets, including measurements issued from the original high-precision program. The second column indicates the number of data points (nightly averages of individual exposures) acquired to date. Thanks to the data-points of the original high-precision program, most of the observations span more than six years for any object. It must be noted however that the first-year observations on HARPS were not carried out using the optimum observational strategy, which resulted in data points  -- actually single exposures -- of poorer precision. Despite this, the RV scatter of all these data points is remarkably low, which justifies our initial choice of the targets. The only target for which the RV scatter exceeds 1.5\,\ms is \object{HD\,192310}. Interestingly, this star has been recently announced by \citet{Howard:2010} to harbor a Neptune-mass planet. The discovery is confirmed by the HARPS data and will be described in more detail in Section\,\ref{se:results}. The RV scatter of Alpha Cen\,B (\object{HD\,128621}) is not indicated in the table, because it is affected by the long-term drift caused by its stellar companion. A detailed discussion specifically of Alpha Cen\,B will be presented in a forthcoming paper. For all targets we give the value of the chromospheric activity indicator $\log(R^{\prime}_{\rm HK})$. As mentioned above, the targets where selected to have  $\log(R^{\prime}_{\rm HK}) < -4.85$. The long-term averages and the dispersion values given in the last column confirm that all targets comply with this requirement.

\begin{table}
\centering
\caption{Observations of the targets with HARPS}
\label{ta:measurements}

\begin{tabular}{l l l l l l l l}
\hline
 Target	&  Data points &Time span &  $RV$ scatter	& $\log(R^{\prime}_{\rm HK})$ \\
		 &		& 	 [days]  		&  [\ms]  		&  \\
	\hline
	\hline
 HD\,1581			&  93 	& 2566 	& 1.26	& $-4.93\pm0.003$\\
 HD\,10700		&  141	& 2190	& 0.92	& $-4.96\pm0.003$\\
 HD\,20794		&  187	& 2610	& 1.20	& $-4.98\pm0.003$\\
 HD\,65907A		&  39 	& 2306	& 1.45	& $-4.91\pm0.006$\\
 HD\,85512		&  185	& 2745	& 1.05	& $-4.90\pm0.043$\\ 
 HD\,109200		&  77 	& 2507	& 1.16	& $-4.95\pm0.018$\\
 HD\,128621	 	&  171	& 2776	& NA 	& $-4.96\pm0.014$\\
 HD\,154577	 	&  99		& 2289	& 1.05	& $-4.87\pm0.026$\\
 HD\,190248	 	&  86		& 2531	& 1.26	& $-5.09\pm0.009$\\
 HD\,192310	 	&  139 	& 2348	& 2.62	& $-4.99\pm0.033$\\
\hline
\end{tabular}
\end{table}

\subsection{The example of Tau Ceti}
\begin{figure}
\includegraphics[bb=0 70 595 390,width=85mm,clip]{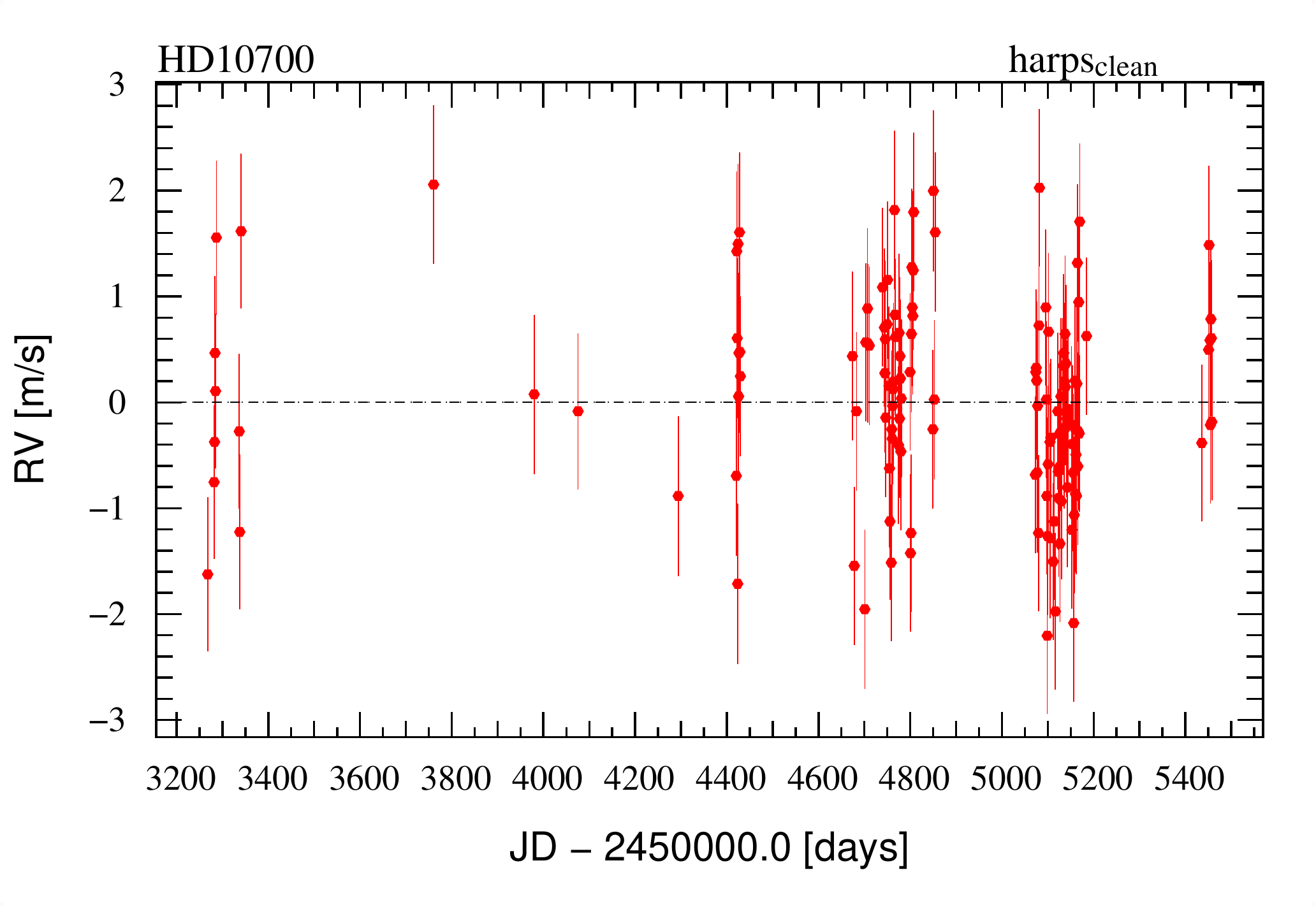}
\includegraphics[bb=0 70 595 390,width=85mm,clip]{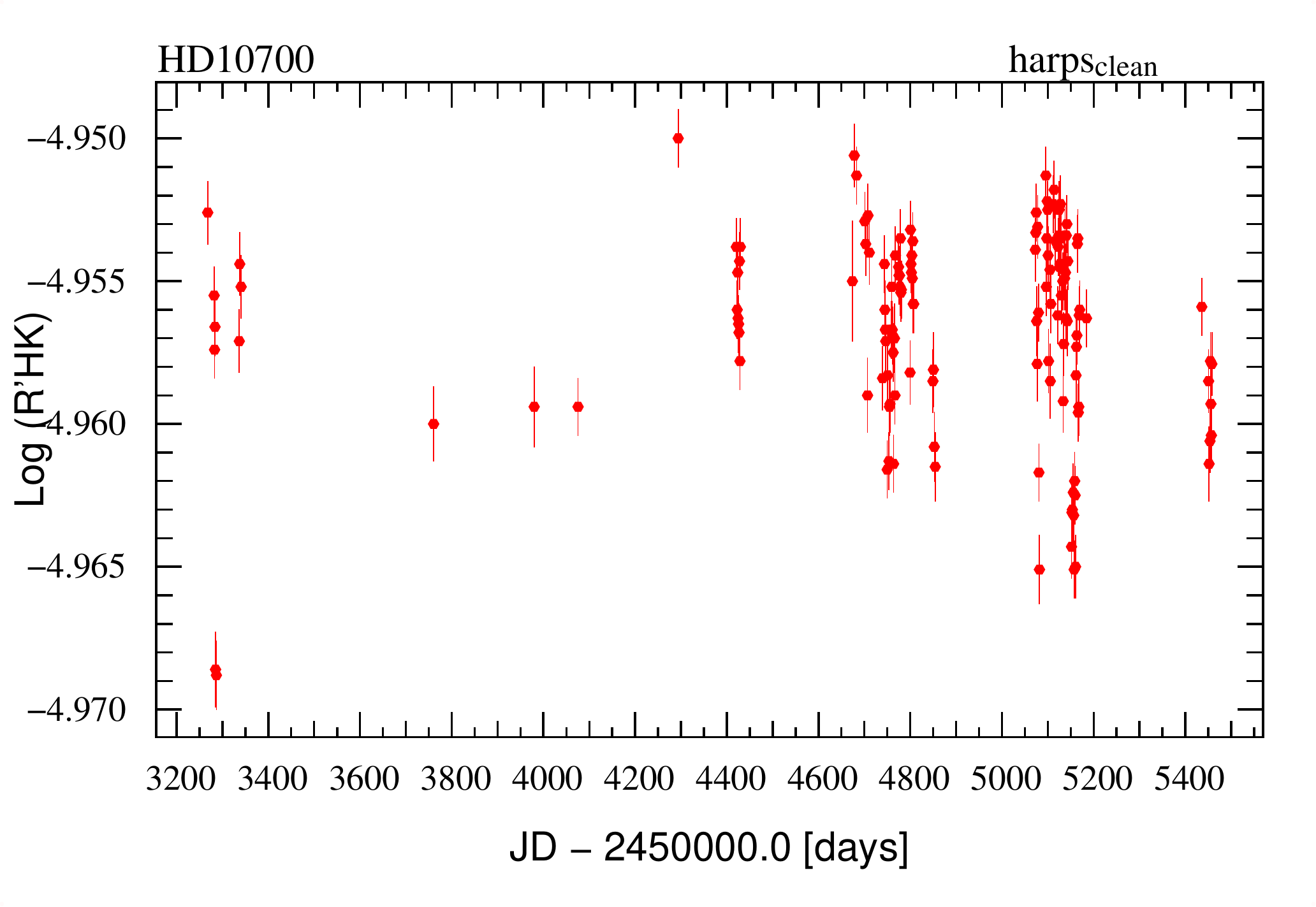}
\includegraphics[bb=0 0 595 390,width=85mm,clip]{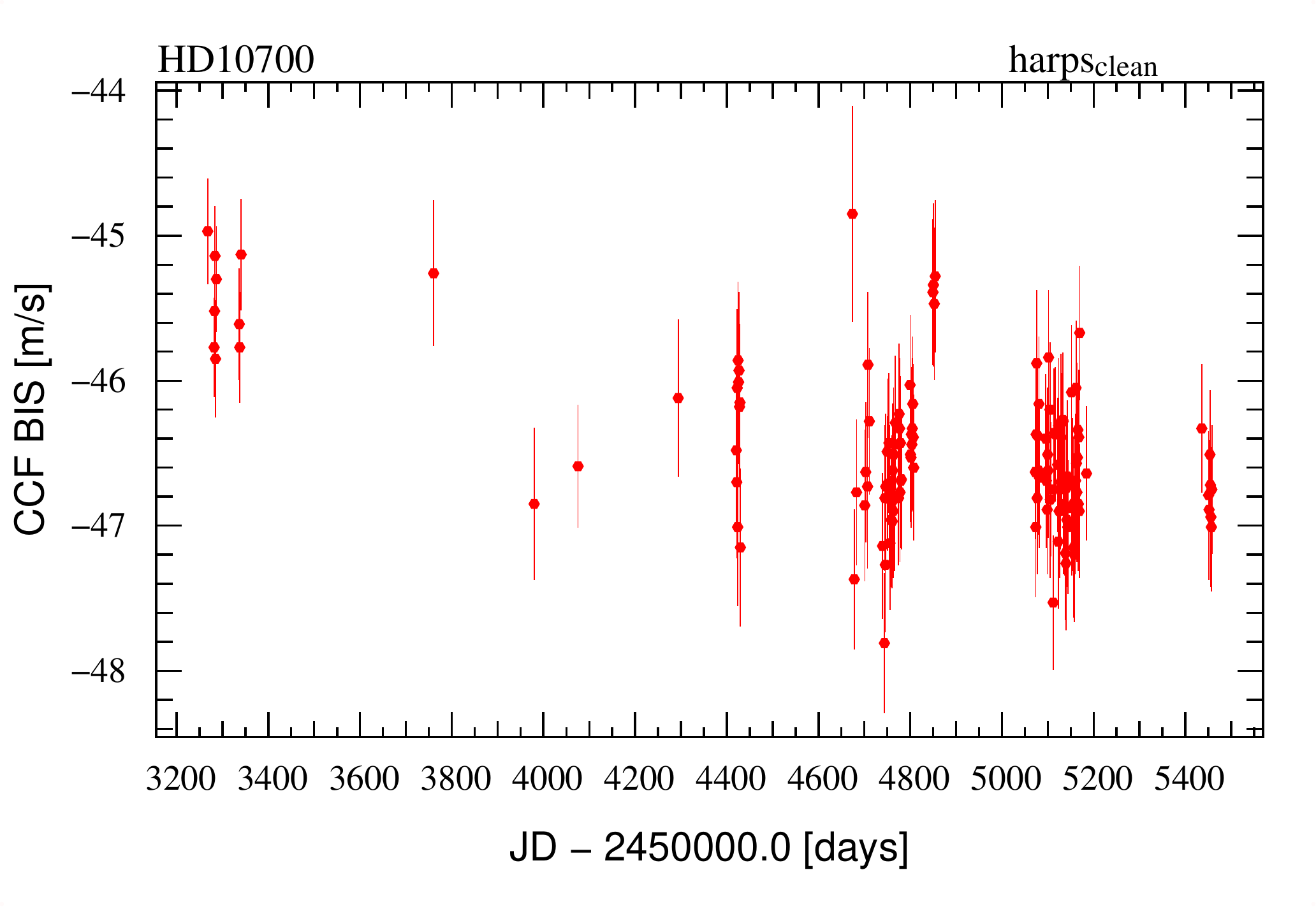}
\centering
\caption{Radial velocities, chromospheric activity indicator \rhk, and line bisector $CCF BIS$ of \object{ HD\,10700} versus time.}
\label{fi:hd10700}
\end{figure}

To illustrate the kind of objects we are dealing with and the level of precision that can be obtained with HARPS, we will focus a moment on \object{HD\,10700}, also known as Tau Ceti. The dispersion over the RVs of the \emph{individual exposures} (typically a few minutes) is 1.5\ms \emph{rms}. As seen in Table\,\ref{ta:measurements}, the dispersion of the data points \emph{(nightly averages)} is instead only 0.92\ms, which proves that the strategy proposed by  \citet{Dumusque:2011a} provides good smoothing of the stellar contributions at the time scale of a night. \object{HD\,10700} turns out to be part of the most quiet stars in our sample despite the large number of data points spanning more than six years of observations. What is even more remarkable is that none of the parameters shown in Figure\,\ref{fi:hd10700}, i.e., the radial velocity, the chromospheric activity indicator \rhk, and the average line bisector of the cross-correlation function $BIS$, show any trend over the six years of observations. For instance, the activity indicator shows a dispersion of only 0.003 dex \emph{rms}. Finally, we underline the stability of the line bisector, which has a dispersion of about 0.5\ms, confirming the high stability of the HARPS' instrumental profile (IP).

\begin{figure}
\centering
\includegraphics[bb=0 47 595 260,width=85mm,clip]{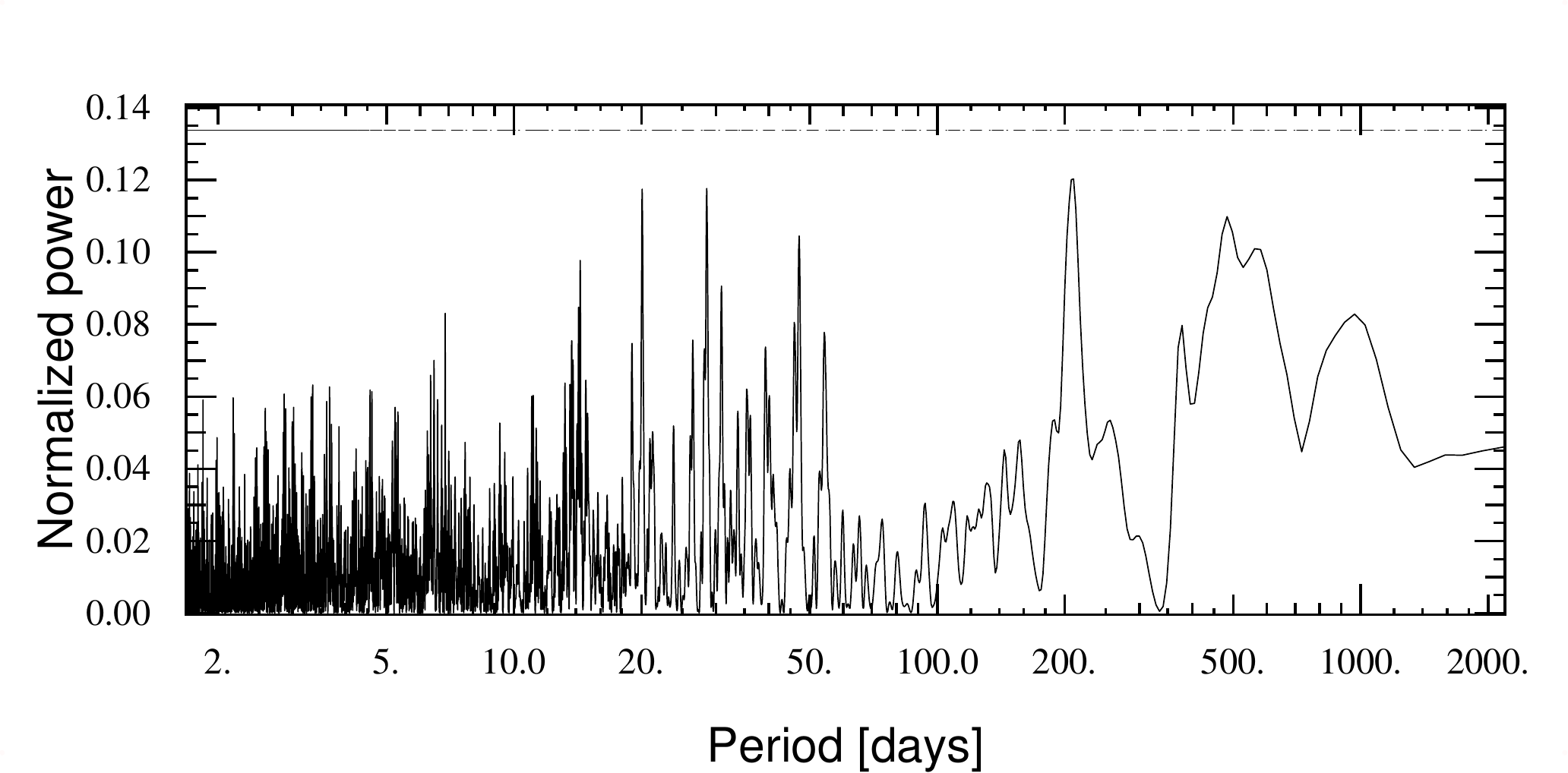}
\includegraphics[bb=0 47 595 260,width=85mm,clip]{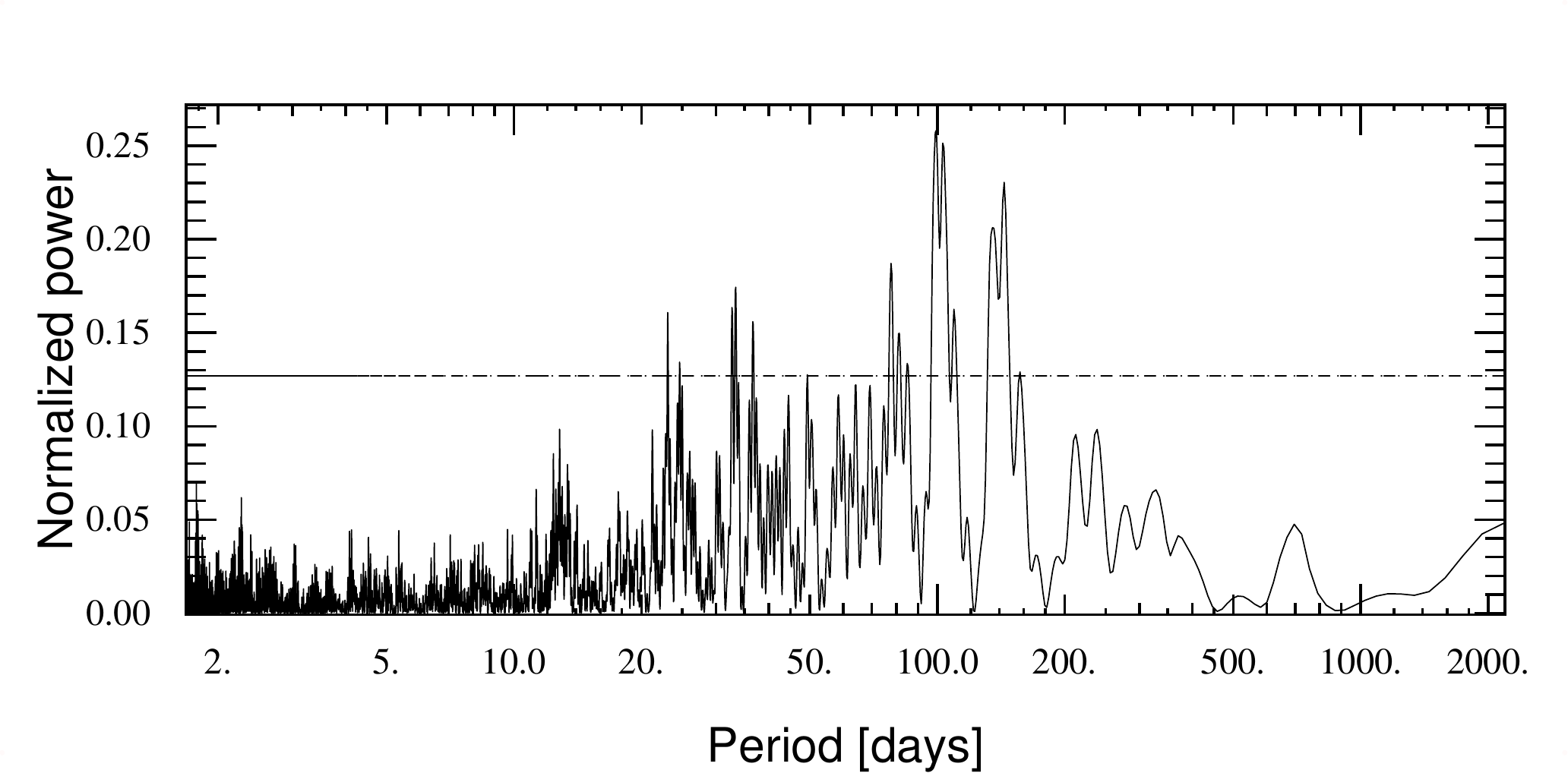}
\includegraphics[bb=0 0 595 260,width=85mm,clip]{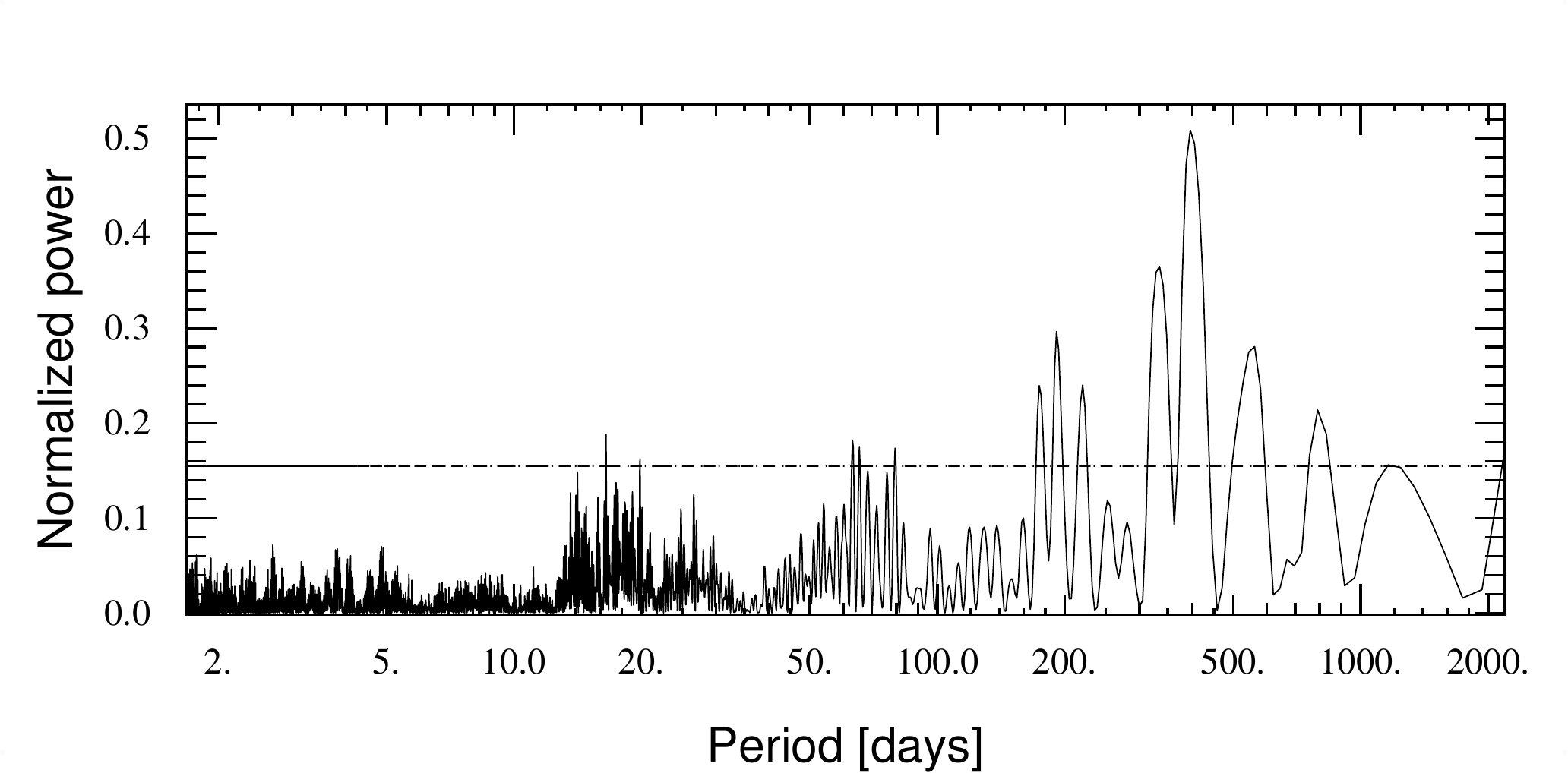}\caption{GLS periodograms of the \object{HD\,10700} radial velocities (upper panel), its \rhk values (central panel) and line bisector $BIS$ (lower panel. The false-alarm probability (FAP) thresholds of 10\% is indicated by a dashed line.}
\label{fi:hd10700_gls_all}
\end{figure}

In the upper panel of Figure\,\ref{fi:hd10700_gls_all} we plotted the generalized Lomb-Scargle (GLS) periodogram of the radial velocity data of \object{HD\,10700}. Note that none of the peaks reaches the 10\% false-alarm probability (FAP) level. This demonstrates that there is no significant planetary signal in the precise HARPS data available to date and we can set a higher limit for the masses of possible planets aroun Tau Ceti as a function of period. For a detailed analysis and discussion we refer to \citet{Lovis:2011b}.
\par
In the central and lower panel of Figure\,\ref{fi:hd10700_gls_all} we present the GLS periodogram of the \rhk and the line bisector $BIS$. The peaks are significant here, in contrast to the RV data, and indicate that there is some 'signal' that clearly is above the noise. Periodograms of activity indicators like \rhk and line bisector $BIS$ often exhibit a complicated pattern of significant peaks around the stellar rotation period and at longer periods. This can probably be explained by the combination of irregular sampling and the presence of signals with short coherence times in the data, as expected from stellar activity and spots. Indeed, solar-type stars are likely covered by a large number of small active regions that rotate with the star and have lifetimes on the order of the rotation period or below. The aspect to be pointed out here, however, is that these peaks do not have any correspondence in the RV GLS periodogram. As remarked above, the (low but significant) activity signal is not correlated to any RV variation.

\begin{figure}
\includegraphics[bb=0 70 595 390,width=85mm,clip]{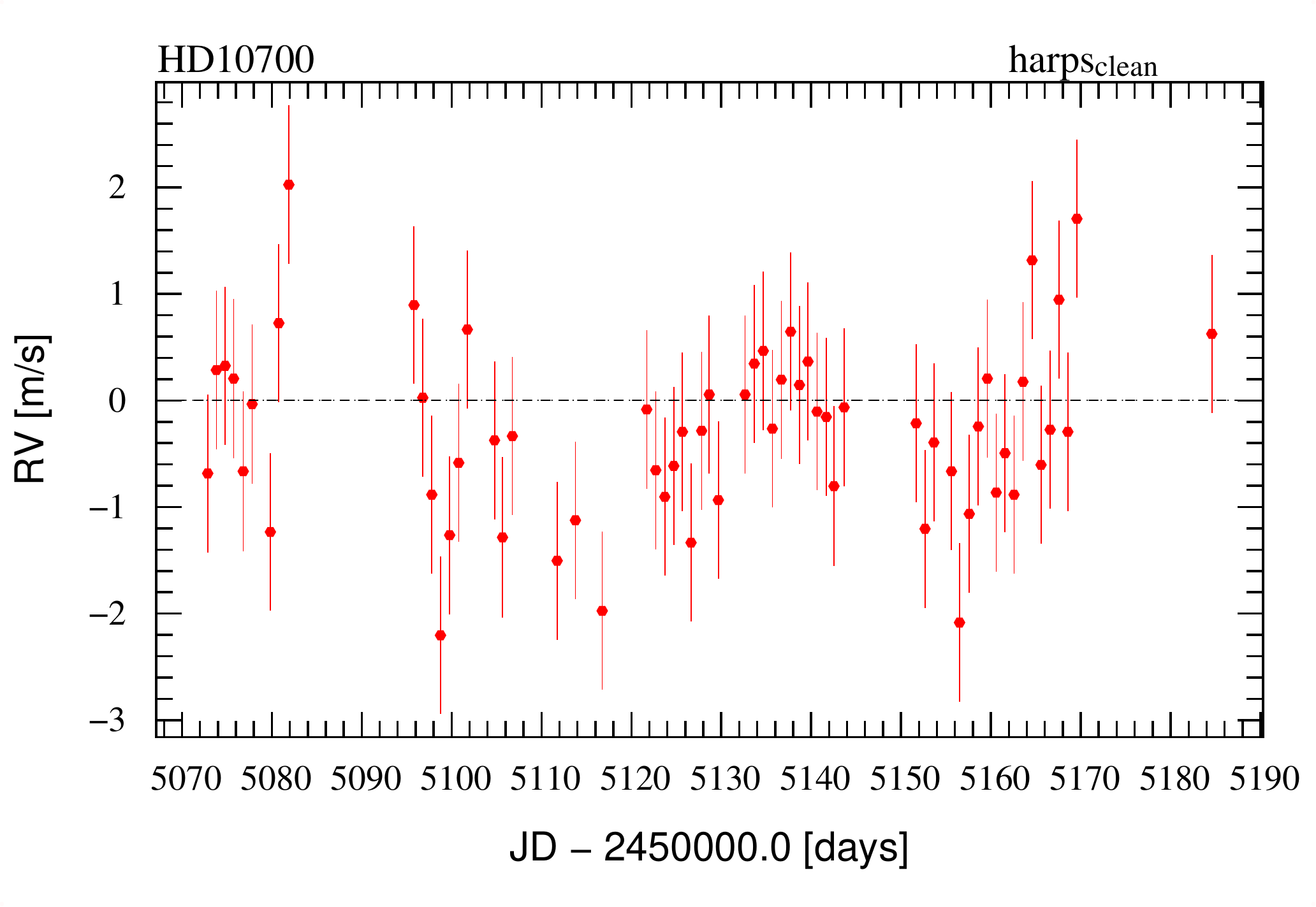}
\includegraphics[bb=0 70 595 390,width=85mm,clip]{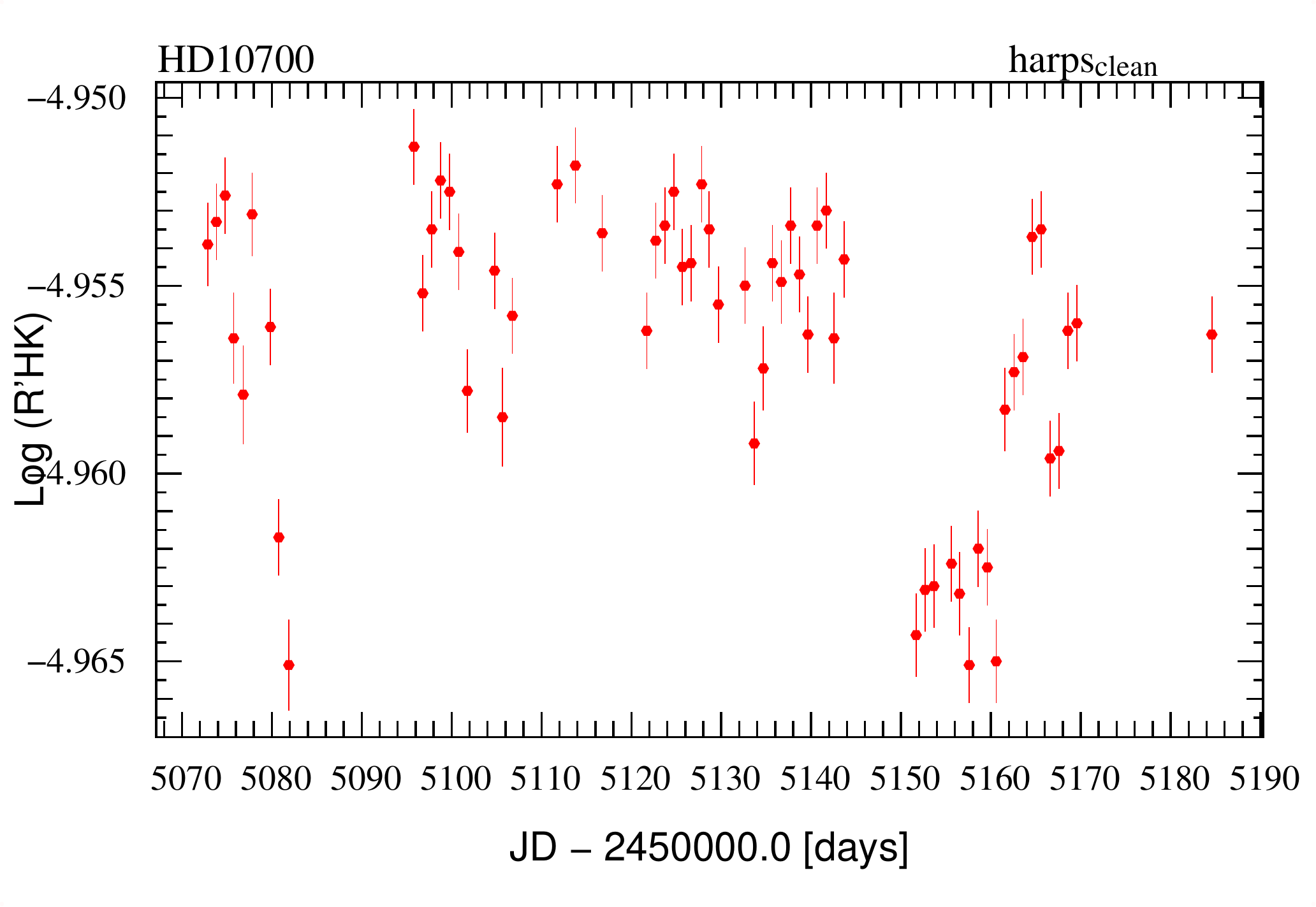}
\includegraphics[bb=0 0 595 390,width=85mm,clip]{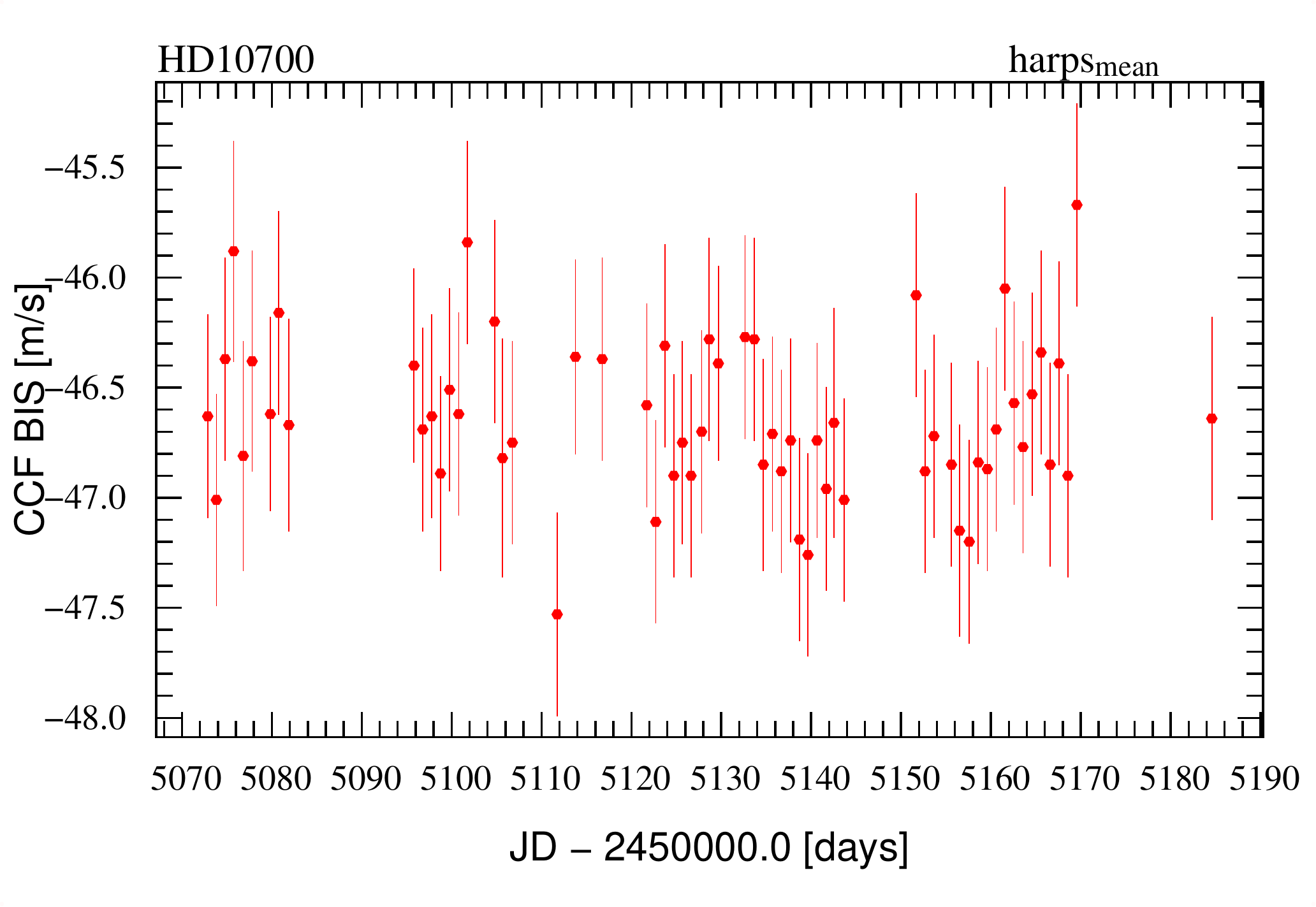}
\centering
\caption{Radial velocitiy, chromospheric activity indicator \rhk and line bisector $BIS$ of \object{ HD\,10700}  over a single season.}
\label{fi:hd10700_zoom}
\end{figure}

\begin{figure}
\includegraphics[width=\columnwidth]{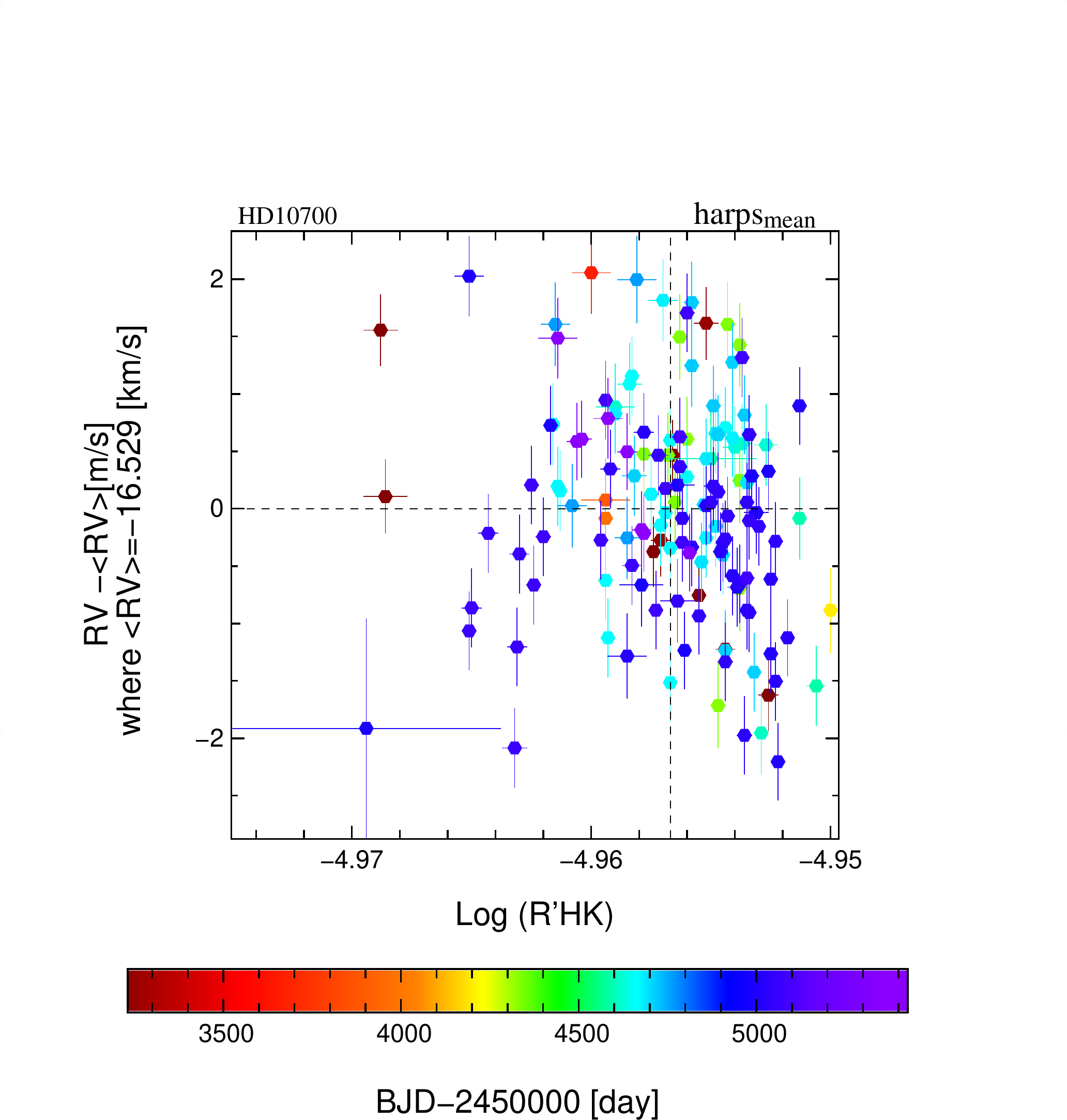}
\centering
\caption{Correlation plot of radial velocity versus chromospheric activity indicator \rhk for \object{ HD\,10700}. The color scale represents the observation date.}
\label{fi:hd10700_corr}
\end{figure}

One may also wonder how strongly stellar activity may influence the radial-velocities. We refer to \citet{Dumusque:2011b} and \cite{Lovis:2011c} for a detailed discussion. In the specific case of Tau Ceti we observe only little RV jitter although the \rhk seems to vary during some seasons, while it remains stable over others. As an example, Figure\,\ref{fi:hd10700_zoom} shows the radial velocity, \rhk and $BIS$ over a single season. Error bars on log(R'HK) values have been calculated from the expected photon statistics, but to take into account the effective noise floor (determined on the most stable stars), an uncertainty of 0.001 was quadratically added. This noise floor is likely caused by residual instrumental systematics. The variation in \rhk is larger than the expected error, but does not seem to follow the radial velocity, and vice-versa. This is better illustrated in Figure\,\ref{fi:hd10700_corr}, where the correlation between the two quantities is plotted. We deduce that in the specific case of Tau Ceti, the radial velocity is not affected by stellar activity.

HD\,10700 therefore is a star that shows no detectable coherent signal in the RV data, while activity indicators exhibit some non-random, low-amplitude patterns that likely convey information on small active regions at the stellar surface. In this sense, these indicators are more sensitive to activity than RV data, which make them useful tools to better understand stellar jitter on quiet solar-type stars. Work is going on to simulate and model these auxiliary data, which may then allow us to eventually correct the RV data themselves from activity effects.

%__________________________________________________________________
%
\section{Parameters of the planet-hosting stars} 
\label{se:hosts} 
%__________________________________________________________________

Of the ten targets presented in the previous section, three presently show planetary signatures. The stars are \object{HD\,20794}, \object{HD\,85512} and \object{HD\,192310}. The target \object{HD\,192310} was already announced recently by \citet{Howard:2010} to harbor a Neptune-mass planet on a 74\,days orbit. Below we will confirm this detection and show that a second planet orbits this star on a much longer orbit.
\par
The photometric and astrometric parameters of the three targets are presented in the first part of Table\,\ref{ta:stellar}. They are all obtained from the HIPPARCOS catalog \citep{ESA-97}. Given that all targets are nearby stars, the astrometric parameters are also very precise.
\par
The spectroscopic parameters presented in the second part of Table\,\ref{ta:stellar} have been collected from \citet{Sousa:2008}, who did an in-depth spectroscopic
analysis of all stars of the HARPS high-precision program. We report them here without any further analysis. We only comment that for the K5V dwarf HD85512 the stellar parameters derived in Sousa et al. may be affected by systematic errors \citep[see e.g.][]{Neves:2009}, given the late spectral type of the star. Indeed, \citet{Sousa:2008} derived an effective temperature of 4715$\pm$102\,K for this star, which disagrees with the value of 4419\,K obtained using the IRFM calibration of \citet[][]{Casagrande:2006}.

\par
To check if this difference could significantly change the derived stellar metallicity and estimated stellar mass, we proceeded with a series of tests. First, fixing the effective temperature to 4419\,K and the gravity and microturbulence values derived by Sousa et al. (4.39\,dex and 0.23\,km\,s$^{-1}$, typical for late K-dwarfs), we derived a new value for the stellar metallicity based on the list of \ion{Fe}{i} lines (and their equivalent widths) measured by Sousa et al. The resulting [Fe/H] is $-0.34$, very close to the original value of Sousa et al. Finally, we estimated the stellar mass using both effective temperature values mentioned above, and making use of the mass--T$_{\mathrm{eff}}$--$\log{g}$--[Fe/H] calibration of \citet[][]{Torres:2010}. In this procedure we fixed the stellar metallicity to $-0.32$ and the gravity to 4.39, while attributing reasonable (though conservative) uncertainties of 0.1 and 0.2\,dex to [Fe/H] and $\log{g}$, respectively. The error in effective temperature was considered to be 100\,K in each case. After a series of 10'000 Monte Carlo trials, using values of T$_{\mathrm{eff}}$, $\log{g}$, and [Fe/H] taken from a Gaussian distribution around the central value and a dispersion of 1-sigma (error), we obtained mass values of 0.73 and 0.65\,M$_\odot$  for input temperatures of 4715 and 4419\,K, respectively (the uncertainly is 0.08\,M$_\odot$, taking into account the scatter of the individual mass values and the dispersion of the Torres et al. calibration). Both values are compatible within the error bars. The former value is close to the 0.74\,M$_\odot$ derived by Sousa et al. through isochrone fitting. We thus conclude that both [Fe/H] and mass values are not strongly sensitive to the choice of the effective temperature scale. We decided to adopt the average value of the two for the remainder of the paper: $-$0.33 for [Fe/H] and 0.69\,M$_\odot$ for the mass of \object{HD85512}.

\par
The third part of Table\,\ref{ta:stellar} shows additional stellar parameters: The projected stellar rotation velocity is issued from the instrumental calibration presented by \citet{Santos:2002}. The chromospheric activity indicator. $\log(R^{\prime}_{\rm HK})$ has been computed in accordance with \citet{Lovis:2011b}. Finally, the stellar rotation period $P_{\rm rot}(R^{\prime}v_{\rm HK})$ and the estimated stellar age are both computed from $\log(R^{\prime}_{\rm HK})$ following the prescription given in \citet{Mamajek:2008}.

\begin{table*}
\centering
\caption{Observed and inferred stellar parameters for the observed targets.}
\label{ta:stellar}
\begin{tabular}{l@{}lccc}
\hline
  \multicolumn{2}{l}{\bf Parameter} 
& \multicolumn{1}{c}{\bf HD\,20794} & \multicolumn{1}{c}{\bf HD\,85512} & \multicolumn{1}{c}{\bf HD\,192310}\\
\hline
\hline
Spectral Type	& 		&	G8V		& K5V	& K3V\\
$V$     		&  		&  4.26		& 7.67 	& 5.73\\
$B-V$ 		& 		& 0.711		& 1.156	& 0.878 \\
$V$ variability 	& 		& $< 0.005$ 	& $< 0.016$ & $< 0.005$ \\
$\pi$ &[mas]   & $165.02 \pm0.55 $ 	& $89.67 \pm0.82 $	& $113.33 \pm0.82 $\\
Distance &[pc]   & 6.06 	& 11.15 & 8.82 \\
$pm$RA &[mas/yr]   & $39.21\pm0.44$ 	& $461.28 \pm0.56 $ 	& $1241.35\pm0.78$ \\
$pm$DEC &[mas/yr]   &$ 27.57\pm0.45$ 	& 	$-472.87 \pm0.57$	& -$181.46\pm0.51$\\
\hline
$M_V$ & & 5.35 	& 7.43	& 6.00 \\
$L$ & [L$_\odot$] & $0.656\pm0.003$ 	& $0.126\pm0.008$	& $0.385\pm0.007$\\ 
$[Fe/H]$ & [dex]  & $ -0.40 \pm 0.01 $  	& $ -0.33 \pm 0.03 $	& $ -0.04 \pm 0.03 $\\ 
$M$  & [M$_\odot$]& $0.70$ 	& $0.69 $	& 0.80\\ 
$T_{\rm eff}$ & [K] & $5401\pm 17$ 	& $ 4715 \pm 102$	& $ 5166\pm 49$\\
$\log g$ & [cgs] & $4.22 \pm 0.11$ 	& $4.39 \pm 0.28$	& $4.51 \pm 0.09$\\
\hline
$v\sin i$ & [km\,s$^{-1}$] & $  < 3  $ 	& $<3$	& $<3$\\
$\log(R^{\prime}_{\rm HK})$ & [dex] & $-4.976\pm0.003$ & $-4.903\pm0.043$  	& $-4.994\pm0.027$\\
$P_{\rm rot}(R^{\prime}v_{\rm HK})$& [days] & $33.19\pm3.61$ 	& $47.13\pm6.98$	& $47.67\pm4.88$\\
age($R^{\prime}_{\rm HK}$) & [Gyr] &   $5.76\pm0.66$	& $5.61\pm0.61$	& $7.81\pm0.90$ \\
\hline
\end{tabular}
\end{table*}

%__________________________________________________________________
%
\section{Results and orbital solutions}
\label{se:results}
%__________________________________________________________________

\subsection{The planetary system of \object{HD\,20794}}              
%__________________________________________________________________

\begin{figure}
\includegraphics[bb=0 70 595 390,width=85mm,clip]{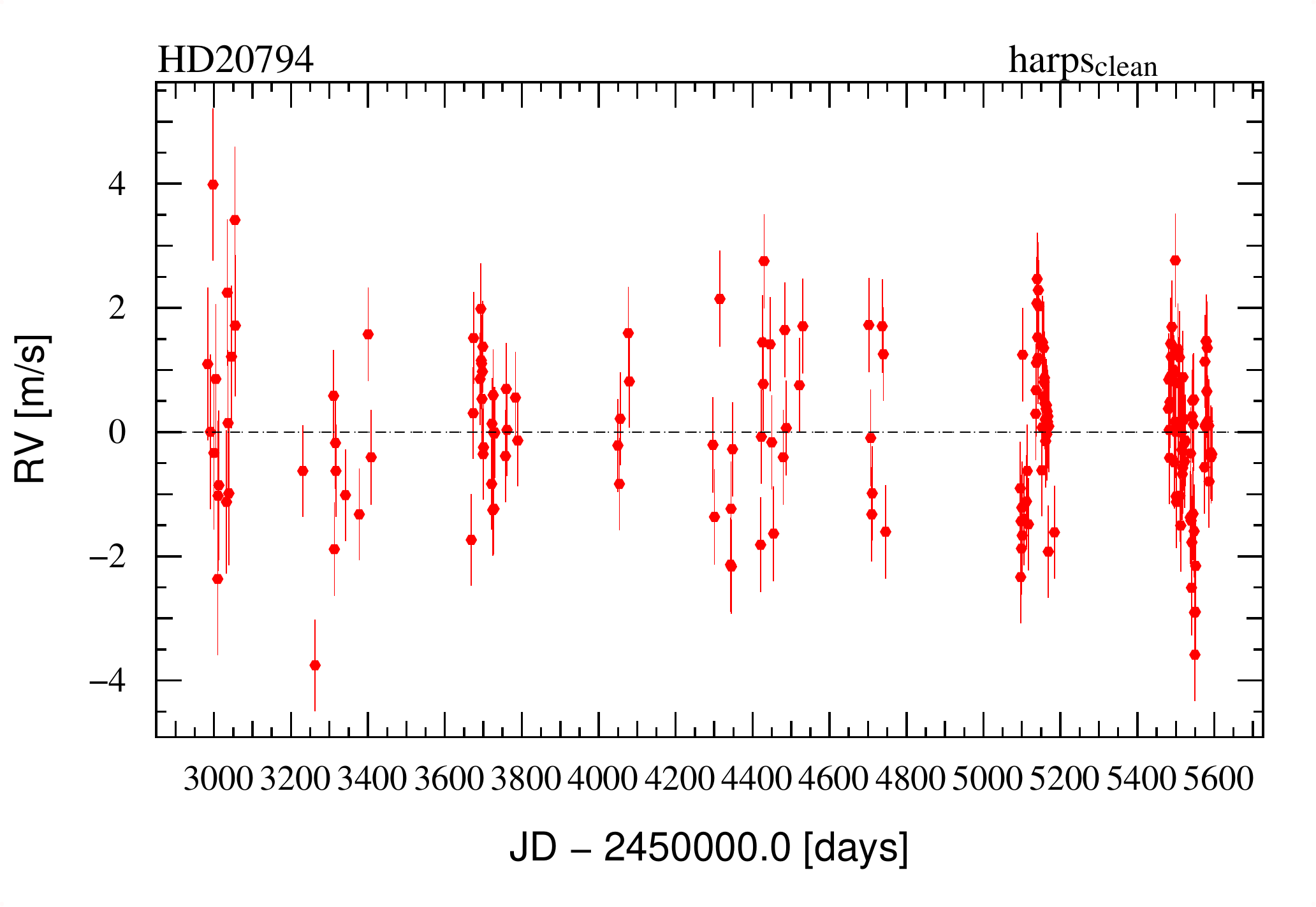}
\includegraphics[bb=0 70 595 390,width=85mm,clip]{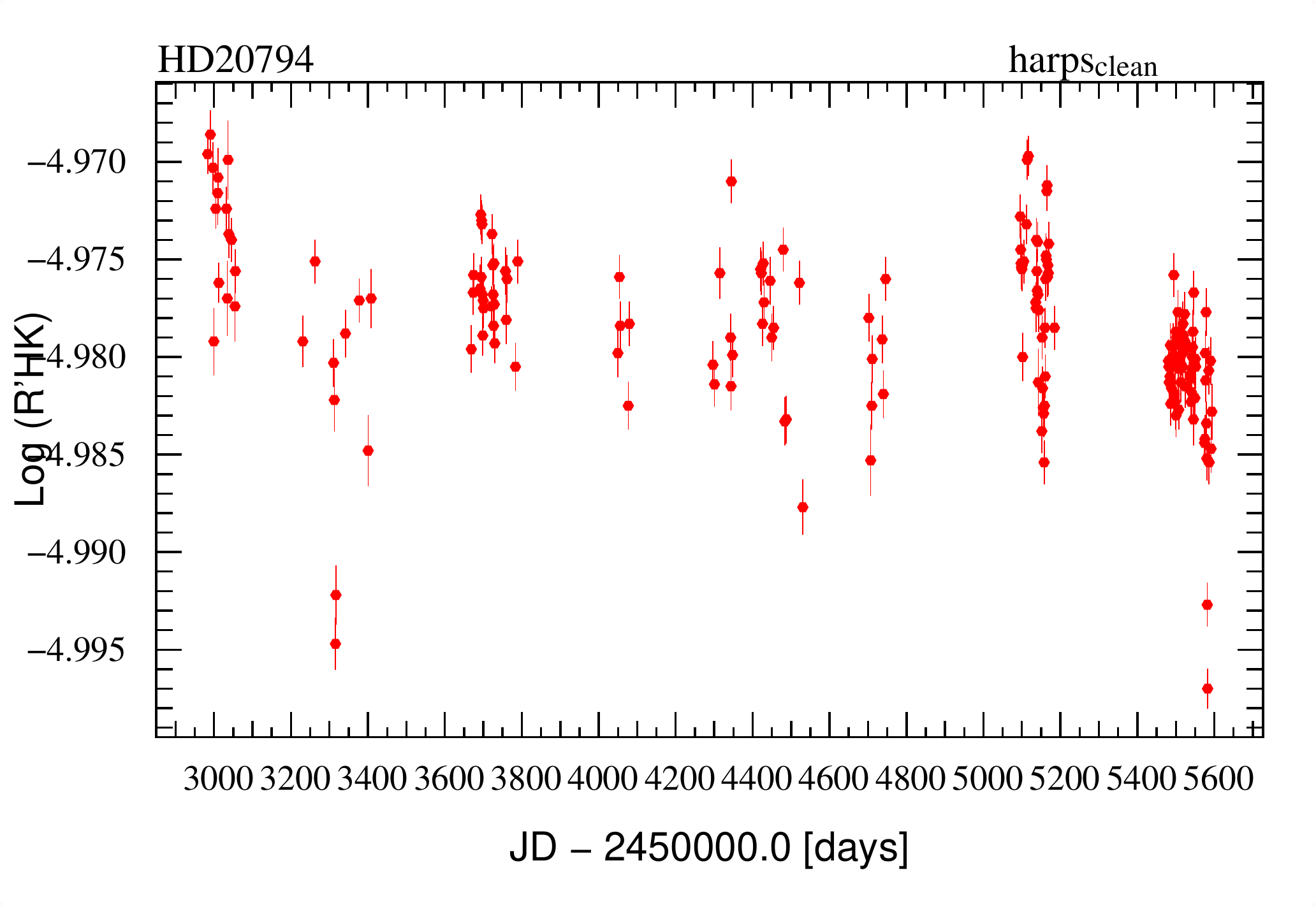}
\includegraphics[bb=0 0 595 390,width=85mm,clip]{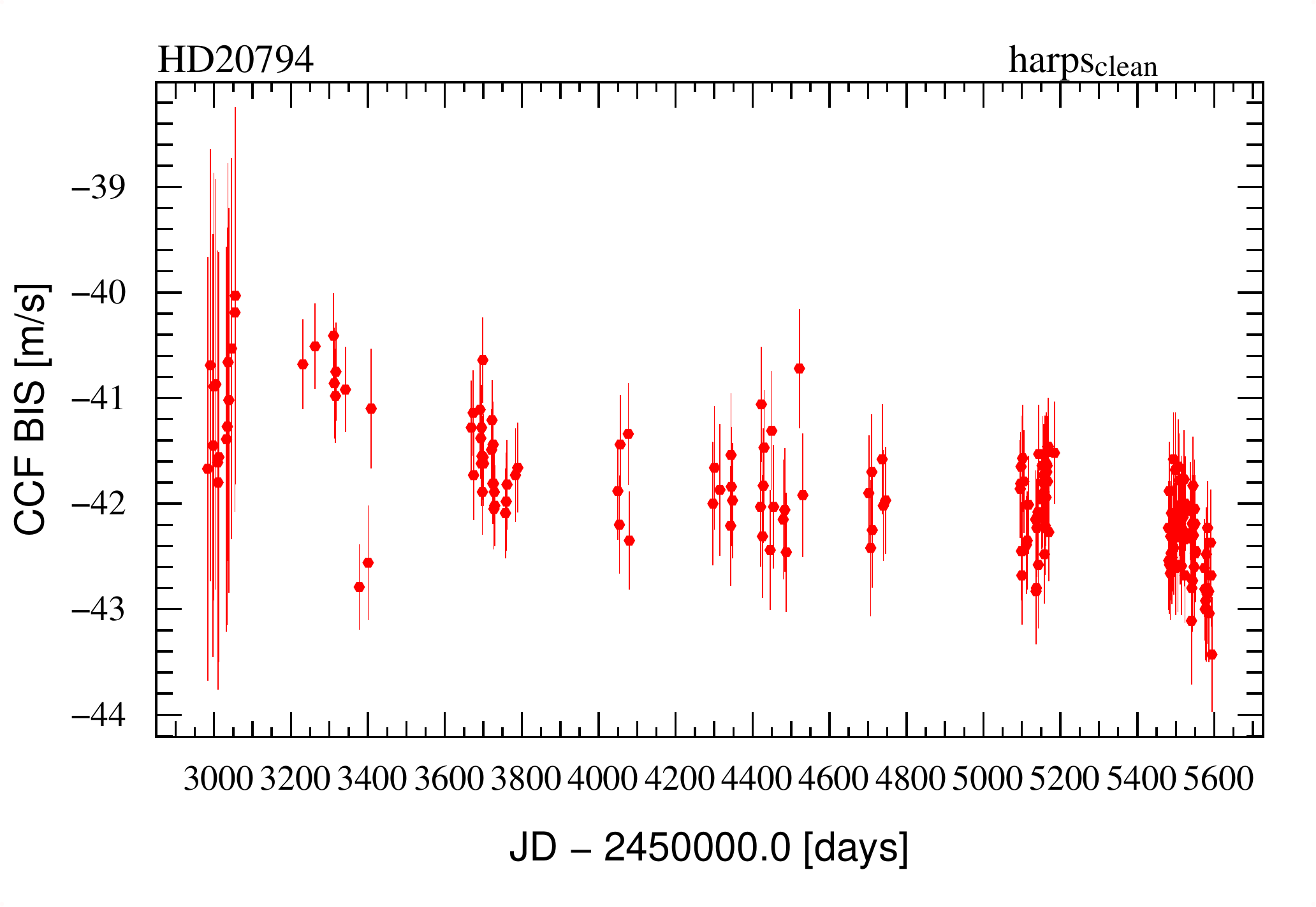}
\centering
\caption{Radial velocity, \rhk and $BIS$ of  \object{HD\,20794}  as a function of time. The RV dispersion is 1.23\ms \emph{rms}.}
\label{fi:hd20794_rv_time}
\end{figure}

The raw radial-velocity data of the star \object{HD\,20794} are shown in Figure\,\ref{fi:hd20794_rv_time}. As can be seen from these raw data, the star belongs to the most stable stars of the original HARPS high-precision sample. The dispersion of the 173 radial-velocity data points is of only 1.2\ms \emph{rms} over the seven-year time span. The typical internal noise on the radial velocity data points, which takes into account photon and calibration noise, is of 0.2 to 0.3\ms. For the analysis, and to take into account any possible external noise such as instrumental, atmospheric and stellar contaminants, as well as stellar jitter, we added quadratically a systematic noise of 0.7\ms to the RV data. This value considers the typical dispersion obtained for the most stable stars of the program and prevents that some high SNR data are erroneously over-weighted with respect to others in the fitting process.

\begin{figure}
\includegraphics[bb=0 47 595 260,width=85mm,clip]{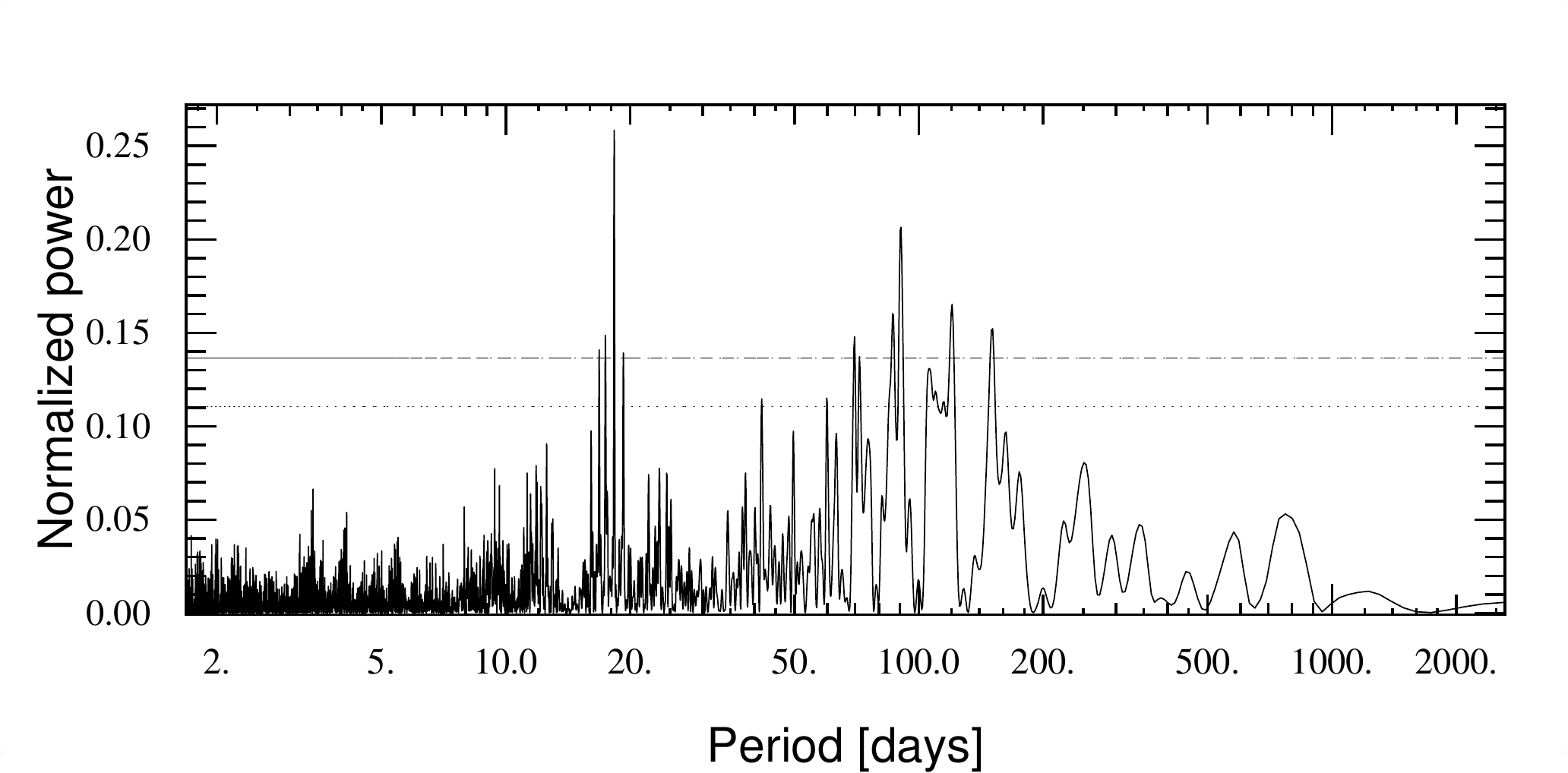}
\includegraphics[bb=0 47 595 260,width=85mm,clip]{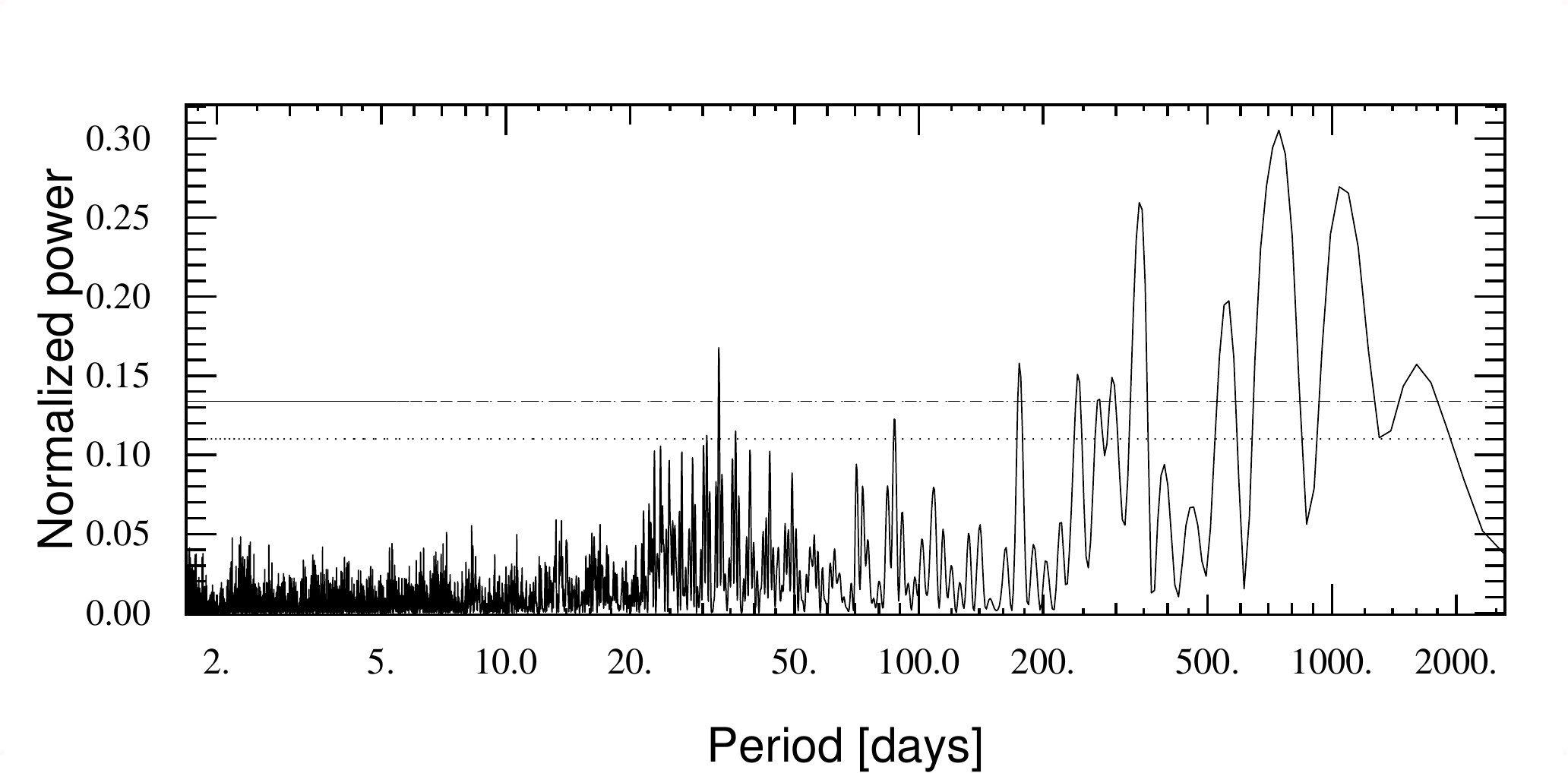}
\includegraphics[bb=0 0 595 260,width=85mm,clip]{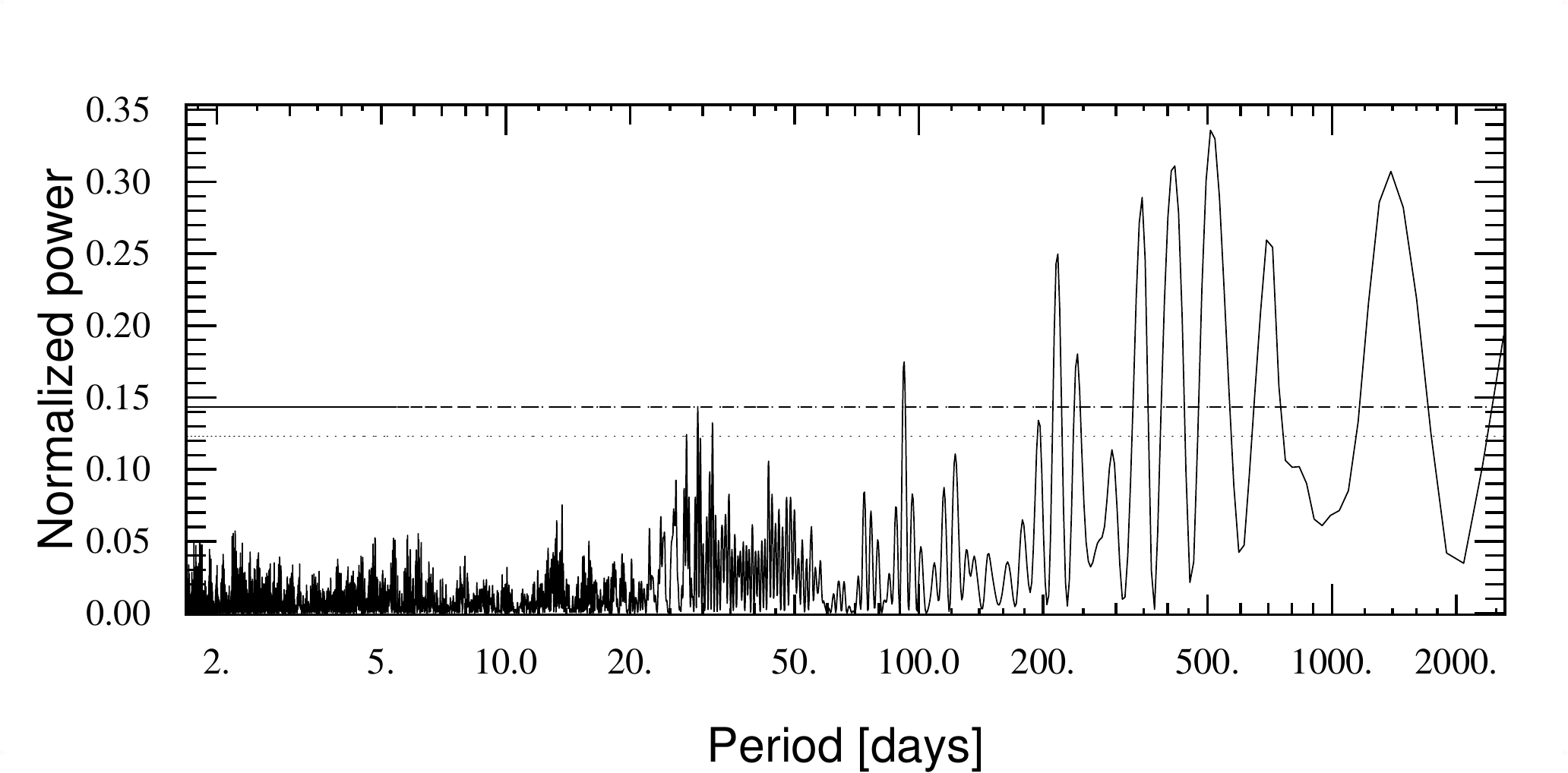}
\centering
\caption{GLS periodogram of the  \object{HD\,20794} radial-velocity data with FAP levels at 10\% and 1\% level, respectively (top), the activity indicator \rhk (center) and the line bisector $BIS$ (bottom). While a very clear peak appears at $P=18$\,days for the radial velocities, no such signal is seen in the two other observables. }
\label{fi:hd20794_rv_gls}
\end{figure}

Although low, the dispersion largely exceeds the estimated total error. The GLS periodogram shows indeed a signal well above the noise level at about 18\,days period (see Figure\,\ref{fi:hd20794_rv_gls}). The dashed line indicates a false-alarm probability level of 1\% but the $P=18$\,days signal exceeds this level by far. In the same figure we plot for comparison the GLS periodogram of the activity indicator $log R'_{HK}$ and of the line bisector $BIS$. None of them reveal any significant power at the period indicated by the radial velocity. The \rhk indicator and the bisector, instead, show an excess of power at periods around 30\,days, which points toward the rotational period of this star. Because the periods are significantly different from the radial-velocity period, we conclude that the two signals are de-correlated. Nevertheless, it must be noted that \rhk has a significant peak at $P=32$\,days, which we suppose reflecting the stellar rotational period. Remarkably, this value perfectly corresponds to the value directly computed from \rhk using the \citet{Mamajek:2008} prescriptions.

\begin{figure}
\centering
\includegraphics[bb=0 47 595 260,width=85mm,clip]{hd20794_rv_gls.pdf}
\includegraphics[bb=0 47 595 260,width=85mm,clip]{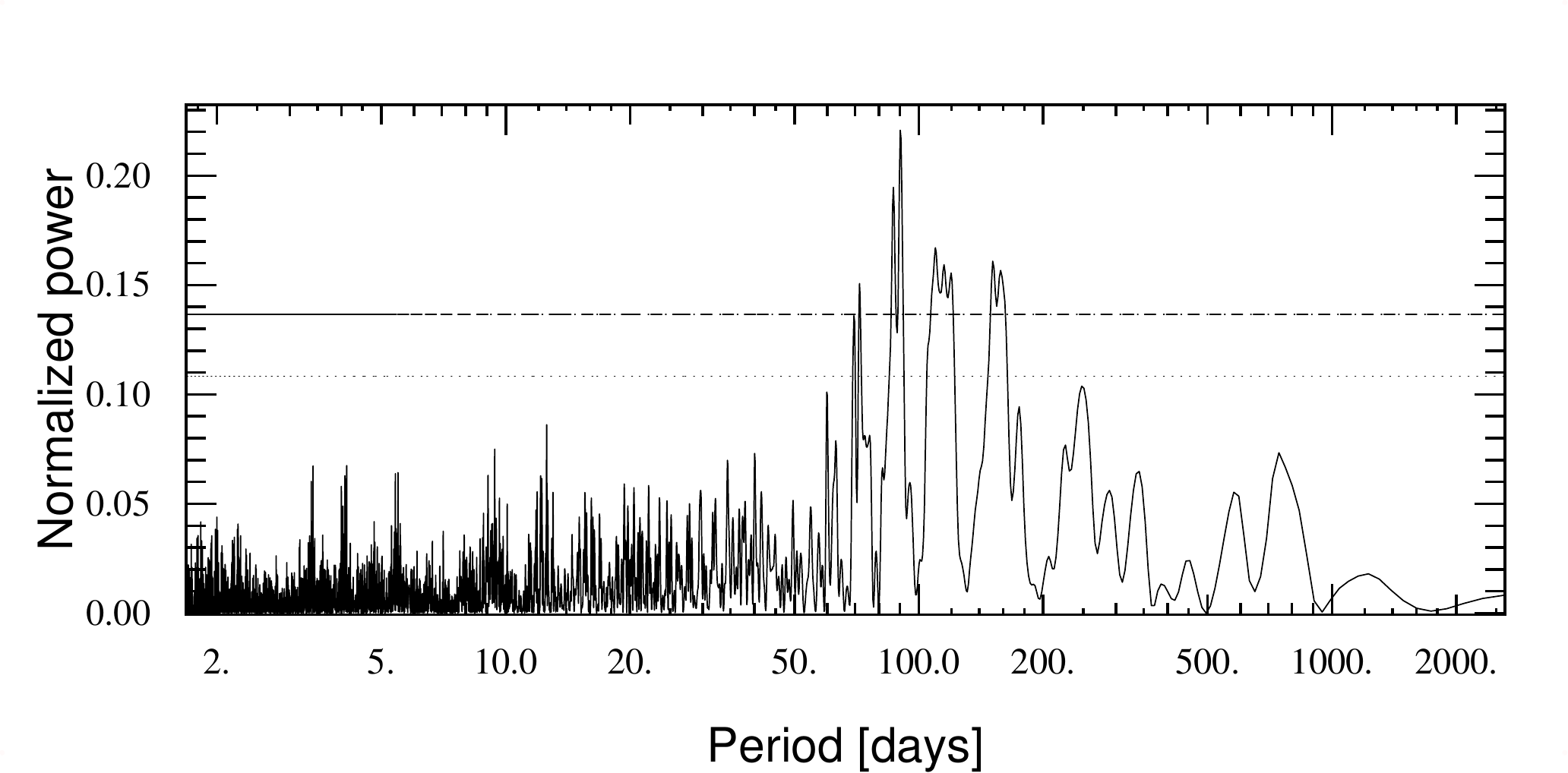}
\includegraphics[bb=0 47 595 260,width=85mm,clip]{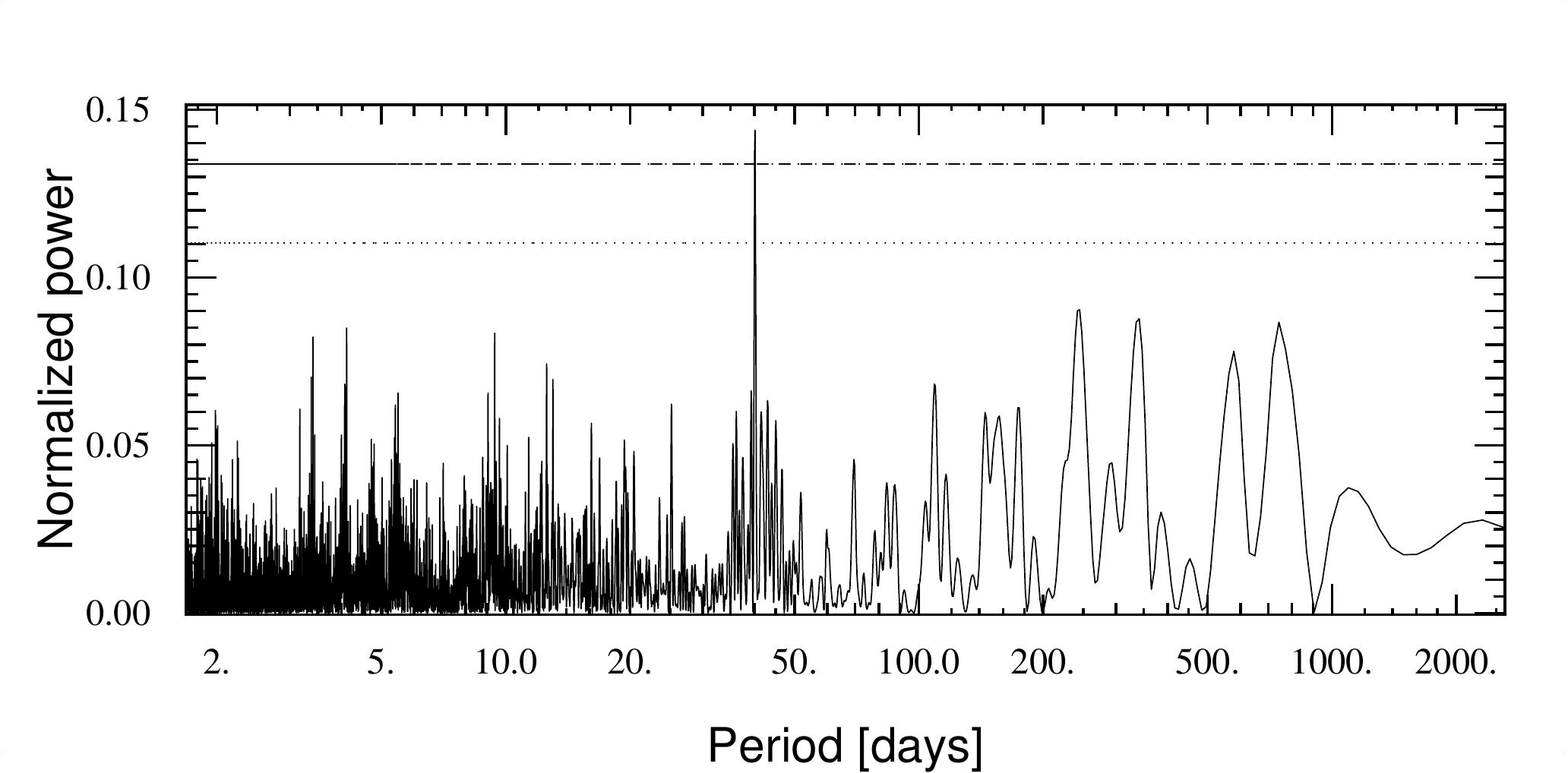}
\includegraphics[bb=0 0 595 260,width=85mm,clip]{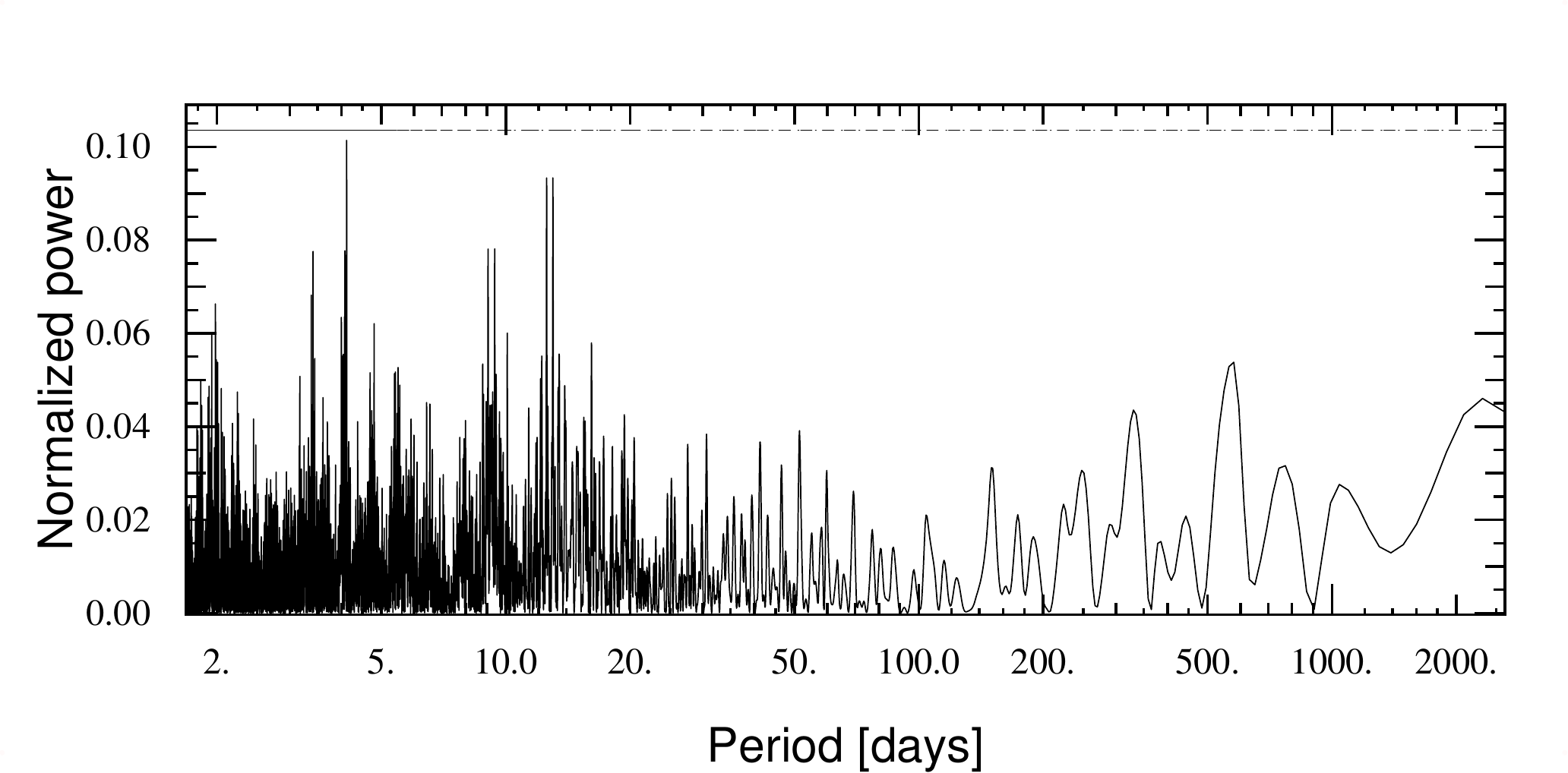}
\caption{Successive GLS periodograms of the \object{HD\,20794} time series, where the main signal is subtracted at each step by adding one more planet in the multi-Keplerian model from top to bottom, starting from the GLS of the raw data and following with the residuals to the one-Keplerian and the two-Keplerian fit. Eccentricities have been set to zero. False-alarm probability thresholds of 10\% and 1\% are indicated. The lowest GLS periodogram of the residuals to the three-Keplerian model shows only the 10\% FAP level.}

\label{fi:hd20794_gls_all}
\end{figure}

Based on the clear $P=18$\,days signal issued from the periodogram, and reassured by the non-coincidence of the \rhk and $BIS$ peaks with the radial-velocity peak, we fitted a single Keplerian to the radial velocities while letting all parameters free. The period of the obtained solution converges very rapidly and unambiguously towards a period of about 18\,days, a semi amplitude around 1\ms, and an eccentricity of 0.25. However, because the eccentricity error is of the same order as its value, and because the residuals almost do not depend on the eccentricity, we have decided to fix it to zero. For the $N=187$ data points we obtain a residual dispersion around the proposed solution of 1.03\ms. The reduced $\chi^2$ is thus of 2.06, indicating that some excess dispersion exists and must be caused by an additional signal or an external noise source, e.g., additional stellar noise, as suggested by the excess power in the \rhk and line-bisector spectra. The GLS periodogram of the residuals to the one-Keplerian fit (center of Figure\,\ref{fi:hd20794_gls_all}) indeed indicates excess power around 90\,days with a false-alarm probability well below 1\%. This signal is not correlated with \rhk or $BIS$ and is confirmed when computing the Pearson's correlation coefficients, which is in both cases close to zero.

\begin{figure}
\includegraphics[bb=0 90 595 510,width=\columnwidth,clip]{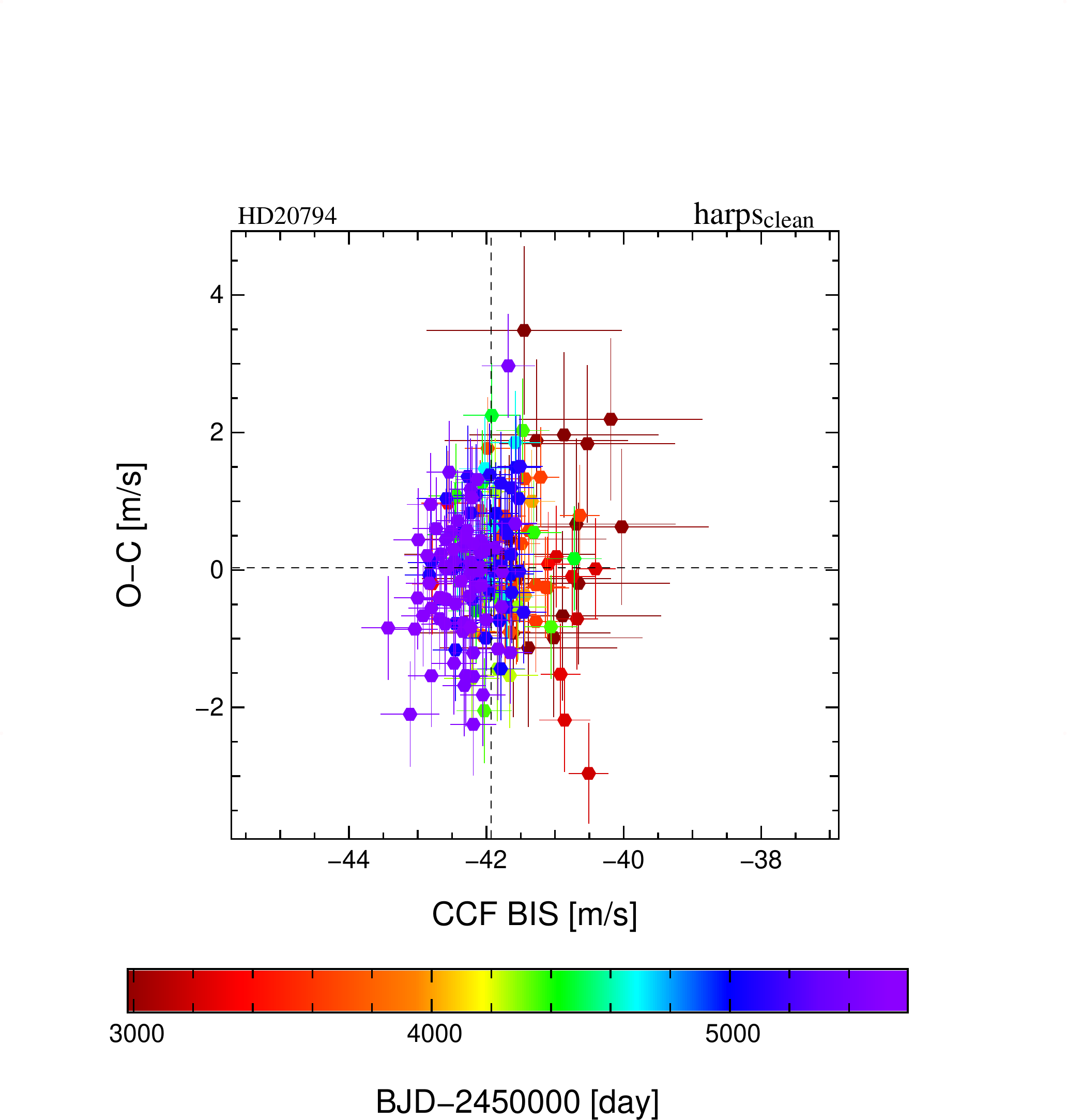}
\includegraphics[bb=0 90 595 510,width=\columnwidth,clip]{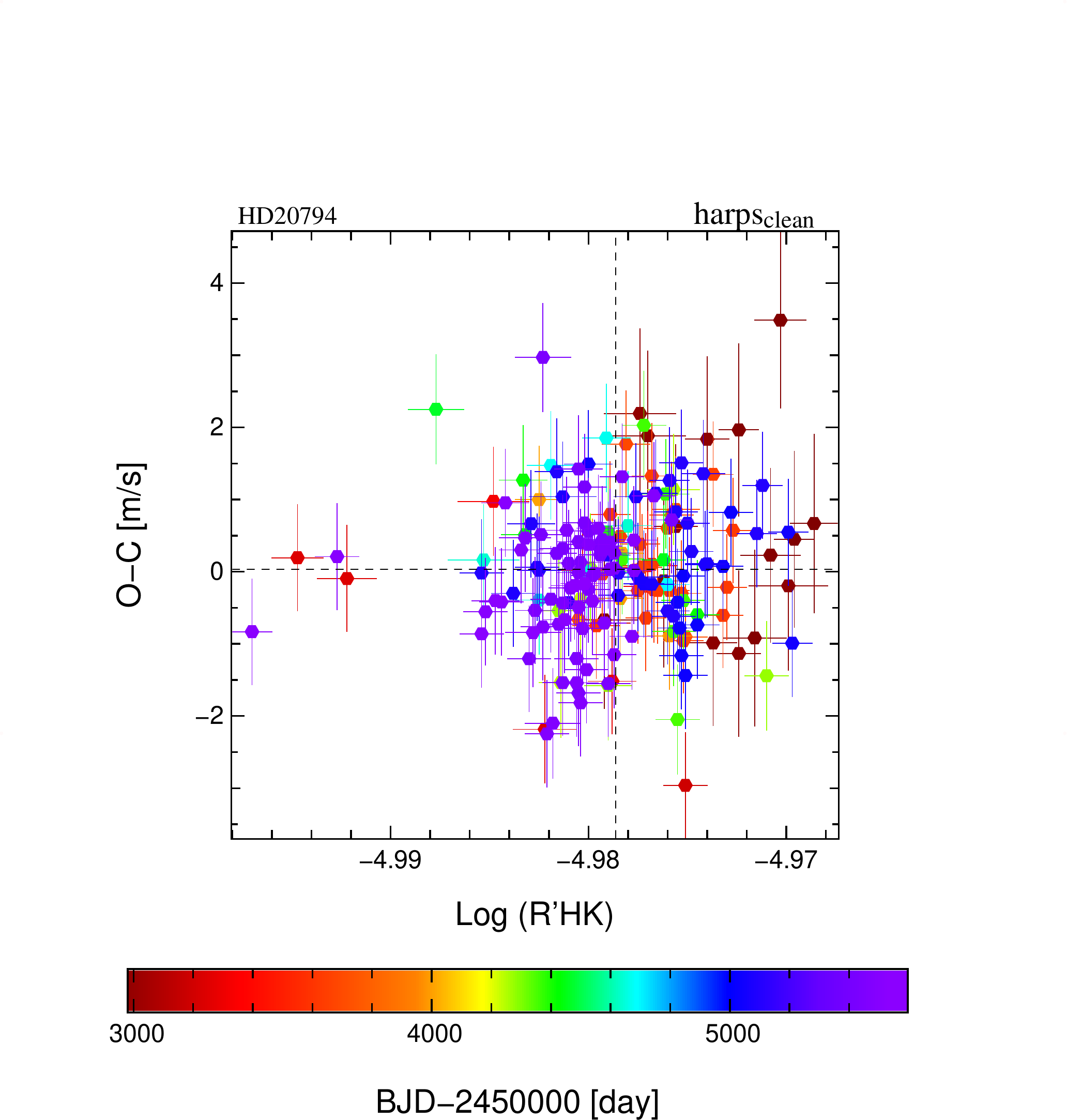}
\centering
\caption{Correlation plot of the residual of the two-Keplerian model to the radial velocity of  \object{HD\,20794} versus the line bisector and the activity indicator \rhk.}
\label{fi:hd20794_corr}
\end{figure}

Encouraged by the significance of the two signals at 18 and 90\,days and the absence of any correlation with the activity indicator and the line bisector, we decided to try a two-Keplerian model to fit our data. For the first run we left all parameters free. The obtained solution is quite convincing but the eccentricities are not significantly different from zero. Therefore, we fixed them to zero in the second run. The two-Keplerian model with zero eccentricities delivers two planets with a period of 18 and 90\,days, both having a semi-amplitude of about 0.8\ms.  The residual dispersion of the data points around the fit is 0.90\ms and  $\chi^2$ is equal to $1.61$. It is interesting to note that the residuals to this solution still contain significant power at $P=40$\,days. This is illustrated by the presence of a significant peak in the third plot of Figure\,\ref{fi:hd20794_gls_all}, which dominates with an FAP of 0.7\%. It must be said, however, that the peak is strongly attenuated if the eccentricities of the two planets are let free, probably because the eccentricity of the two Keplerians 'absorb' part of the 40\,day-period signal. On the other hand, one could argue that a real non-zero eccentricity of the longer-period planet, if not taken into account in the fitted model, would produce harmonics. We would expect a signal around half the longer period, i.e., 45\,days. The period of the peak in the residuals is significantly different, however. We conclude therefore that some additional contribution at $P=40$\,days may still be present in the data, which has not been taken into account by the two-Keplerian model. Furthermore, the 40-day peak in the GLS periodogram does not appear to have a corresponding peaks in either the activity indicator or in the line bisector. This is convincingly confirmed by the correlation plots in Figure\,\ref{fi:hd20794_corr}, which show a Pearson correlation coefficient of only 0.06 between the RV and activity indicator and of 0.15 between the RV and line bisector.
\par
We studied the distribution of the RV residuals to both models (eccentric two-planet and circular three-planet) as a possible way to distinguish between, as done e.g. by \citet{Andrae:2010}. Unfortunately, both models yield distributions of normalized residuals close to Gaussian, and the Kolmogorov-Smirnov test gives similarly high probabilities that the residuals are drawn from a normal distribution (97\% and 80\% for the three-planet and two-planet models, respectively). Note that this result depends on the assumed level of jitter, and that reducing the error bars increases the difference between the models in favor of the three-planet model. However, it is hardly possible to choose between the models on this basis. Gathering more data may be the only way to confirm the existence of the third planet.
 
\begin{figure}
\includegraphics[bb=0 75 595 375,width=70mm,clip]{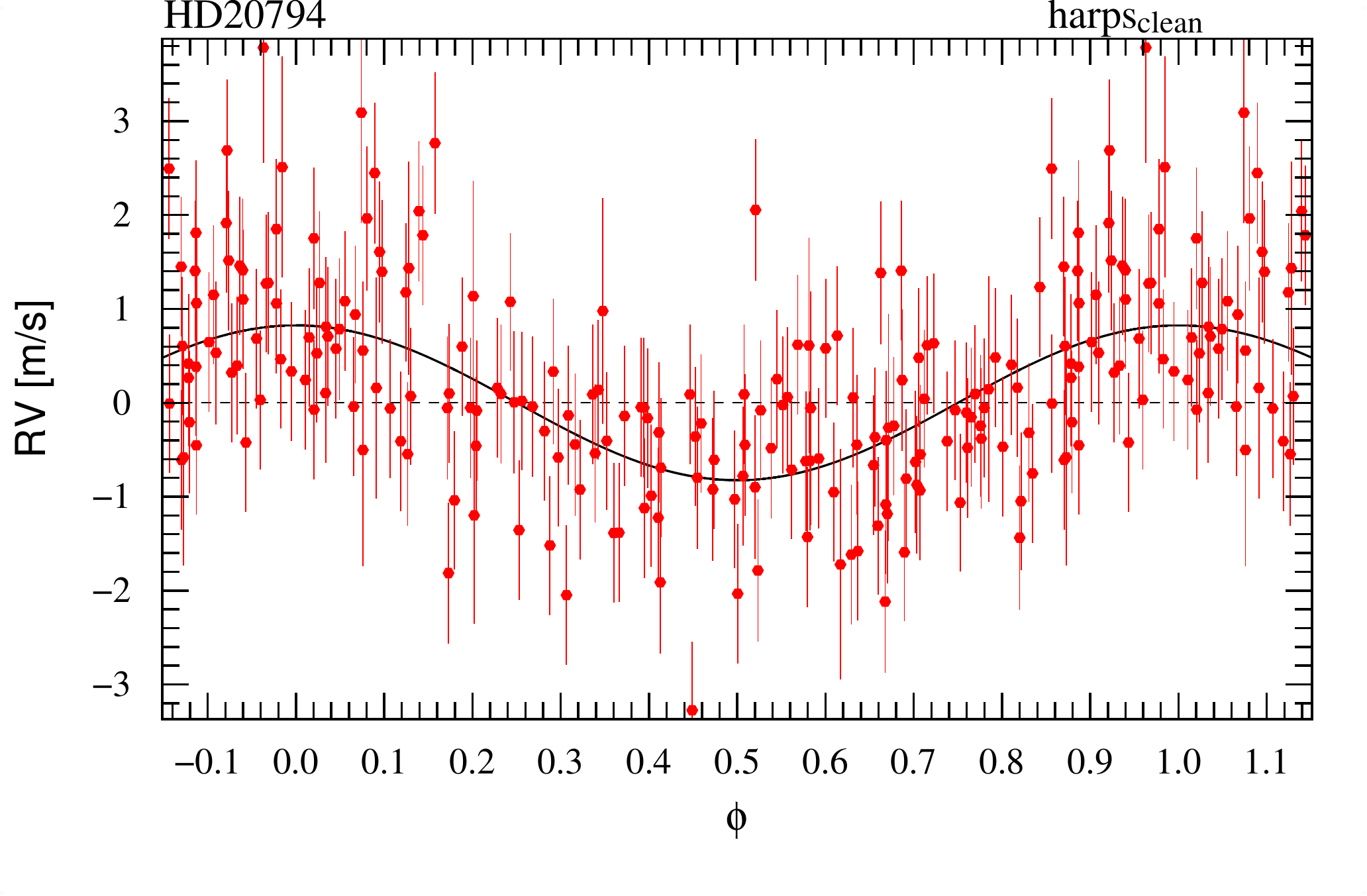}
\includegraphics[bb=0 75 595 375,width=70mm,clip]{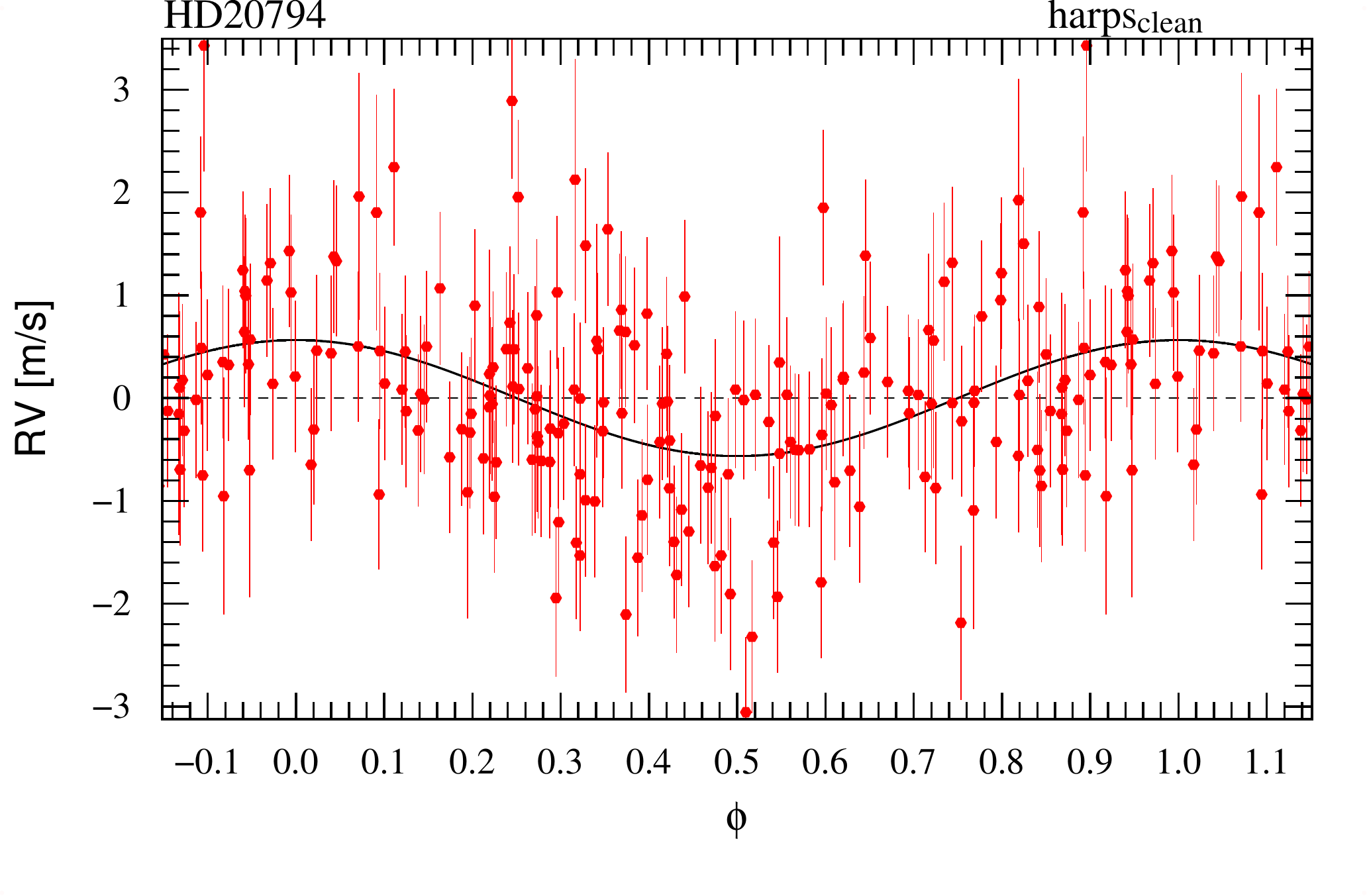}
\includegraphics[bb=0 20 595 375,width=70mm,clip]{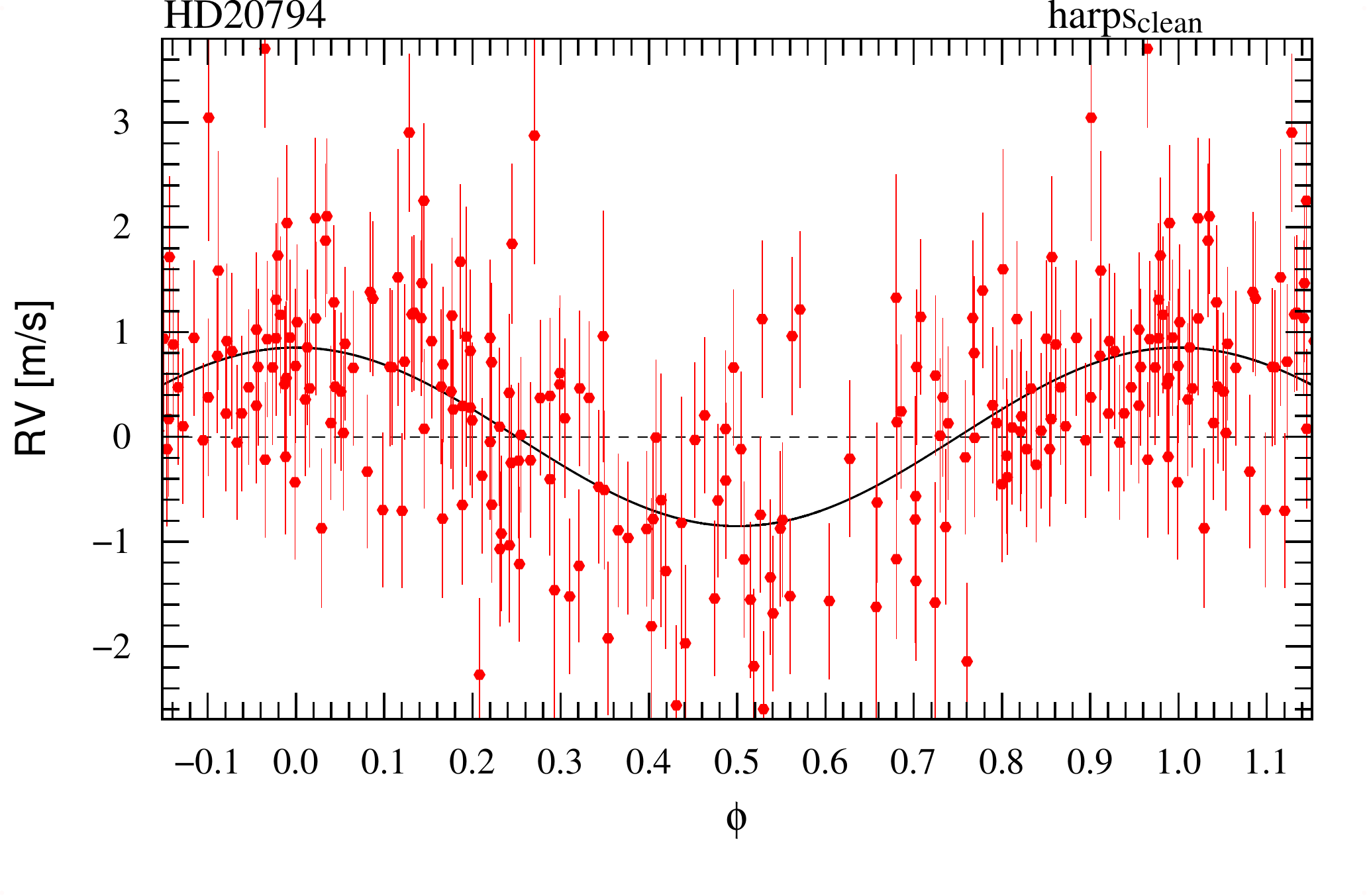}
\centering
\caption{Phase-folded RV data and fitted Keplerian solution for the three planetary components b, c, and d of HD\,20794 with $P=18$\,days, $P=40$\,days, and $P=90$\,days, respectively. The dispersion of the residuals is 0.82\ms \emph{rms}.}
\label{fi:hd20794_k3_phase}
\end{figure}

All these considerations lead us to adopt a three-Keplerian model. The resulting orbital parameters are presented in Table\,\ref{ta:hd20794_k3_par}. The corresponding phase-folded radial-velocity curves are shown in Figure\,\ref{fi:hd20794_k3_phase}.  As mentioned above, the eccentricities were fixed to zero, but only after searching for generic solutions and verifying that the eccentricity values were not significant. The semi-amplitudes of the induced radial-velocity variation are \emph{all} lower than 1\,\ms.  The orbital periods of the three components are at 18, 40, and 90\,days.
\par
The shortest and longest-period signal are very significant, while the $P=40$\,days period is confidently detected in the periodogram but is less apparent in the phase-folded RV plot. Given the low amplitude, which would be the lowest ever detected radial-velocity variation induced by a planet, and because the orbital period is close to the supposed rotational period of the parent star, we remain cautious about this third candidate. In the frame of our program we will continue observing this star to confirm the planetary nature of the c component if possible.

\begin{table*}
\caption{Orbital and physical parameters of the planets orbiting HD\,20794 as obtained from a three-Keplerian fit to the data. Error bars are derived from the covariance matrix. $\lambda$ is the mean longitude ($\lambda$ = $M$ + $\omega$) at the given epoch.}
\label{ta:hd20794_k3_par}
\begin{center}
\begin{tabular}{l l c c c}
\hline \hline
\noalign{\smallskip}
{\bf Parameter}	& {\bf [unit]}		& {\bf HD 20794 b}	& {\bf HD 20794 c}	& {\bf HD 20794 d} \\
\hline 
\noalign{\smallskip}
Epoch		& [BJD]			& \multicolumn{3}{c}{2'454'783.40362208}    \\ 
$i$			& [deg]			& \multicolumn{3}{c}{$ 90 $ (fixed) }  \\  
$V$			& [km\,s$^{-1}$]		& \multicolumn{3}{c}{$ 87.9525\,(\pm 0.0001) $}  \\
\hline 
\noalign{\smallskip}
$P$			& [days]			& $18.315$		& $40.114$		& $90.309$		 \\ 
			&				& $(\pm 0.008)$	& $(\pm 0.053)$	& $(\pm 0.184)$	 \\ 
$\lambda$	& [deg]			& $ 169.0 $		& $ 149.4 $		& $ 16.2$		 \\ 
			&				& $(\pm 6.7) $		& $ (\pm 10.0)  $	& $(\pm 6.8)  $		 \\ 
$e$			&				& $ 0.0 $			& $ 0.0 $			& $ 0.0 $		 \\ 
			&				& (fixed)			& $ (fixed)$		& $(fixed)$	 \\ 
$\omega$		& [deg]			& $  0.0 $			& $ 0.0 $			& $ 0.0 $			 \\ 
			&				& (fixed)			& $(fixed) $		& $(fixed) $	 \\ 
$K$			& [m\,s$^{-1}$]		& $   0.83 $		& $   0.56 $		& $   0.85 $		 \\  
			&				& $(\pm 0.09)   $	& $(\pm 0.10)  $	& $(\pm 0.10)  $	 \\
\hline
\noalign{\smallskip}
$m \sin i$		& [$M_\oplus$]		& $ 2.7$			& $ 2.4 $			& $ 4.8 $		 \\
			&				& $(\pm 0.3)   $		& $(\pm 0.4) $	& $(\pm 0.6)  $	 \\
$a$			& [AU]			& $ 0.1207 $		& $ 0.2036 $		& $ 0.3499 $		 \\
			&				& $(\pm 0.0020)   $	& $(\pm 0.0034)$	& $(\pm 0.0059)  $	 \\
$T_{\rm eq}$	& [K]    			& $ 660 $ 			& $ 508 $ 			& $ 388 $ 			\\
\hline
\noalign{\smallskip}
$N_\mathrm{meas}$ &			& \multicolumn{3}{c}{187}  \\
Span		& [days]			& \multicolumn{3}{c}{2610} \\
rms			& [m\,s$^{-1}$]		& \multicolumn{3}{c}{0.82} \\
$\chi_r^2$	&				& \multicolumn{3}{c}{1.39} \\
\hline
\end{tabular}
\end{center}
\end{table*}

The semi-major axis of the planetary orbits are 0.12\,AU, 0.20\,AU and 0.35\,AU. For the minimum mass of the planets we obtain $m_b \sin i = 2.68\mathrm{[M_{\oplus}]}$, $m_c \sin i = 2.38\mathrm{[M_{\oplus}]}$ and $m_d \sin i = 4.72\mathrm{[M_{\oplus}]}$. We end up with three super-Earths in the same system. The error on the mass owing to only the fit uncertainties is on the order of 10\% and 15\%, respectively, very similar to the error set by the uncertainty on the stellar mass. It is worth noting that this system is of exactly the nature we were looking for, i.e., low-mass (possibly rocky) planets on not too close orbits. However, even for the furthermost planet d, the equilibrium temperature defined by a Bond albedo of 0.3 is on the order of 388\,K (see \citet{Selsis:2007b} for details on the definition and estimation of equilibrium temperature).

\subsection{The planetary system around \object{HD\,85512}}              
%__________________________________________________________________

\begin{figure}
\includegraphics[bb=0 70 595 390,width=85mm,clip]{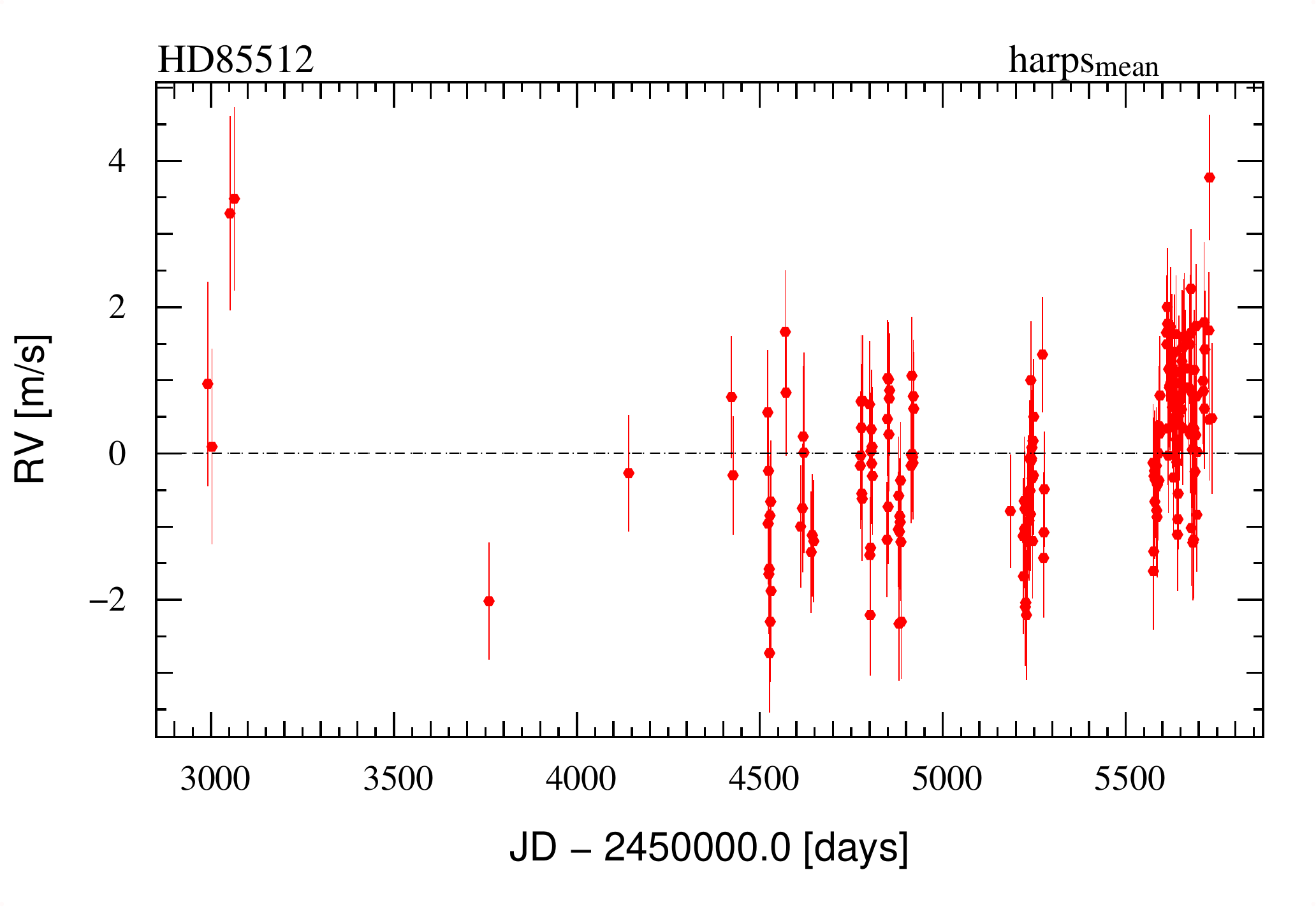}
\includegraphics[bb=0 70 595 390,width=85mm,clip]{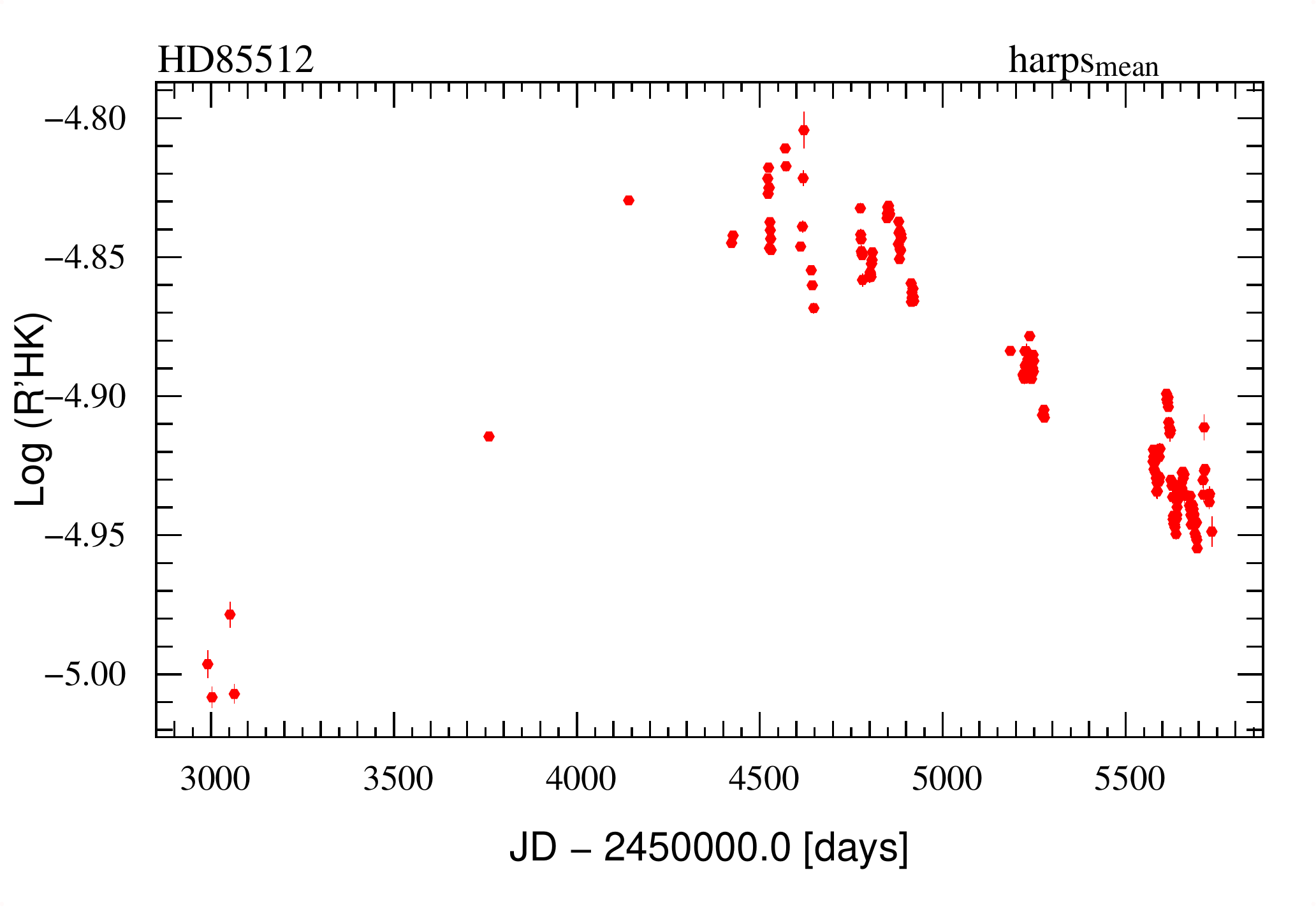}
\includegraphics[bb=0 0 595 390,width=85mm,clip]{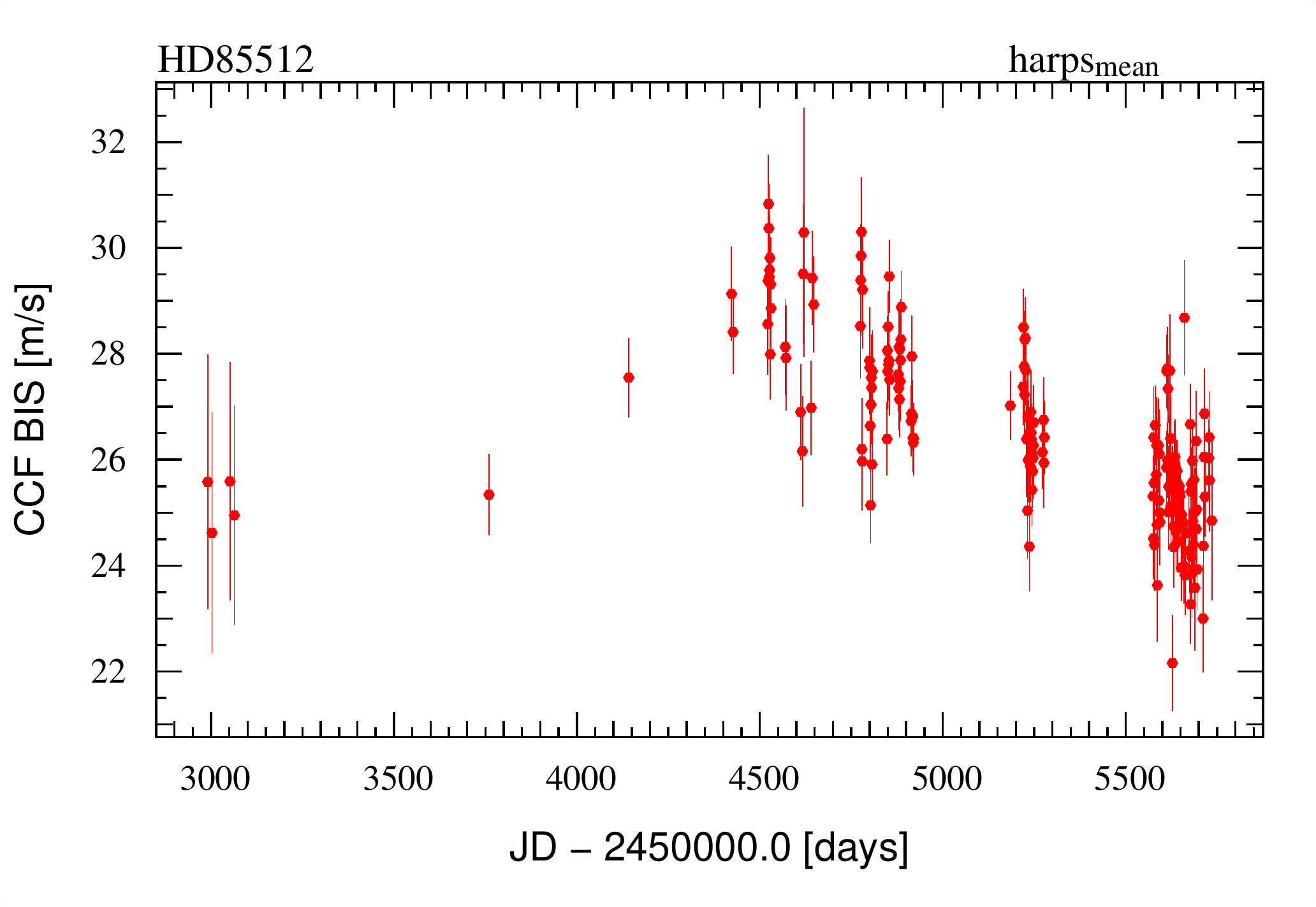}
\centering
\caption{Radial velocity, \rhk and $BIS$ of  \object{HD\,85512}  as a function of time. The dispersion of the 185 data points is only 1.05\ms \emph{rms}.}
\label{fi:hd85512_rv_time}
\end{figure}

The raw radial-velocity data of the star \object{HD\,85512} are shown in Figure\,\ref{fi:hd85512_rv_time}. By initial choice this star is among the most stable stars of the original HARPS high-precision sample. The dispersion of the 185 radial-velocity data points is, with 1.05\ms \emph{rms} over the 7.5-years time span, below that of  \object{HD\,20794}. The typical internal error on the radial velocity data points is of 0.3\ms, to which 0.7\ms have been added quadratically to take into account possible instrumental and stellar noise.

\begin{figure}
\includegraphics[bb=0 47 595 260,width=85mm,clip]{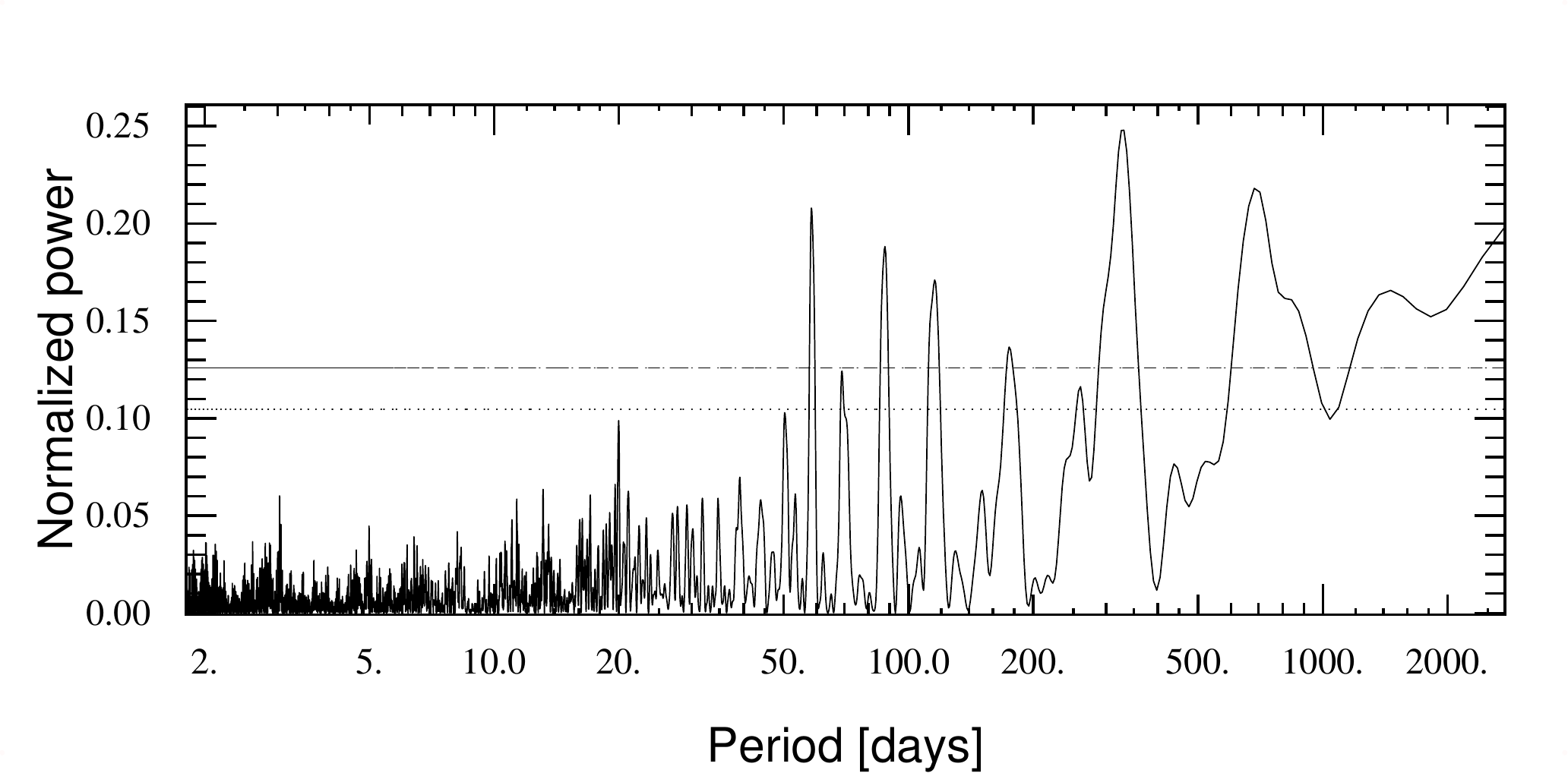}
\includegraphics[bb=0 47 595 260,width=85mm,clip]{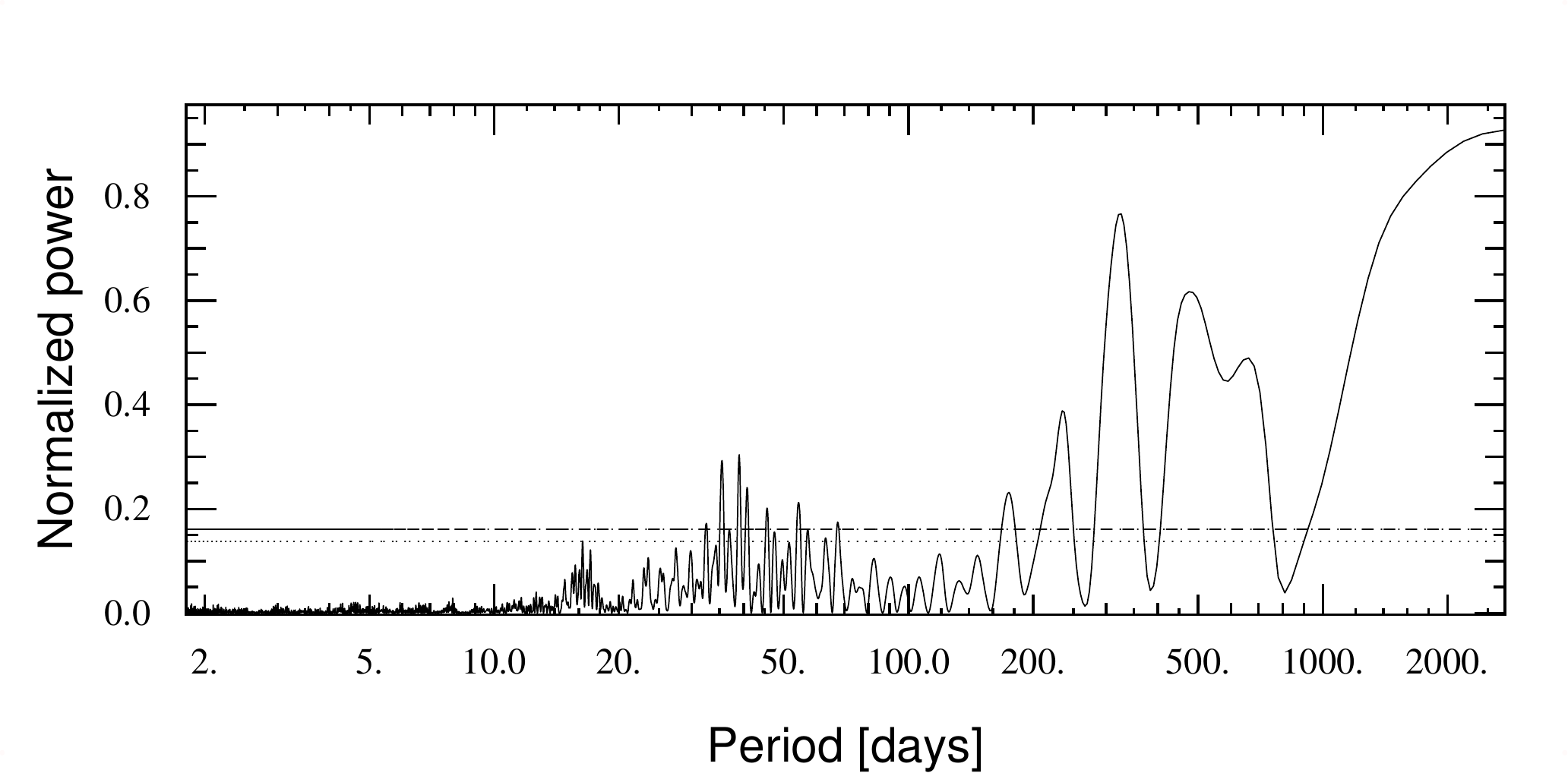}
\includegraphics[bb=0 0 595 260,width=85mm,clip]{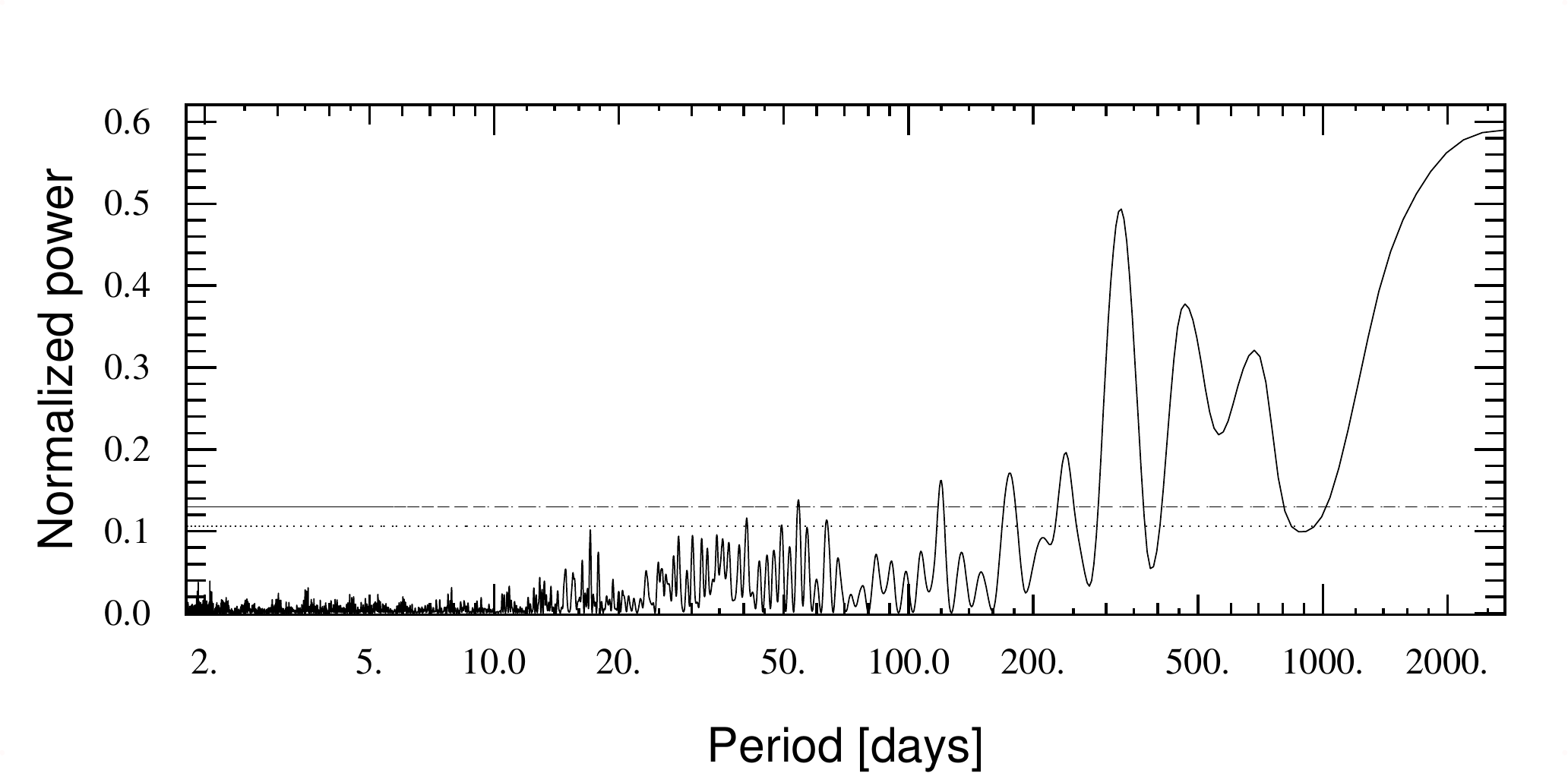}
\centering
\caption{GLS periodogram of the \object{HD\,85512} radial-velocity data with FAP levels at 10\% and 1\% level, respectively (top), the de-trended activity indicator \rhk (center) and the line bisector (bottom). }
\label{fi:hd85512_rv_gls}
\end{figure}

The GLS periodogram shows several peaks well above noise at periods around 60 - 70 \,days (see Figure\,\ref{fi:hd85512_rv_gls}). The dashed and the dotted lines indicate a false-alarm probability of 1\% and 10\%, respectively. The mentioned peaks clearly exceed this level. The three highest peaks are found at 58, 69, and 85\,days period, which are the yearly alias of each other. The peak at 58 days is the strongest of the series, however. Furthermore, there is significant excess power around 300\,days period at above. In the same figure we plot for comparison the GLS periodogram of the activity indicator $log R'_{HK}$ and of the line bisector $BIS$. The former does not show any excess power at periods between 50 and 100\,days. Instead, the excess power observed at periods above 300\,days in the radial velocities seems to follow the same pattern. The reason for this pattern can be directly seen in the time domain (see Figure\,\ref{fi:hd85512_rv_time}). In both $log R'_{HK}$ and line bisector $BIS$ a long-term trend is observed, which most probably reflects the magnetic cycle of the star. A similar trend, although with opposite sign and smaller amplitude, is observed in the radial-velocity data. From this we deduce that the power observed in the radial velocity at long periods is most probably caused by the stellar activity. We refer to \citet{Lovis:2011c} for a general discussion, in which the specific case of HD\,85512 is treated as well.

\begin{figure}
\includegraphics[width=\columnwidth]{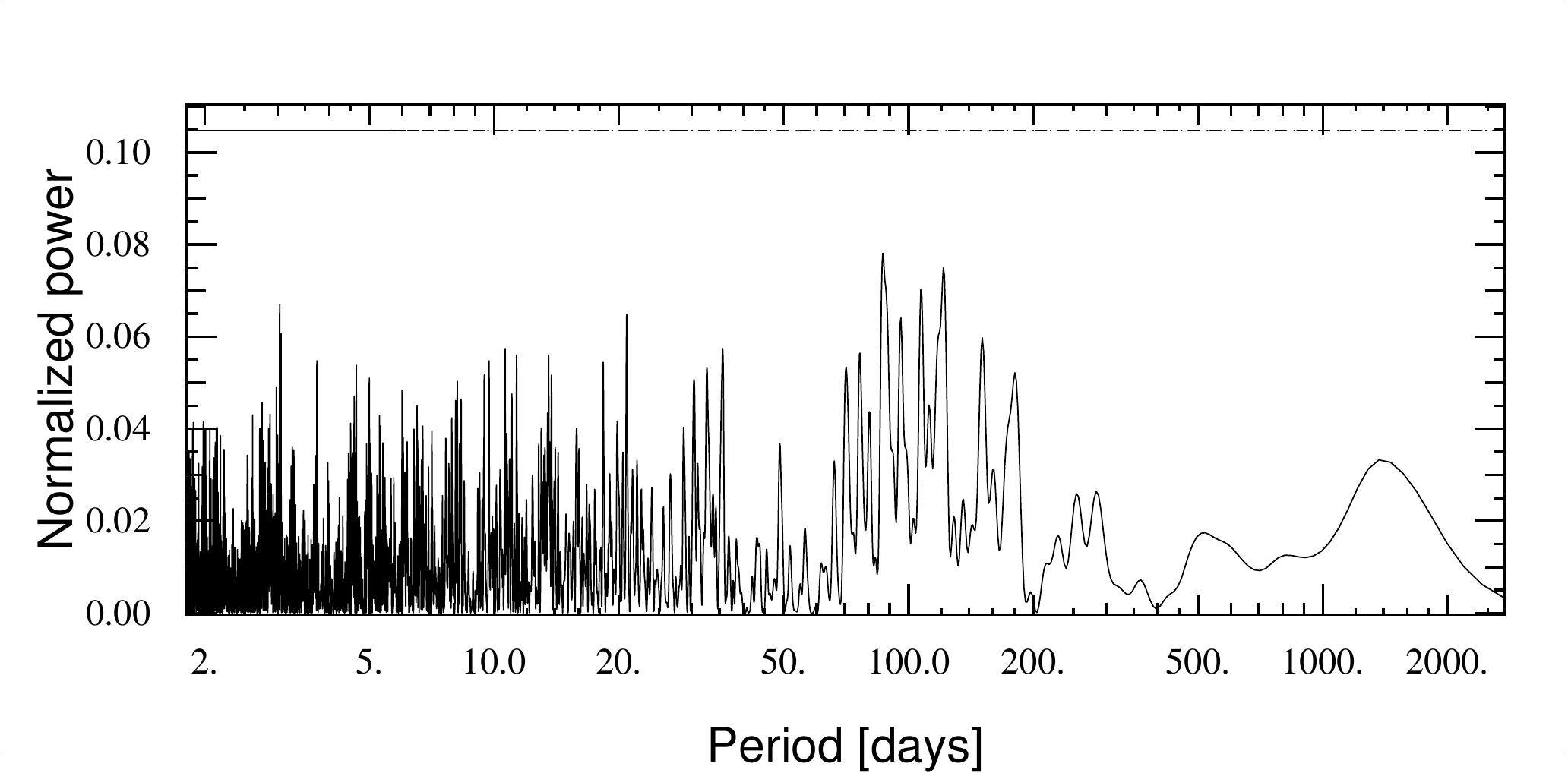}
\centering
\caption{GLS  periodogram of the residuals to the $P=58$\,days best Keplerian fit to the \object{HD\,85512} RV data.}
\label{fi:hd85512_omc_gls}
\end{figure}

Given the absence of corresponding peaks in the activity indicator and the line bisector, we assume that the radial velocity signals at periods around 60\,days are induced by a planetary companion. We therefore fitted a single Keplerian to the radial-velocity data, but also added a second-degree polynomial to take into account the long-term variation induced by the stellar activity. The solution converges very rapidly and unambiguously towards a 58.4\,days orbit. The activity signal is fitted well by a linear component of 0.5\ms per year and a quadratic component of 0.14\ms per year$^2$. The GLS periodogram of the residuals is very clean, as shown in Figure\,\ref{fi:hd85512_omc_gls}, because none of the peaks reaches the 10\% false-alarm probability indicated by the dashed line. It must be noted that all original side peaks have disappeared and that our assumption that they are aliases must be correct.

\begin{figure}
\includegraphics[width=\columnwidth]{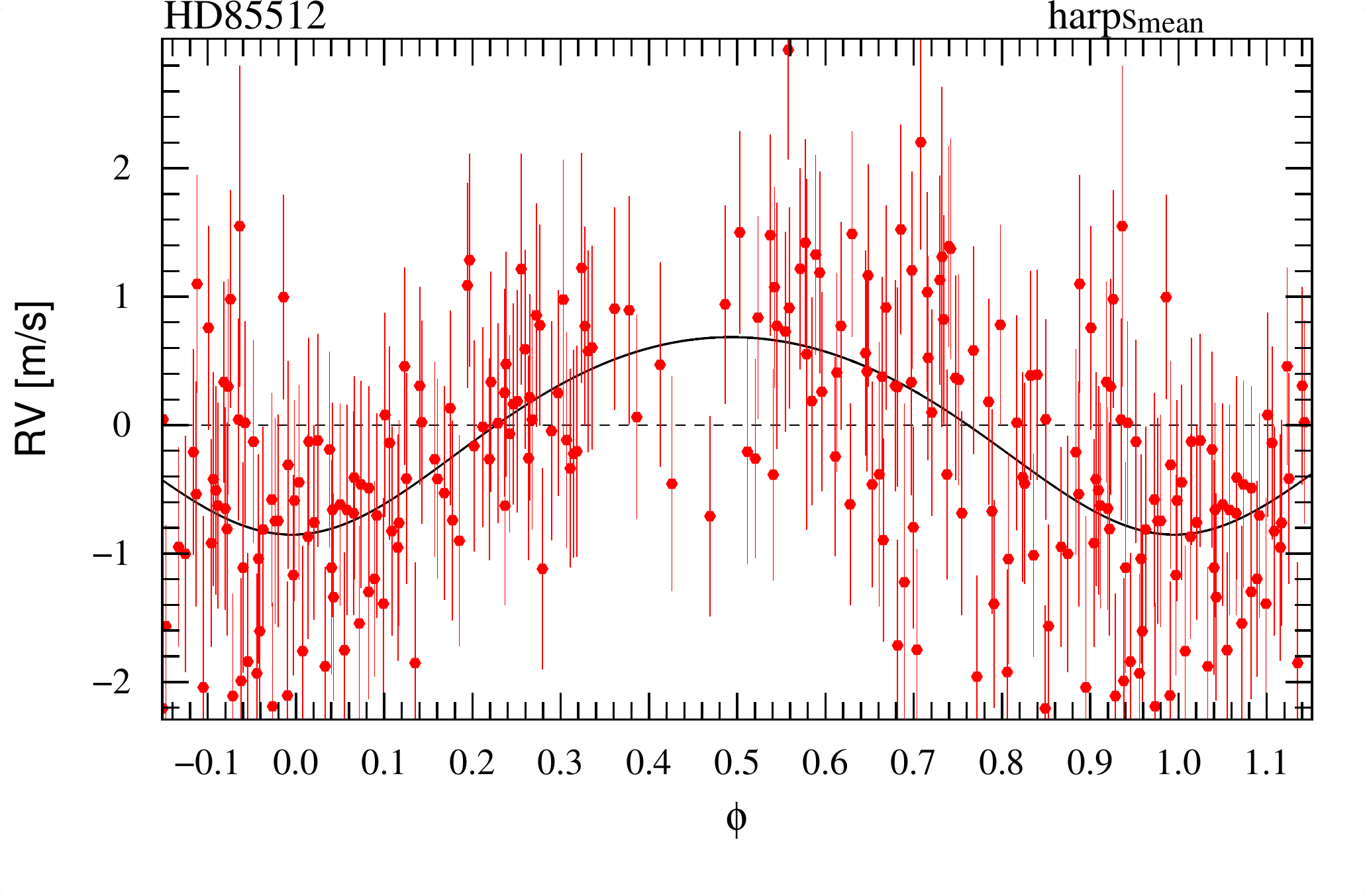}
\centering
\caption{Phase-folded RV data of  \object{HD\,85512} and fitted Keplerian solution. The dispersion of the residuals is 0.75\ms \emph{rms}.}
\label{fi:hd85512_k1d2_phase}
\end{figure}

The eventually obtained orbital parameters of HD\,85512\,b are given in Table\,\ref{ta:hd85512_k1d2_par} and the corresponding phase-folded radial-velocity curve is shown in Figure\,\ref{fi:hd85512_k1d2_phase}. The adopted period is $P=58.4$\,days. The eccentricity is clearly not significant, but given the large number of data points and the good convergence of the fit, we preferred to leave it as free parameter. The obtained value is of 0.11, but we recall that setting it to zero would not change the quality of the solution at all. For the semi-amplitude we obtain the value of 0.77\ms. The residual dispersion of the $N=185$ data points around the fitted solution is only 0.77\ms. The reduced $\chi^2$ is thus close to 1, indicating that the 0.7\ms systematic error we added quadratically to the error bars may be a good estimate.
\par
Using the parameters determined above we can set the semi-major axis of the planetary orbit to 0.26\,AU. For the minimum mass of the planet we obtain $m \sin i = 3.6\mathrm{[M_{\oplus}]}$. The error on the mass due to only the fit uncertainties is on the order of 15\%, despite the very low semi-amplitude. As in the case of the planets around HD\,20794, this planet is possibly a super-Earth, i.e, a planet consisting to high degree of rocky material. Opposite to HD\,20794, however, its parent star is much less luminous. The computed equilibrium temperature is only 298\,K if we again assume an albedo of 0.3. This sets this possibly rocky planet at the inner edge of the habitable zone of its parent star.

\begin{table}
\caption{Orbital and physical parameters of the planets orbiting \object{HD\,85512} as obtained from a Keplerian fit to the data. Error bars are derived from the covariance matrix. $\lambda$ is the mean longitude ($\lambda$ = $M$ + $\omega$) at the given epoch.}
\label{ta:hd85512_k1d2_par}
\begin{center}
\begin{tabular}{l l c}
\hline \hline
\noalign{\smallskip}
{\bf Parameter}	& {\bf [unit]}		& {\bf HD 85512 b}	\\
\hline 
\noalign{\smallskip}
Epoch		& [BJD]			& \multicolumn{1}{c}{2'455'239.70888680}    \\ 
$i$			& [deg]			& \multicolumn{1}{c}{$ 90 $ (fixed) }  \\  
$V$			& [km\,s$^{-1}$]		& \multicolumn{1}{c}{$ -9.4913\,(\pm 0.0038) $}  \\
linear		& [m\,s$^{-1}/$yr]		& \multicolumn{1}{c}{$ +0.5085\,(\pm 0.0003) $} \\
quadratic		& [m\,s$^{-1}/$yr$^2$]	& \multicolumn{1}{c}{$ +0.1423\,(\pm 0.0002) $} \\
\hline 
\noalign{\smallskip}
$P$			& [days]			& $  58.43$		 \\ 
			&				& $(\pm 0.13)$		 \\ 
$\lambda$	& [deg]			& $ 114.5 $		 \\ 
			&				& $(\pm 6.3) $		 \\ 
$e$			&				& $ 0.11 $			 \\ 
			&				& $ (\pm 0.10) $	 \\ 
$\omega$		& [deg]			& $ 178 $			 \\ 
			&				& $ (\pm 0.58)	 $	 \\ 
$K$			& [m\,s$^{-1}$]		& $   0.769 $		 \\  
			&				& $(\pm 0.090)   $	 \\
\hline
\noalign{\smallskip}
$m \sin i$		& [$M_\oplus$]		& $ 3.6$			 \\
			&				& $(\pm 0.5)   $	 \\
$a$			& [AU]			& $ 0.26 $			 \\
			&				& $(\pm 0.005)   $	 \\
$T_{\rm eq}$	& [K]    			& $298$ 			\\
\hline
\noalign{\smallskip}
$N_\mathrm{meas}$ &			& \multicolumn{1}{c}{185}  \\
Span		& [days]			& \multicolumn{1}{c}{2745} \\
rms			& [m\,s$^{-1}$]		& \multicolumn{1}{c}{0.77} \\
$\chi_r^2$	&				& \multicolumn{1}{c}{1.00} \\
\hline
\end{tabular}
\end{center}
\end{table}

\begin{figure}
\includegraphics[bb=0 47 595 260,width=85mm,clip]{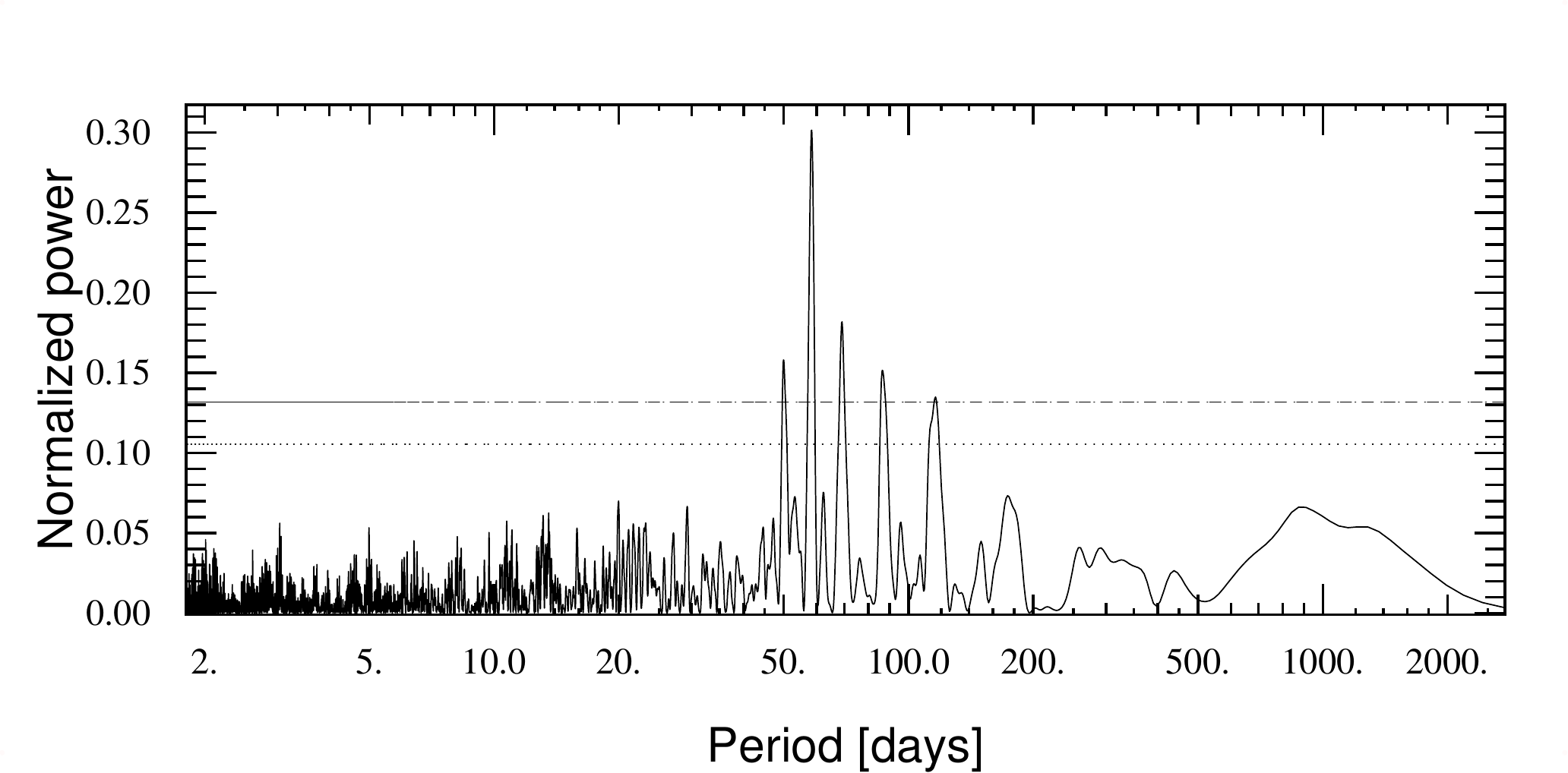}
\includegraphics[bb=0 0 595 260,width=85mm,clip]{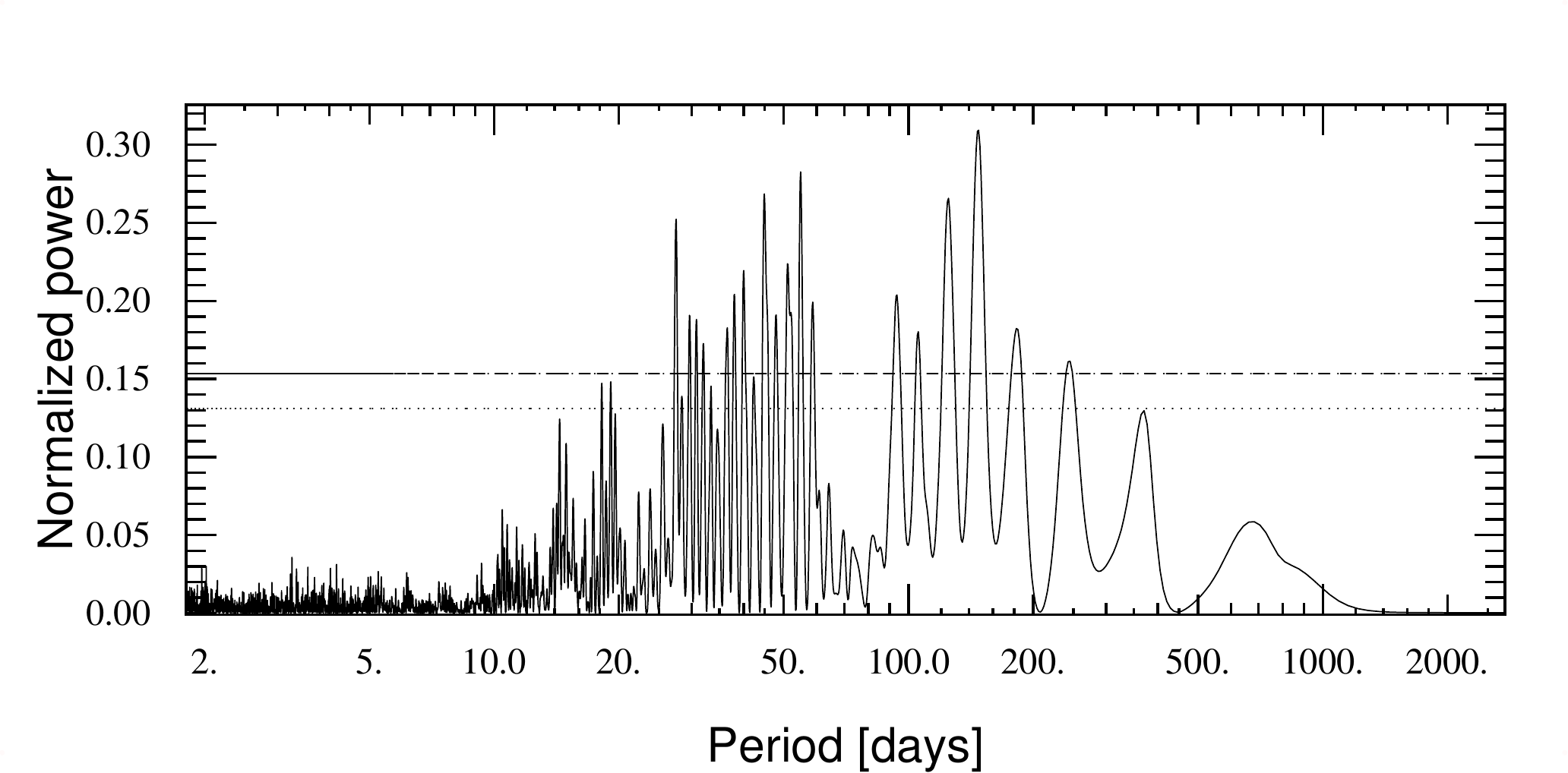}
\centering
\caption{Top: GLS periodogram of the de-trended radial velocities of HD\,85512. Bottom: GLS periodogram of the detrended activity indicator \rhk.}
\label{fi:hd85512_gls_clean}
\end{figure}

To illustrate the cleanliness of the planetary radial-velocity signal and its poor correlation with the activity indicator, we detrended both series from the respective long-period signals. The results are best represented by the GLS periodogram in Figure\,\ref{fi:hd85512_gls_clean}. The top panel shows the radial-velocity signal, which appears even more clearly after detrending. The peak at 58 days is by far the strongest and the aliases show up at the expected periods with decreasing amplitude the farther away from the main peak they are. No such 'coherent' signal is observed in the \rhk at the mentioned periods, although significant power is still present at various (shorter) periods. This excess power is also seen in the time series of \rhk, but the signal's amplitude, frequency and its phase is varying as a function of the observation epoch. We conclude that this signal is the expression of stellar rotation coupled with appearing and disappearing spots and plages, but that it cannot explain the presence of a a strong and coherent radial-velocity. This latter is therefore most likely caused by a planetary companion.

\subsection{The planetary system around \object{HD\,192310}}     
%__________________________________________________________________

\begin{figure}
\includegraphics[bb=0 70 595 390,width=85mm,clip]{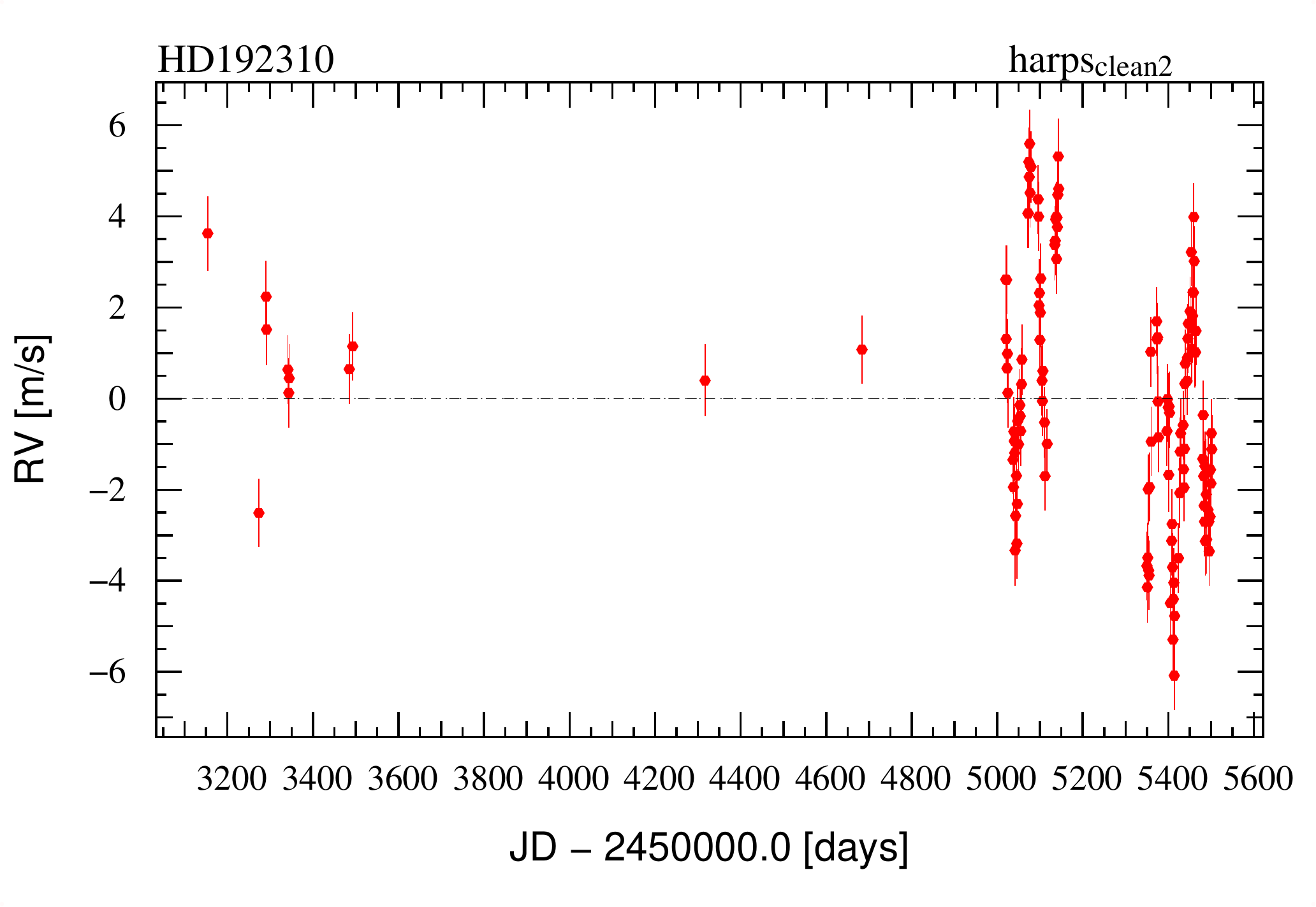}
\includegraphics[bb=0 70 595 390,width=85mm,clip]{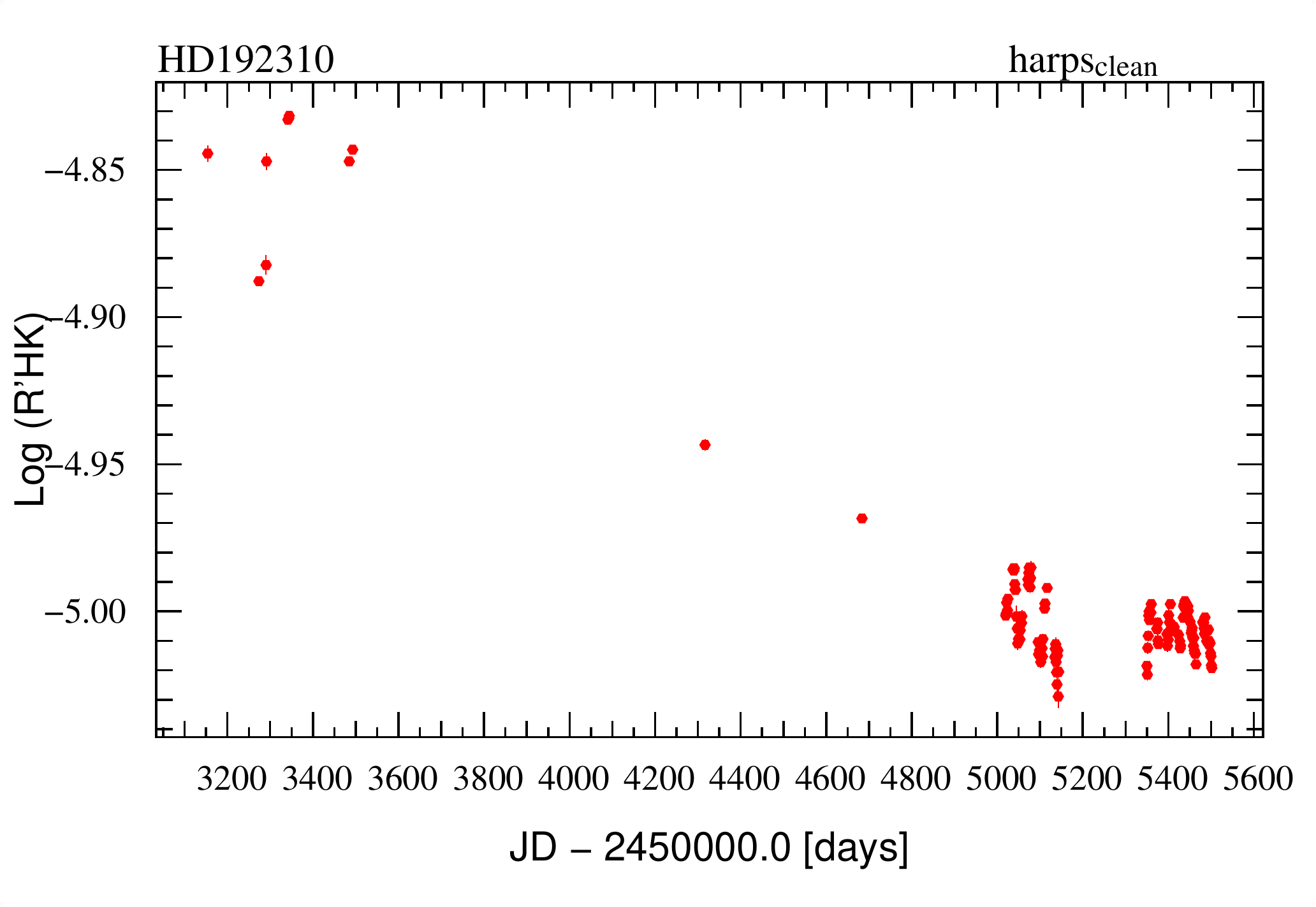}
\includegraphics[bb=0 0 595 390,width=85mm,clip]{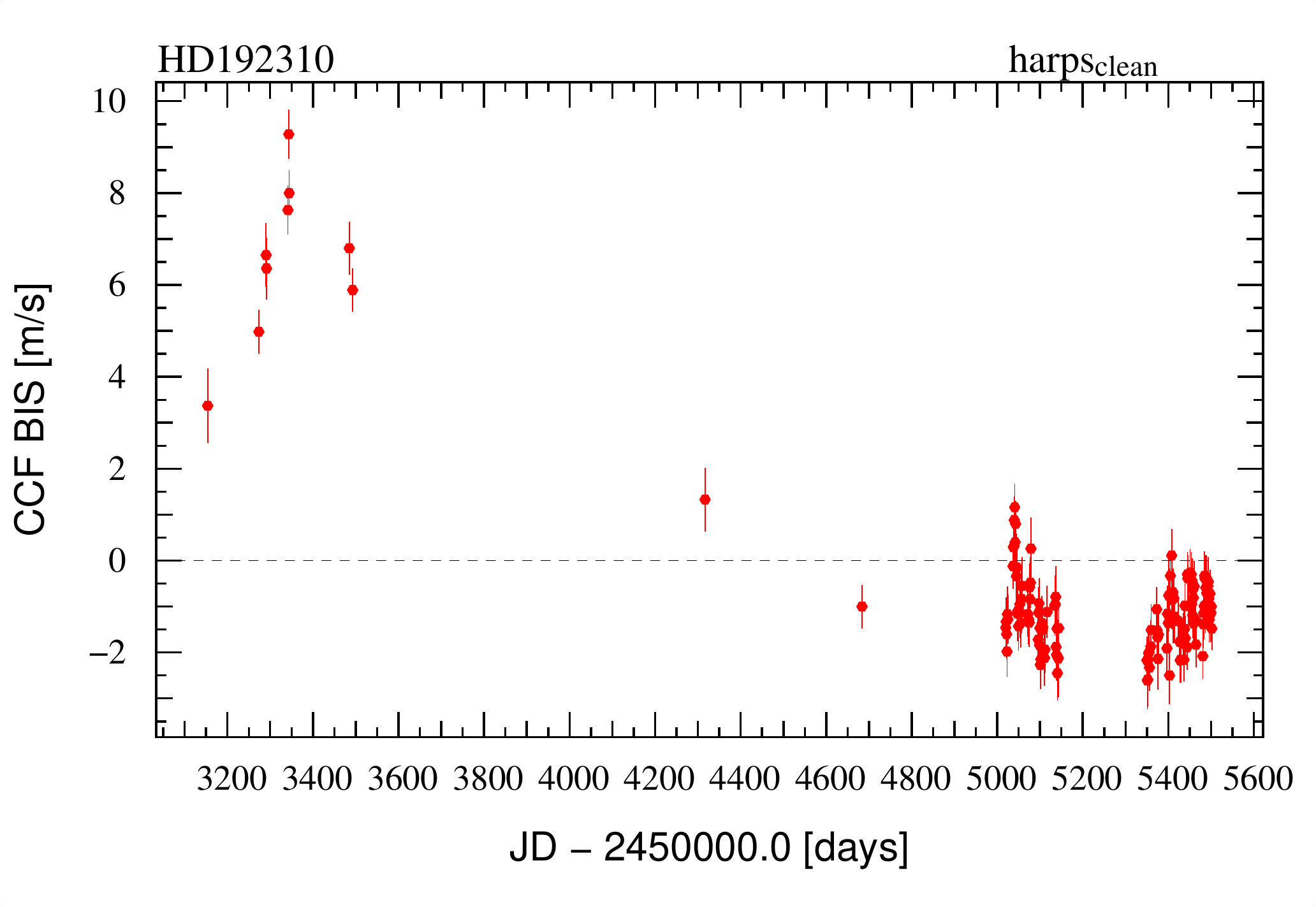}
\centering
\caption{Radial velocity, \rhk and $BIS$ of  \object{HD\,192310}  as a function of time. The dispersion of the raw data is 2.6\ms \emph{rms}.}
\label{fi:hd192310_rv_time}
\end{figure}

Recently, \citet{Howard:2010} have announced the discovery of a Neptune-mass planet around  \object{HD\,192310}. Because it is part of the original HARPS GTO program and of our program, we have been observing this star for several seasons. The raw radial-velocity data of \object{HD\,192310} are shown in Figure\,\ref{fi:hd192310_rv_time}. What strikes one immediately is that, contrary to the previous objects presented in this paper, the dispersion of the 139 radial-velocity data points is 2.6\ms \emph{rms} over the 6.5-year time span. The typical photon noise on the radial velocity data points is 0.25\ms, to which, for the analysis, we added quadratically a noise of 0.7\ms to account for possible instrumental and stellar noise.

\begin{figure}
\includegraphics[bb=0 47 595 260,width=85mm,clip]{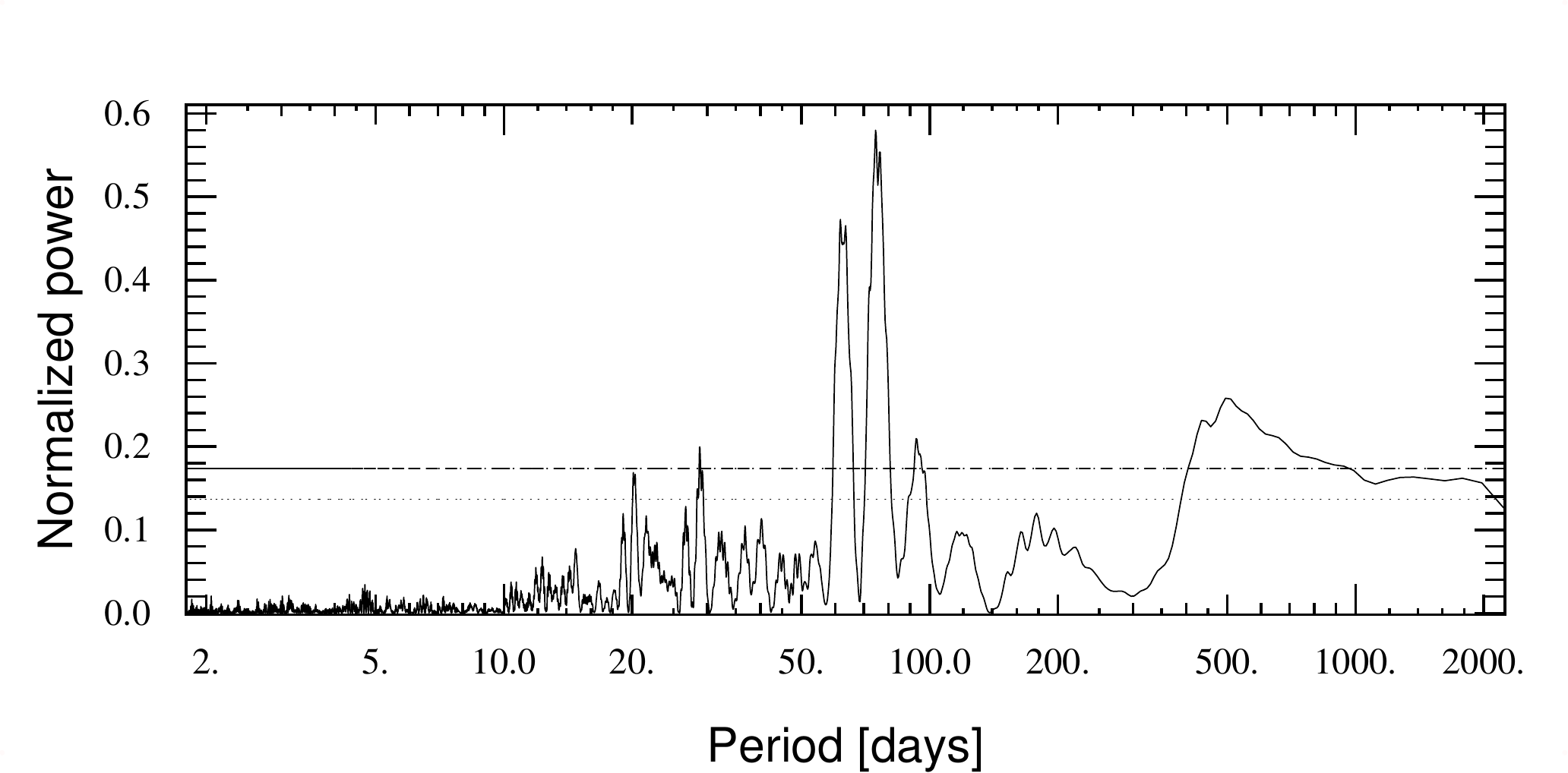}
\includegraphics[bb=0 47 595 260,width=85mm,clip]{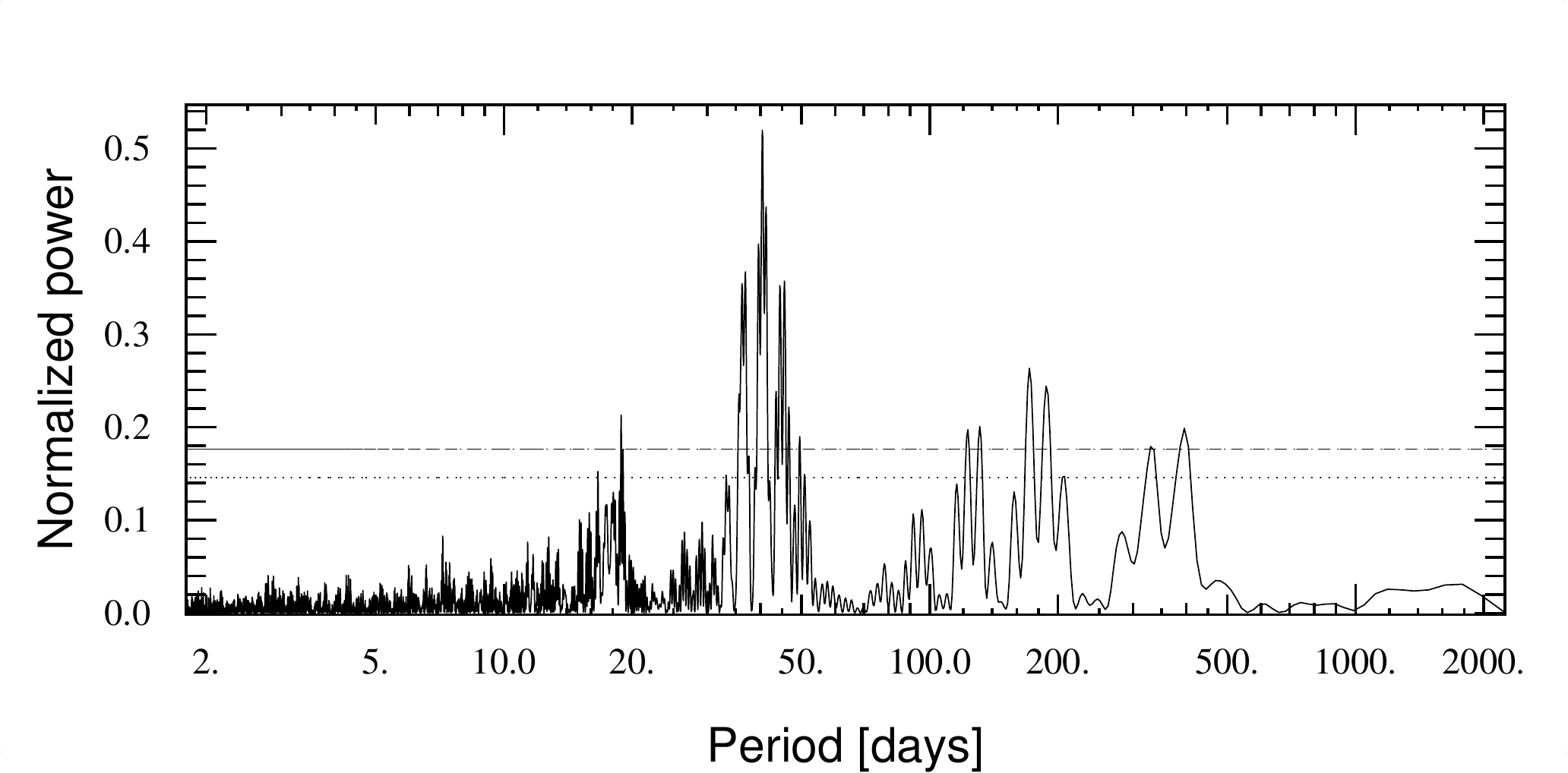}
\includegraphics[bb=0 0 595 260,width=85mm,clip]{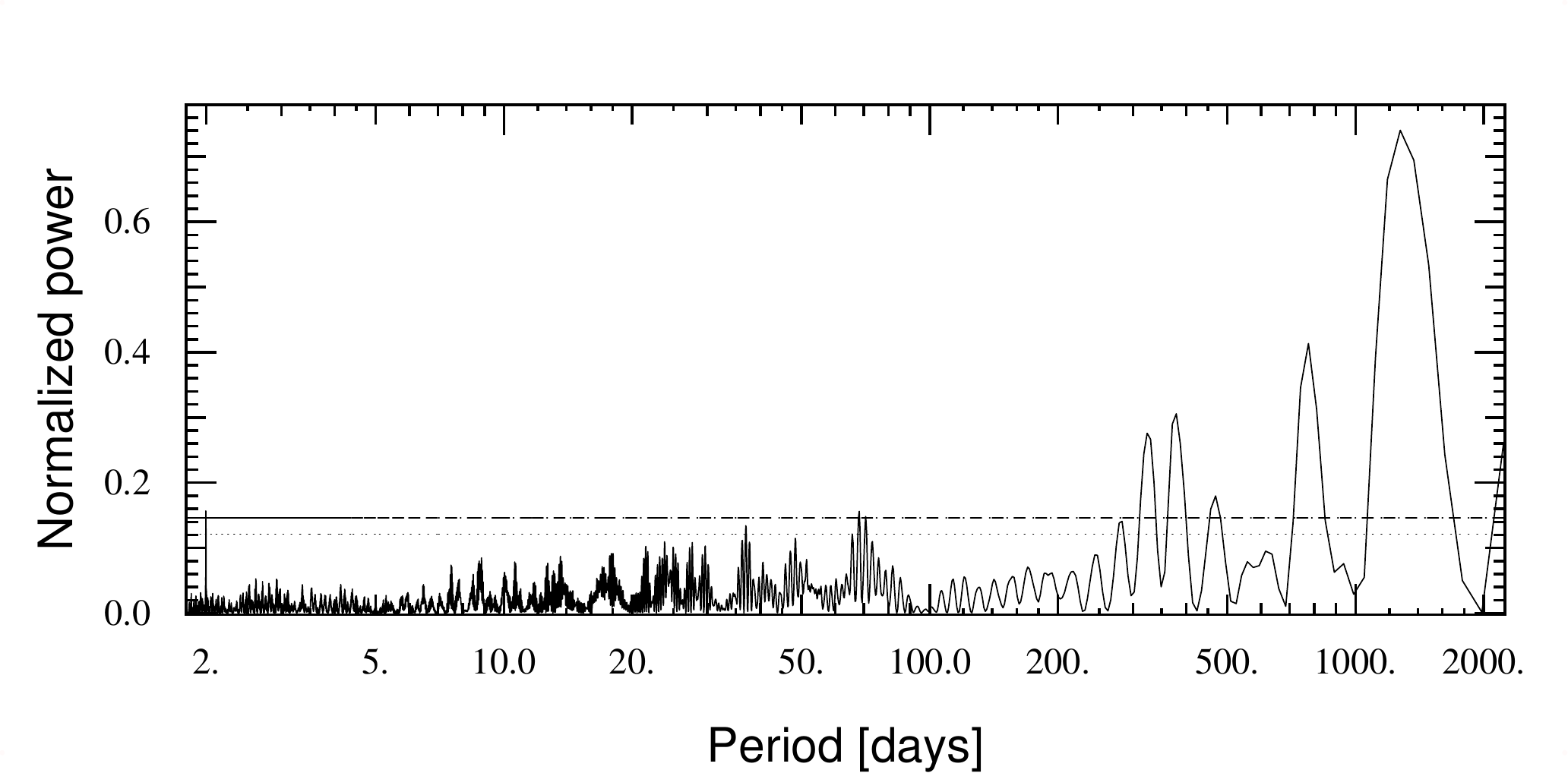}
\centering
\caption{GLS periodogram of the \object{HD\,192310} radial-velocity data with FAP levels at 10\% and 1\% level, respectively (top), the de-trended activity indicator \rhk (center) and the line bisector (bottom). While a very clear peak appears at $P=74$\,days for the radial velocities, no such signal is present in the other two observables. }
\label{fi:hd192310_rv_gls}
\end{figure}

The dispersion clearly exceeds the internal error and the stellar jitter expected for this quiet K3 dwarf. Even by eye one can recognize a rapid radial-velocity variation. The GLS periodogram (Figure\,\ref{fi:hd192310_rv_gls}) confirms the presence of a strong signal at about 74\,days, which is exactly the period of the planet announced by \citet{Howard:2010}. The dashed line indicates a false-alarm probability level of 1\% but the $P=74$\,days signal exceeds this level by far. In the same figure we plot for comparison the GLS periodogram of the activity indicator $log R'_{HK}$ and of the line bisector $BIS$. None of them reveals any significant power at the period indicated by the radial velocity.

\begin{figure}
\includegraphics[bb=0 0 595 390,width=85mm,clip]{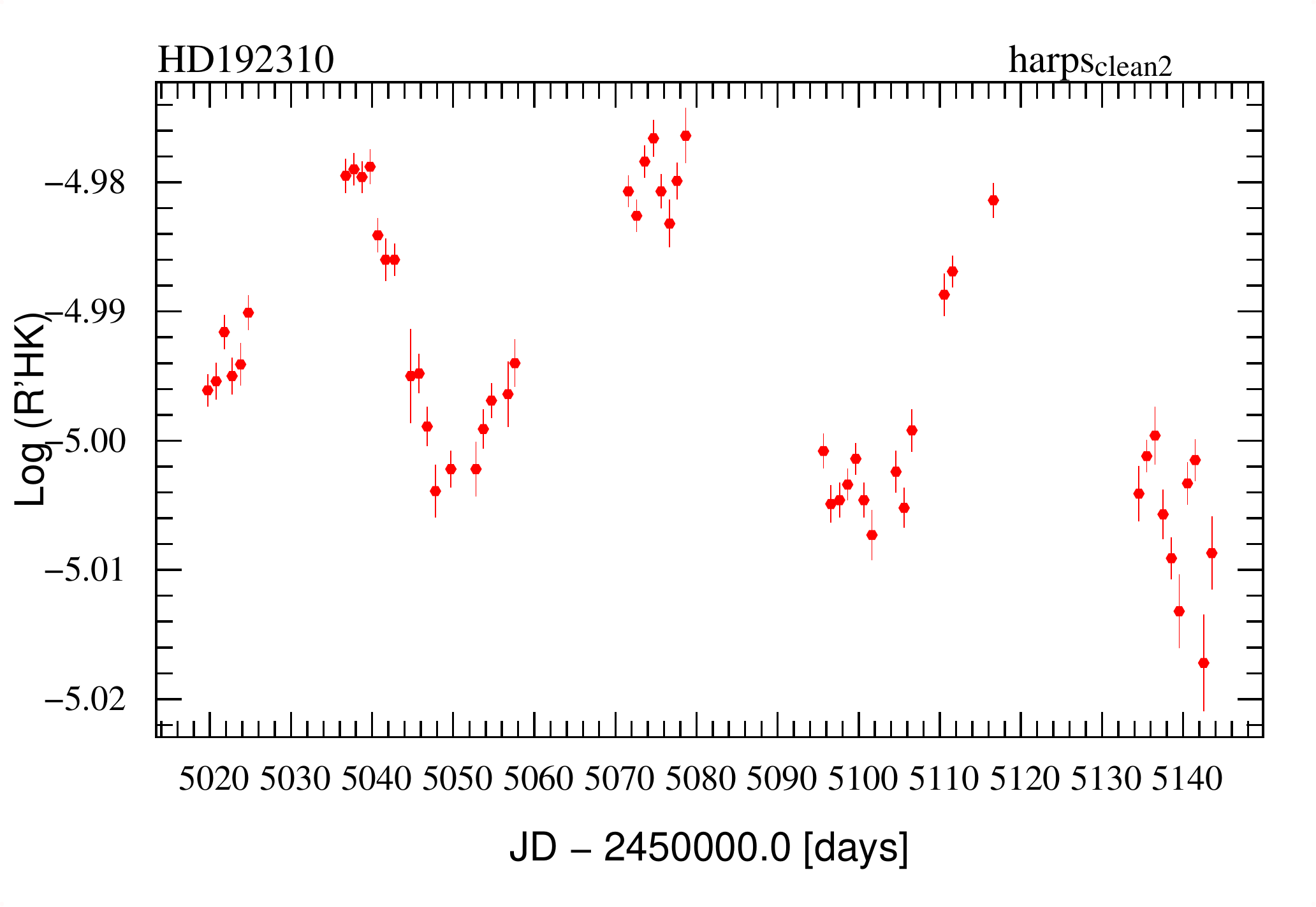}
\includegraphics[bb=0 0 595 260,width=85mm,clip]{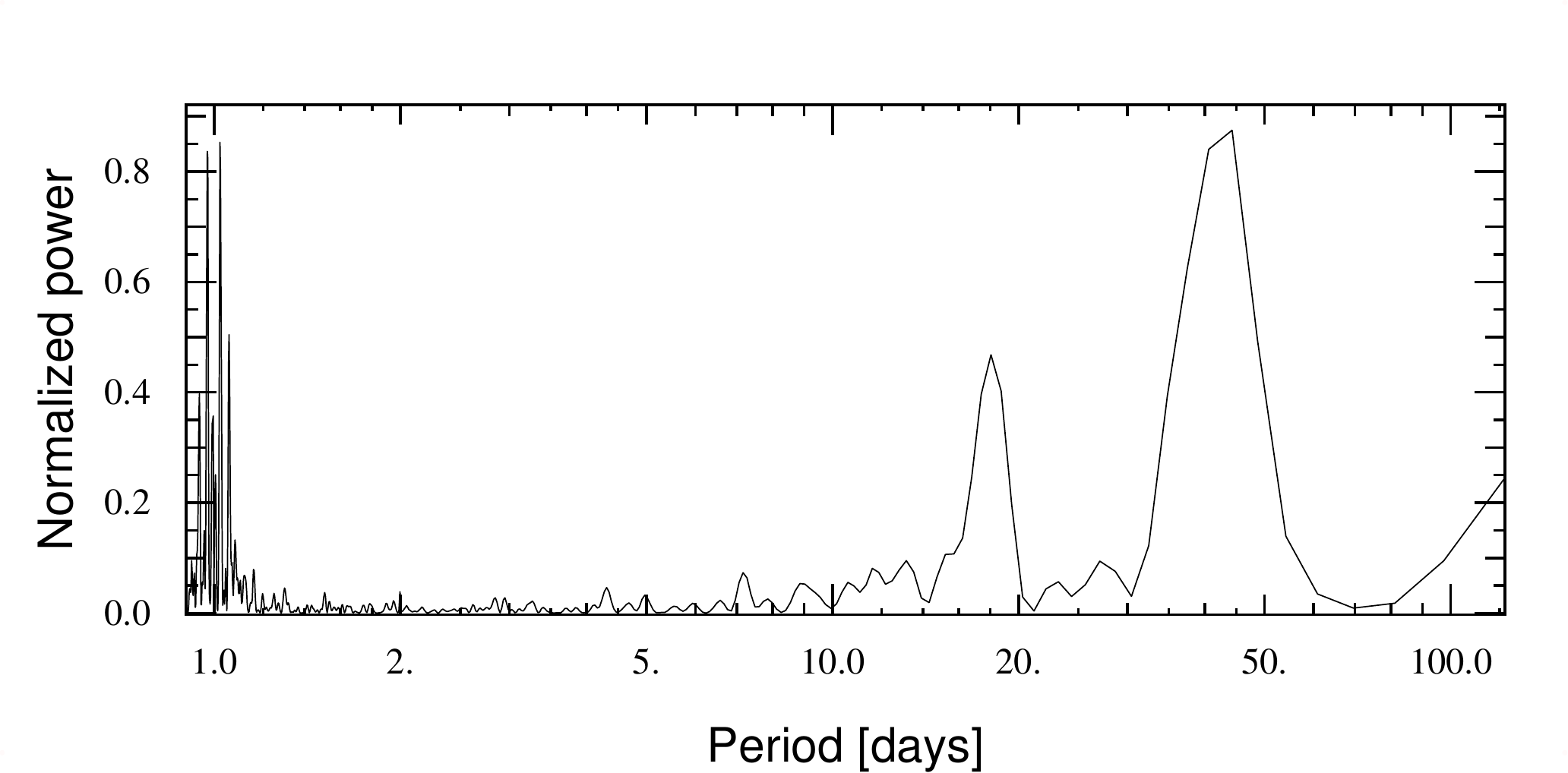}
\centering
\caption{Top: \rhk time series of the two last observational seasons of  \object{HD\,192310} . Bottom: Corresponding GLS periodogram.}
\label{fi:hd192310_rhk_zoom}
\end{figure}

The highest peak in the GLS periodogram for \rhk would be observed at very long periods ($>1000$\,days) and is produced by the stellar magnetic cycle. We therefore de-trended the data to make short-period variations appear more clearly. Some excess power then becomes visible around 40\,days. A closer look at the last observational season indeed reveals a well-established periodicity in the \rhk time series (Figure\,\ref{fi:hd192310_rhk_zoom}, top). The GLS periodogram of only these two seasons indeed reveals a series of significant peaks centered at 40 to 45\,days (Figure\,\ref{fi:hd192310_rhk_zoom}, bottom). It should be noted that this peak is most likely linked to the rotational period of the star and therefore delivers a direct and precise measurement of the stellar rotational period. Indeed, the period obtained in this way matches the value that is computed by directly using the value of \rhk and that is indicated in Table\,\ref{ta:stellar}. On the contrary, a complete lack of power is observed at the $P=74$\,day period, so that we can exclude any relation with the radial-velocity peak.

\begin{figure}
\includegraphics[bb=0 90 595 510,width=\columnwidth,clip]{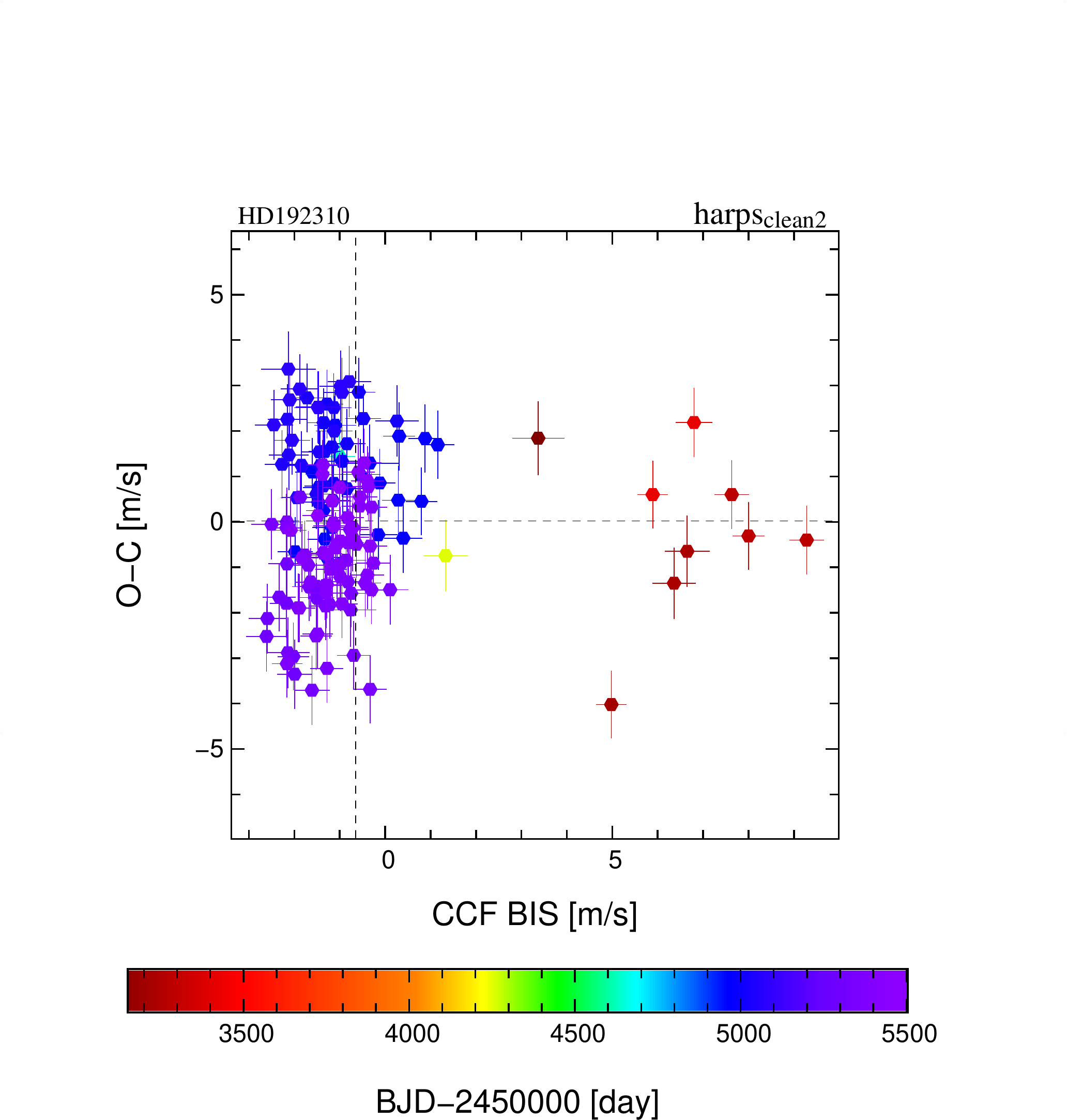}
\includegraphics[bb=0 90 595 510,width=\columnwidth,clip]{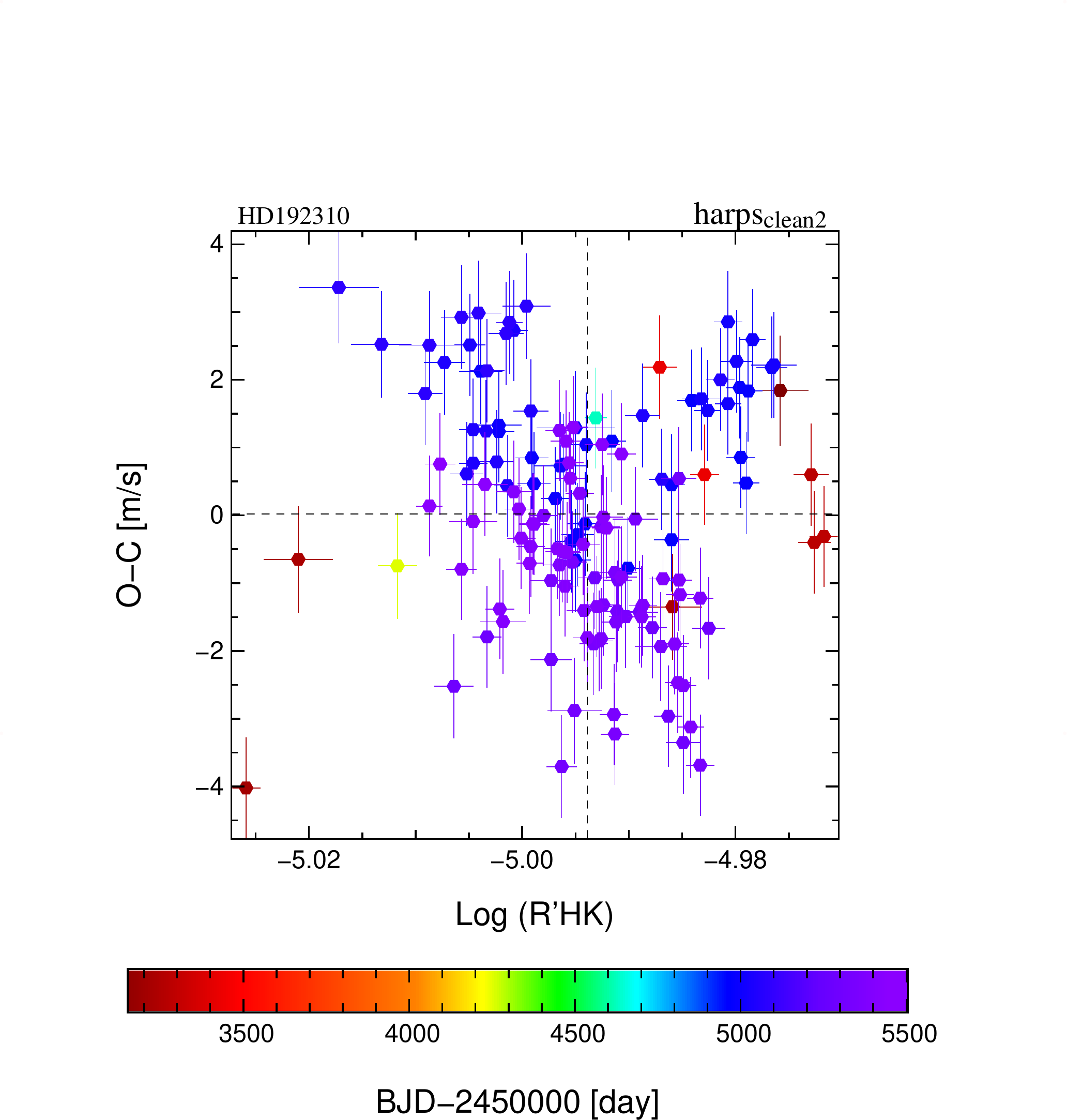}
\centering
\caption{Correlation plot of the residuals of the one-Keplerian fit to the radial velocities of  \object{HD\,192310}  versus the line bisector and the activity indicator \rhk.}
\label{fi:hd192310_k1_corr}
\end{figure}

\begin{figure}
\includegraphics[bb=0 47 595 260,width=85mm,clip]{hd192310_rv_gls.pdf}
\includegraphics[bb=0 47 595 260,width=85mm,clip]{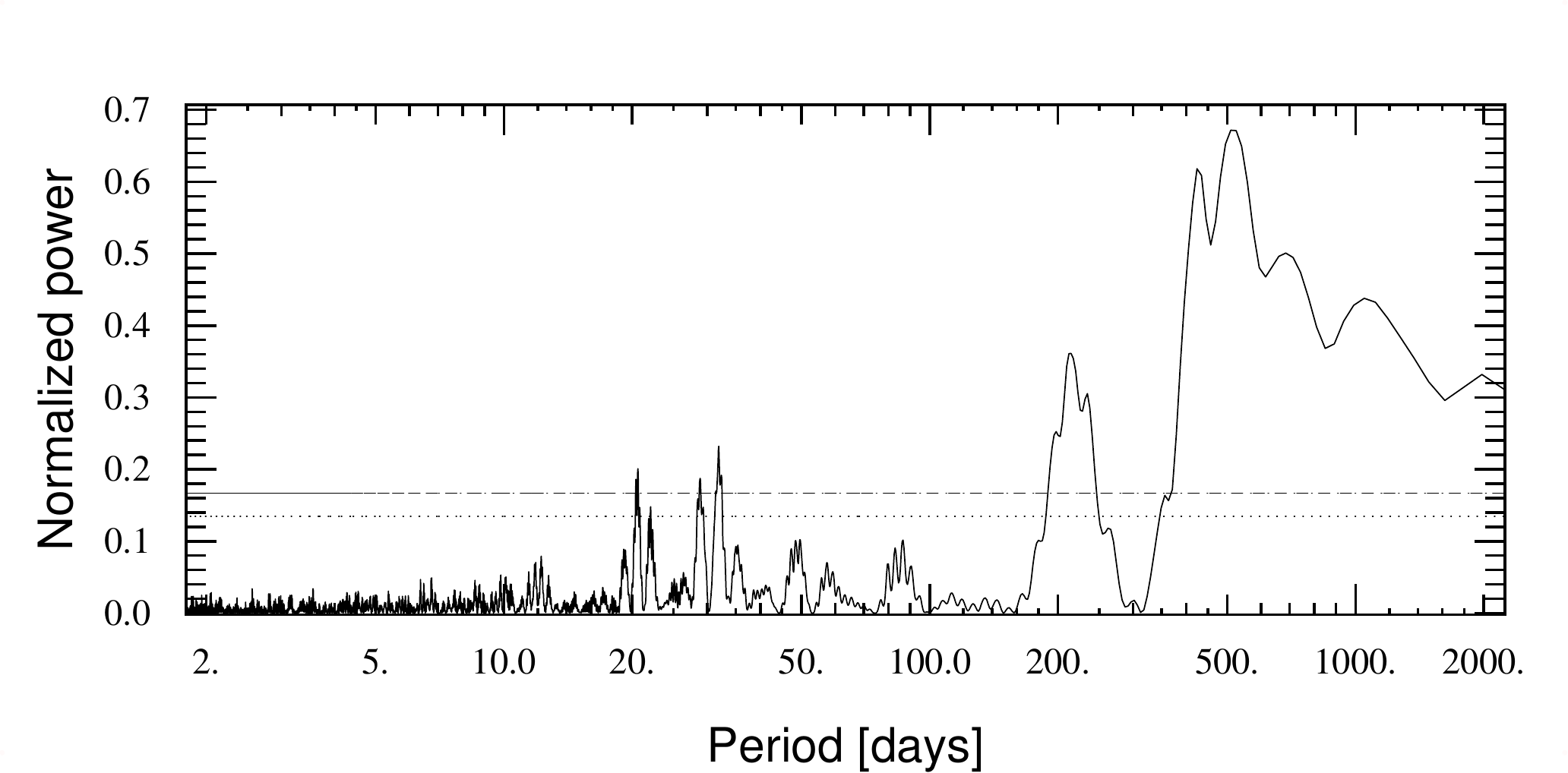}
\includegraphics[bb=0 0 595 260,width=85mm,clip]{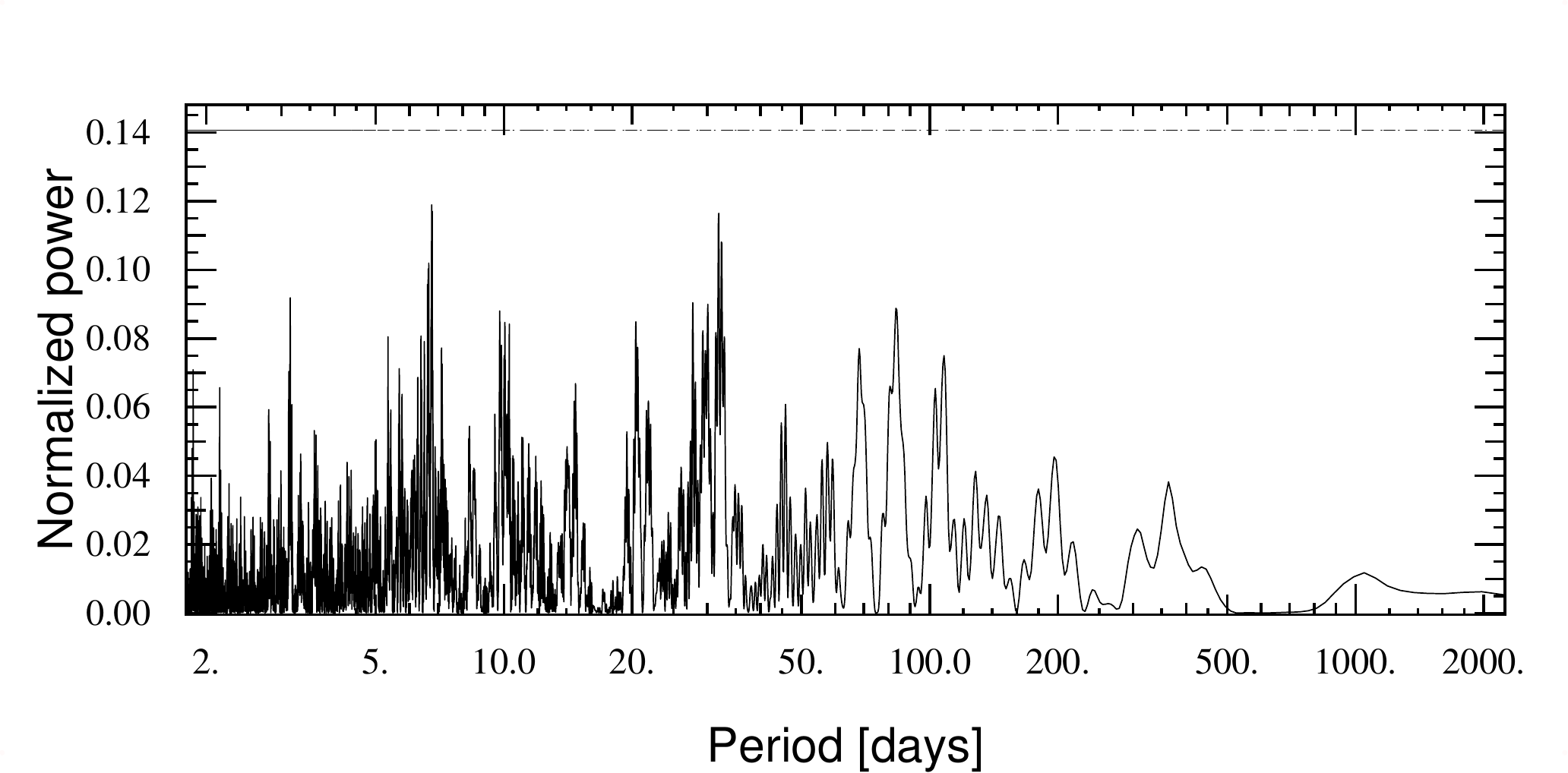}
\centering
\caption{From top to bottom the GLS periodograms of a) the RVs of  \object{HD\,192310}, b) the residuals to the one-Keplerian solution and c) the residuals to the two-Keplerian solution. FAP levels of 10\% and 1\% level are indicated except for the last plot, where only the 10\% level is shown.}
\label{fi:hd192310_omc_gls}
\end{figure}

We therefore fitted a Keplerian signal to the raw data and obtained a solution almost identical to that presented by \citet{Howard:2010}. However, we considered this to be an intermediate solution, as suggested by the GLS periodogram of the residuals to the one-Keplerian solution (Figure\,\ref{fi:hd192310_omc_gls}). Indeed, a significant power peak is left over at a period around 500\,days. Although there is no direct correspondence with the peaks in the bisector and \rhk power spectra, one may suspect that the residual radial-velocity signal is produced by stellar activity. Therefore, we tested the correlation between the residuals to the one-Keplerian solution, on the one hand, and the line bisector and the \rhk, on the other hand. This time the raw (not de-trended) \rhk data where used to identify possible long-period correlations. Figure\,\ref{fi:hd192310_k1_corr} shows the complete absence of correlation between these quantities, the Pearson correlation coefficient being close to zero in both cases. We conclude that the residual radial velocity variations are probably caused by a second planetary companion at longer period. A closer look at the raw radial-velocity data of Figure\,\ref{fi:hd192310_rv_time} reveals a significant decrease in radial-velocity between the two last seasons, which, on the other hand, both show a trend upward. This should have suggested a second periodicity of 400 to 600\,days.

\begin{figure}
\includegraphics[width=\columnwidth]{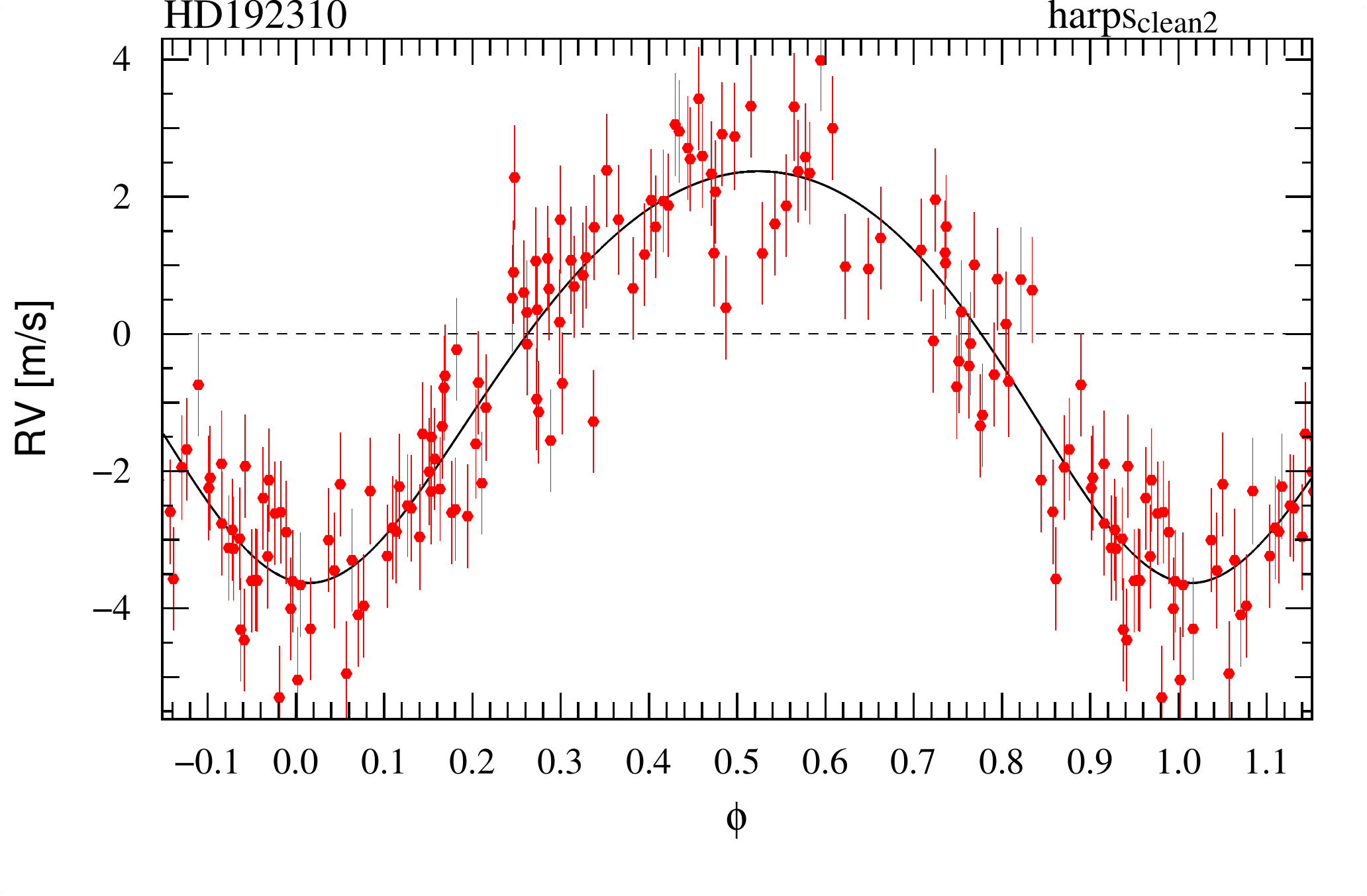}
\includegraphics[width=\columnwidth]{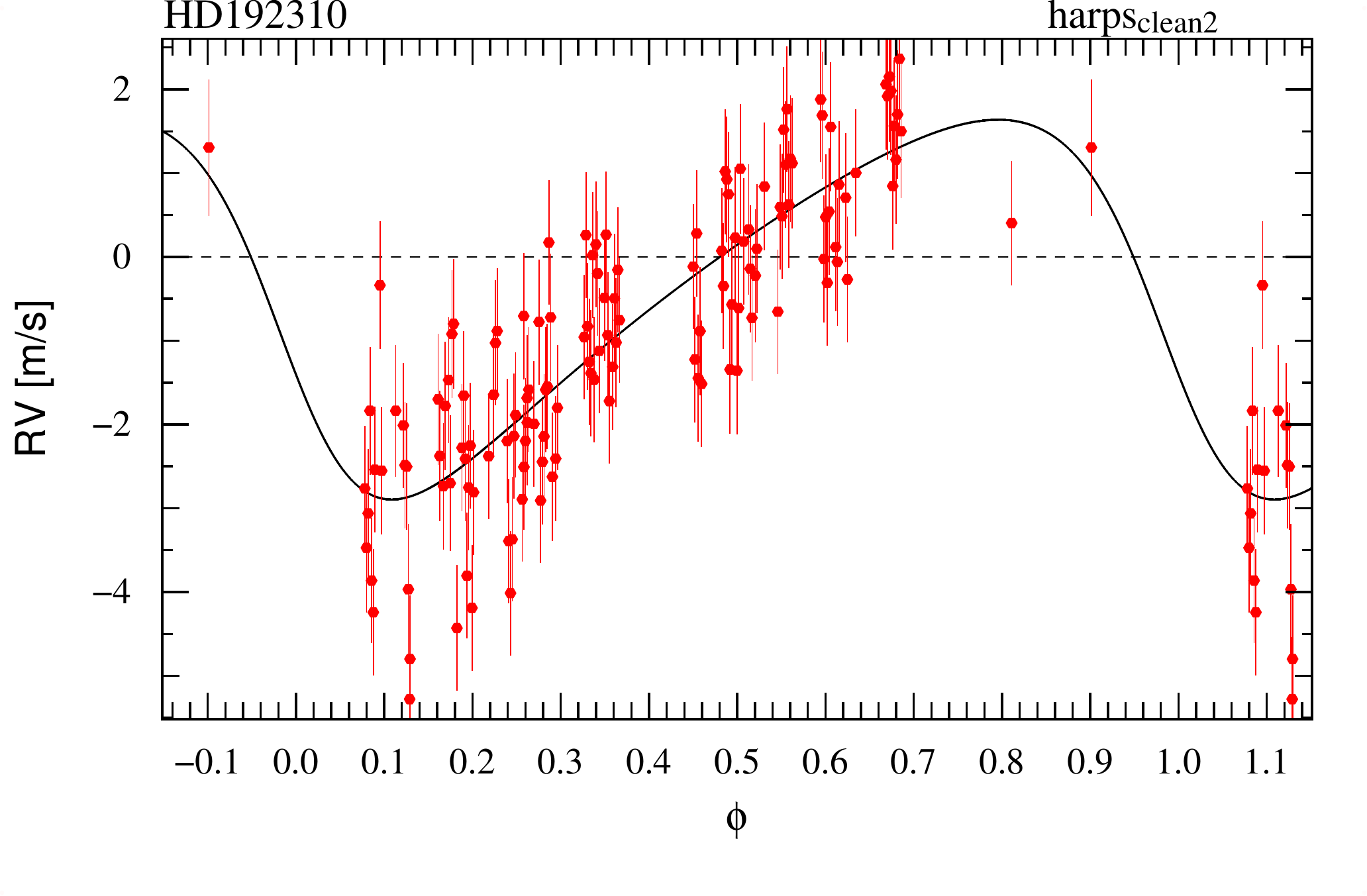}
\centering
\caption{Phase-folded RV data of  \object{HD\,192310} and fitted Keplerian solution for the two planetary companions with $P=75$\,days and $P=526$\,days, respectively. The dispersion of the residuals is of 0.91\ms \emph{rms}.}
\label{fi:hd192310_k2_phase}
\end{figure}

Based on the above considerations, we fitted two Keplerians to the radial velocities of \object{HD\,192310} letting all the parameters free. The solution converges very rapidly and unambiguously towards a period of about 74.7\,days and another of 526\,days. The respective semi-amplitudes are 3.0\ms and 2.3\ms, and the eccentricities 0.13 and 0.32. The eccentricity of the second planet is significant at a 3-$\sigma$ level, and we therefore decided to let both the eccentricities as free parameter.  The eventually obtained orbital parameters are given in Table\,\ref{ta:hd192310_k2_par} and the corresponding phase-folded radial-velocity curve is shown in Figure\,\ref{fi:hd192310_k2_phase}. The GLS of the residuals to the two-Keplerian does not show any significant peak (lower panel of Figure\,\ref{fi:hd192310_omc_gls}).

\begin{table}
\caption{Orbital and physical parameters of the planets orbiting HD\,192310 as obtained from a two-Keplerian fit to the data. Error bars are derived from the covariance matrix. $\lambda$ is the mean longitude ($\lambda$ = $M$ + $\omega$) at the given epoch.}
\label{ta:hd192310_k2_par}
\begin{center}
\begin{tabular}{l l c c}
\hline \hline
\noalign{\smallskip}
{\bf Parameter}	& {\bf [unit]}		& {\bf HD 192310 b}	& {\bf HD 192310 c}	 \\
\hline 
\noalign{\smallskip}
Epoch		& [BJD]			& \multicolumn{2}{c}{2'455'151.02574596}    \\ 
$i$			& [deg]			& \multicolumn{2}{c}{$ 90 $ (fixed) }  \\  
$V$			& [km\,s$^{-1}$]		& \multicolumn{2}{c}{$ -54.2232\,(\pm 0.0003) $}  \\
\hline 
\noalign{\smallskip}
$P$			& [days]			& $74.72$			& $525.8$		 \\ 
			&				& $(\pm 0.10)$		& $(\pm 9.2)$	 \\ 
$\lambda$	& [deg]			& $ 340.8 $		& $ 1.9 $		 \\ 
			&				& $(\pm 2.3) $		& $ (\pm 12.3)  $	 \\ 
$e$			&				& $ 0.13 $			& $ 0.32 $			 \\ 
			&				& $(\pm 0.04)$		& $ (\pm 0.11)$		 \\ 
$\omega$		& [deg]			& $  173$			& $ 110 $			 \\ 
			&				& $(\pm 20)$		& $ 21 $		 \\ 
$K$			& [m\,s$^{-1}$]		& $   3.00 $		& $   2.27 $		 \\  
			&				& $(\pm 0.12)   $	& $(\pm 0.28)  $	 \\
\hline
\noalign{\smallskip}
$m \sin i$		& [$M_\oplus$]		& $ 16.9 $			& $ 24 $			\\
			&				& $(\pm 0.9) $		& $(\pm 5)  $	 \\
$a$			& [AU]			& $ 0.32 $			& $ 1.18 $		\\
			&				& $(\pm 0.005)   $	& $(\pm 0.025)  $	 \\
$T_{\rm eq}$	& [K]    			& $ 355 $ 			& $ 185 $ 			\\
\hline
\noalign{\smallskip}
$N_\mathrm{meas}$ &			& \multicolumn{2}{c}{139}  \\
Span		& [days]			& \multicolumn{2}{c}{2348} \\
rms			& [m\,s$^{-1}$]		& \multicolumn{2}{c}{0.92} \\
$\chi_r^2$	&				& \multicolumn{2}{c}{1.66} \\
\hline
\end{tabular}
\end{center}
\end{table}

For the $N=139$ data points we obtain a residual dispersion around the two-Keplerians solution of 0.92\ms. The reduced $\chi^2$ is 1.66. The uncertainly on the orbital period and the semi amplitude for the $P=74$\,days planet result to be very low. For the longer-period planet, the error is somewhat larger on the semi-amplitude, which is mainly linked to the uncertainty on the eccentricity, which is in turn given by the lack of data points in the (short) descending slope of the radial-velocity curve. Period and eccentricity of the b component are, within the error bars, identical to the values given \citet{Howard:2010}. The semi-amplitude -- and thus the companion's mass -- is different at a level of 2-3 sigmas though. This difference may be explained, on the one hand, because the second planet had not been detected by \citet{Howard:2010}. On the other hand, it may simply be caused by the higher uncertainty (0.4\ms on $K$ given by \citet{Howard:2010}), which is in turn due to the higher errors on the HIRES data: Indeed, when fitting two planets to both data sets simultaneously by adding to both a systematic error of 1.5\ms, the residual dispersion of the HIRES data is 1.8\ms, while for the HARPS data the dispersion is only 1\ms. We conclude that we can undoubtedly confirm the planet \object{HD\,192310\,b}  announced by  \citet{Howard:2010} but suggest a correction of the semi-amplitude and thus its minimum mass.

The semi-major axis of the planetary orbits are 0.32\,AU and 1.18\,AU. For the minimum mass of the planets we obtain $m_b \sin i = 16.9\mathrm{[M_{\oplus}]}$ and $m_c \sin i = 23.4\mathrm{[M_{\oplus}]}$. The error on the mass due to only the fit uncertainties is on the order of 6\% and 20\%, respectively. While for the 74-day period planet the mass is quite well-determined thanks to the significant RV amplitude, for the long-period planet the uncertainty is fairly large, mainly because of a lack of coverage over the steeply descending orbital phase portion. Given their mass and period, both planets are probably of Neptune's composition. The equilibrium temperature defined by a Bond albedo of 0.3 is on the order of 355\,K and 185\,K, which places the two planets at the inner, respectively outer edge of the habitable zone of \object{HD\,192310}.

%__________________________________________________________________
%
\section{Discussion}
\label{se:discussion}      
%__________________________________________________________________

\subsection{Frequency of low-mass planets}
By down-selecting our small sample of stars in terms of measured 'raw' RV dispersion, we have explicitly excluded the possibility of finding planets typically found by ongoing RV surveys. We have reduced the probability of finding any kind of planet in our program. On the other hand, the absence of more massive planets on short and intermediate orbits leaves free space for low-mass planets, possibly in the habitable zone, which is precisely what we are looking for. We then optimized the observations to increase the probability for success by choosing the best possible instrument, selecting bright and inactive stars, and by optimizing the observation sampling and time span.  Finally it should be noted that when down-selecting from the original HARPS high-precision sample, we did not select on spectral type or other criteria that may favor or penalize the presence of planets.
\par
Under these conditions we have found that at least three out of the ten stars observed within our program harbor low-mass planets. Although statistics is poor over only ten targets, it is interesting to note that this 30\% value was already announced by \citet{Lovis:2009}, who based their analysis on the larger ($<200$ stars) HARPS high-precision program. Theoretical works by \citet{Mordasini:2009} actually forecasted that the frequency of small Neptunes and super-Earths on short and intermediated orbits would be considerably higher than that of Saturns and Jupiters.  The recent amazing discoveries made by the KEPLER satellite using the transit technique further strengthen this.\citet{Borucki:2011} report that the probability of finding low-mass planets is considerably higher than for Jupiter or Saturn-mass planets. Furthermore, when summing up the frequency of finding a planet of any mass, these authors end up with a probability of about 30\%, again in perfect agreement with the results of \citet{Lovis:2009}. On the other hand, there is partial disagreement between the HARPS results on the one side and the results by \citet{Howard:2010b} and \cite{Howard:2011} on the other side, who both claim that the total planet occurrence is on the order of 15\%. It has to be mentioned, however, that the studies were restricted to stars with a given spectral type and to planets with an upper period limit, and that the detection biases for low planetary masses and radii leave quite some room for making the different values compatible.
 
\par
The fact that these low-mass planets were hardly ever found by the RV technique before was probably simply due to a 'threshold effect' arising from the reduced precision and sometimes poor time sampling of existing instruments, which had a hard time in detecting signals of few \ms or even sub-\ms. HARPS has produced impressive results by exploring this previously unexplored domain. The number and quality of new low-mass planets is amazing, and many more are still to come. Indeed, the detection of these low signals is not only a matter of precision, but also of observation strategy and time span. Furthermore, complementary analysis is particularly important if one wants to exclude that the RV signal is produced by stellar noise. Systematic long-term follow-up is required. We refer here to \citet{Mayor:2011} for an overview and a summary of the outcomes of the HARPS program as a whole.

\subsection{Nature of found planets}
Radial velocities only provide minimum mass of the planets and contain no information on their radii. The mean density - and consequently an indication on the composition of the planets - can only be determined by combining RVs and transit measurements. Some specific discoveries, as for instance GJ\,1214\,b \citep{Charbonneau:2009}, COROT-7\,b \citep{Leger:2009,Queloz:2009}, KEPLER-10\,b \citep{Batalha:2011}, have shown that these low-mass planets have mean densities similar to our Earth and that they are probably made of rocky material. On the other hand, the atmosphere of planets on such short orbit (days) may have been evaporated by the star, contrary to planets on wider orbits. KEPLER discoveries of small planets on longer-period orbit \citep{Lissauer:2011} have indeed shown that their average density is generally lower than in the previously mentioned cases. One possible explanation is that they may have at least kept a thick gaseous envelope. By construction of our program, we have found planets of low-mass and on not too short-period orbits. Using the previous references we conclude that the presented planets are probably rocky or at least composed to a large extent of rocky material, but may have a thick atmosphere. An exception in our sample is \object{HD\,192310}, which represents instead a 'classical' system of two Neptune-mass planets on long-period orbits.

\par
In the case of \object{HD\,20794} all found planets have masses in the domain of super-Earths and are probably rocky. Unfortunately, the period of the outermost planet is slightly too short to place it in the habitable zone of its star. Nevertheless, this particular example illustrates best the potential of our program and the fact that already today it is possible to find rocky planets in the habitable zone of solar-type stars.

\par  The system of \object{HD\,85512} is much simpler in terms of dynamics, given the single planet, and does therefore not deserve any particular discussion. For this K5 dwarf the habitable zone is much closer in than for \object{HD\,20794}. The planet's equilibrium temperature supposing an albedo of 0.3 is on the order of 298\,K. This value is slightly higher than that of the Earth and places the planet just at the inner edge of the habitable zone \citep{Selsis:2007b}. On the other hand, we might expect a thick atmosphere for this planet. If the albedo were higher, the equilibrium temperature would decrease. However, depending on the atmosphere's composition, the planet may also run into a runaway greenhouse effect \citep{Selsis:2007b}. Opposite, if its atmosphere is water-poor (in the case of a 'dry planet'), the habitable zone happens to be much closer in \citep{Abe:2011}. Because the detail depends on many unknown parameters, it would be pure speculation to conclude on the habitability of this planets. Nevertheless, we note that  \object{HD\,85512} is the lowest-mass planet discovered by RVs so far possibly lying in the habitable zone of its star.

\subsection{Limits and possibilities of the ongoing program}
The HARPS spectrograph allows us to find planetary companions to solar-type stars that induce sub meter-per-second radial velocity signatures. This proves two facts: a) Instrument precision is on short \emph{and} long term below 1\ms. b) For well-selected stars the stellar 'jitter', i.e., the sum of pulsation, granulation at all time scales, and activity, does not exceed 1\ms.
\par
\begin{figure}
\includegraphics[width=\columnwidth]{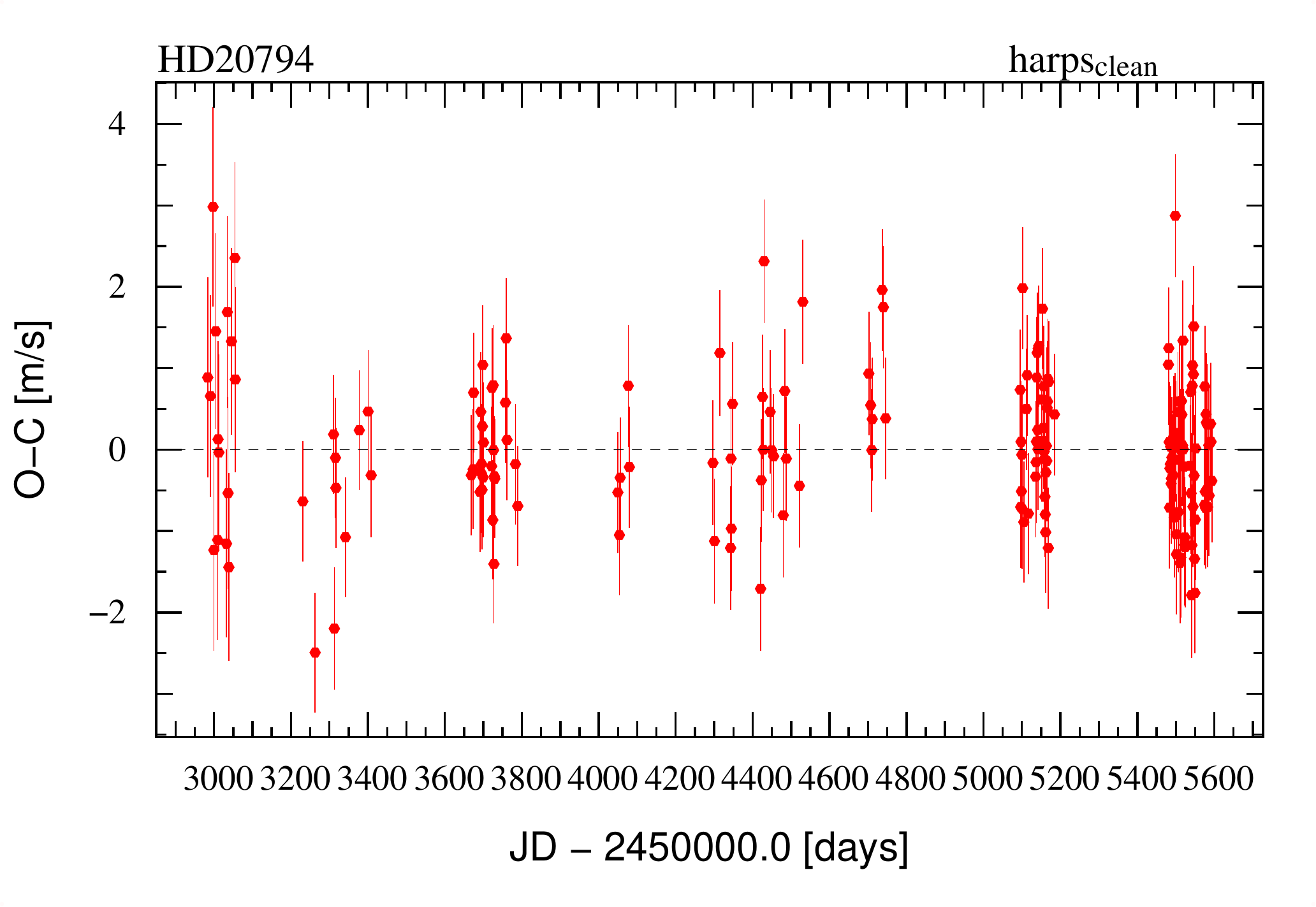}
\includegraphics[width=\columnwidth]{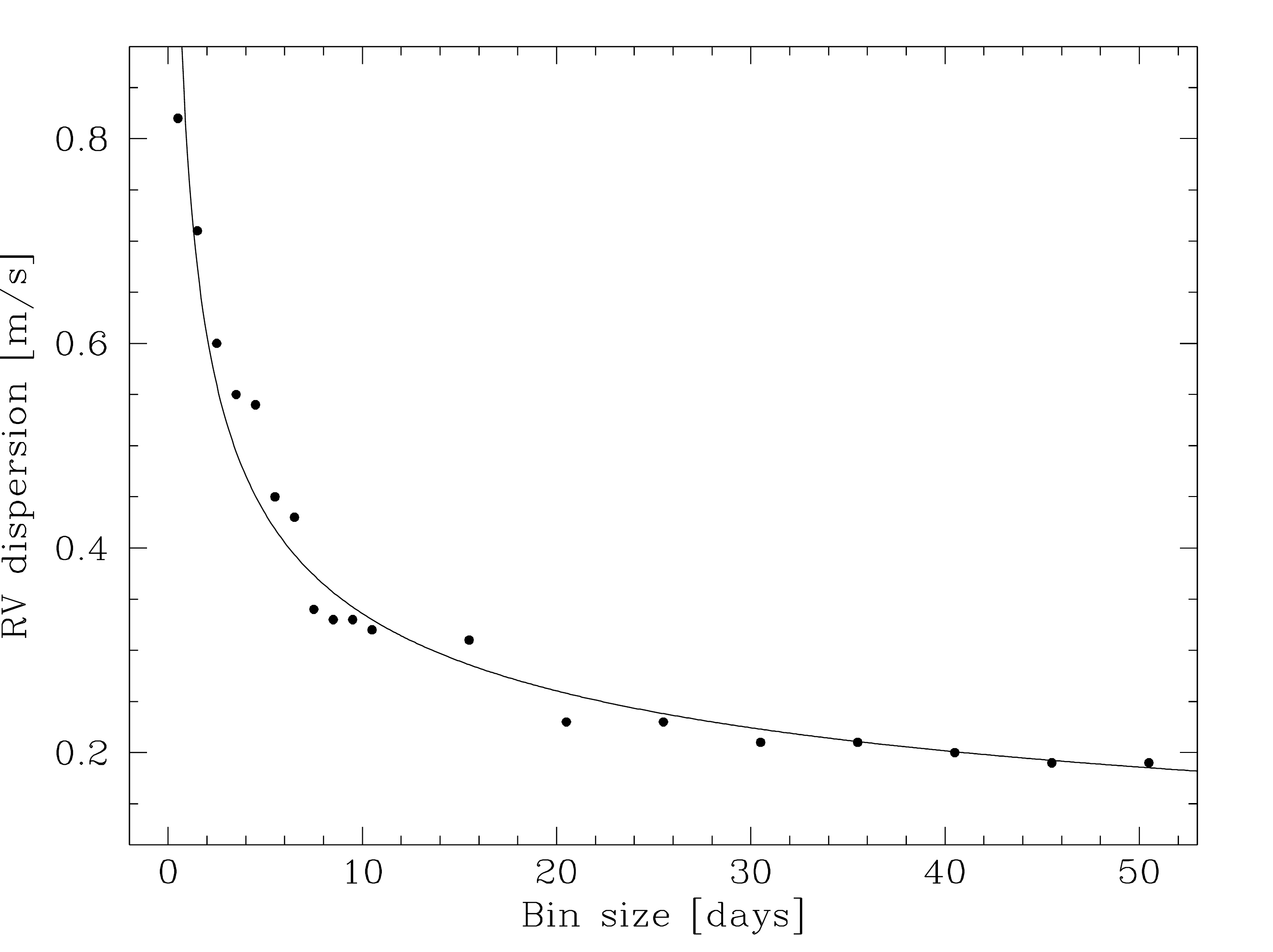}
\centering
\caption{ Top: Residuals of the radial velocities of HD\,20794 to the three-Keplerian fit as a function of time. Bottom: Allen analysis of the dispersion of these same residuals over the past two years.}
\label{fi:allen}
\end{figure}

To illustrate the achieved overall precision, we have analyzed the residuals to the orbital solution of \object{HD\,20794} in more detail. Figure\,\ref{fi:allen} (upper panel) shows the residuals of the data to the three-Keplerian solution. Although some structure is identified in the early phases (data from the original HARPS GTO program), no secular trend at the sub-\ms level is found over seven years of observations. It has to be mentioned that the observation strategy has been improved and the time sampling increased on the mentioned object in the frame of the HARPS-Upgrade GTO, which was specifically designed to obtain the best possible RV performance. The results are best illustrated by (Figure\,\ref{fi:allen}, lower panel), where the dispersion of the residuals is shown as a function of binning period for the last two years of observation (optimized strategy). For no binning (one-day bin) we obtain the mentioned scatter of 0.82\ms \emph{rms}. When the binning is increased, the dispersion decreases with the bin size, demonstrating that most of the noise is indeed averaged away. The noise decreases with the power of -0.37, however. Therefore it is different from 'white noise', which is expected to decrease as the square root of the bin size. This shows that the noise is not completely uniformly distributed over the frequency domain. Nevertheless, we end up with a scatter of only 0.2\ms when averaging over the timescale of a month.

In interesting to note that the stellar noise has a 'color', which reveals itself best in the GLS periodograms of \rhk and $BIS$. If detected in the RV data (by comparing the frequency with \rhk and $BIS$ data, it could be 'filtered', or, in the worst case, used to identify radial-velocity signals of same frequency that are possibly induced by the star. In the former case one could benefit from the fact that the noise is colored to improve the RV data and reduce the error bars on the planets fitted parameters. This is easily done when the frequency of the planet signal is significantly different from the frequency detected in \rhk and $BIS$, supposing that possible stellar jitter would appear at similar frequencies (or harmonics of them) as for the activity indicator and the line bisector. Filtering of the unwanted frequencies could be performed in the Fourier domain, for instance. Alternatively, we propose to fit the planet orbital signal convolved with the window function directly in the Fourier domain.
\par
As mentioned above and as we did in the present paper, the activity indicator and the line bisector may be used for 'monitoring' and detecting possible correlations among the various observables. If we assume that RV jitter caused by stellar noise should appear at the same frequencies (or harmonics) as for the activity indicator and the line bisector, we can use this information to detect 'suspect' signals. It must be pointed out, however, that both indicators remain a diagnostic rather than a proof for the presence or absence of induced RV signal, and this is best illustrated by two extreme examples: a) If no signal is detected in the $BIS$, one cannot exclude that the observed RV signal would \emph{not} be induced by the star, as demonstrated by \citet{Santos:2003}. b) The cases of \object{HD\,85512} and \object{HD\,192310} show very nicely that despite the presence of a long-term signal in the \rhk, no correlated signal is observed in the RV. This confirms the conclusions drawn by \citet{Santos:2010}, who did not find any evidence for radial-velocity variations induced by variations of the stellar magnetic cycle with amplitudes significantly above 1\ms for early K-type stars.
\par
\rhk and $BIS$ are thus excellent diagnostic parameters, provided that they are determined to sufficient precision level. Thanks to the spectroscopic stability of HARPS, both can be determined to extreme accuracy. For the line bisector we demonstrated that it can be even more stable than the RV itself because it is only affected by asymmetry changes. The \rhk, on the other hand, can be measured to the milli-dex precision, a level at which it is hard to distinguish stellar effects from instrument effects.  An interesting side product of this is the direct determination of the stellar rotational velocity, which, for all the three cases presented here, turned out to agree excellently with the value directly derived from \rhk using the procedures given by \citet{Mamajek:2008}. The changes made recently on the HARPS data-reduction pipeline have greatly contributed to the improved spectroscopic performances of HARPS \citep{Lovis:2011c}.

%______________________________________________________________
%
\section{Conclusions}
%______________________________________________________________
\label{se:conclusions} 
Of the original HARPS high-precision program we have chosen ten bright and quiet stars to be followed-up in the frame of the HARPS-Upgrade GTO program. These stars were furthermore selected because they were showing very low radial-velocity dispersion, from which we could exclude the presence of massive and/or short-period planets. The goal of this program is to search for low-mass, possibly rocky planets in the habitable zone.
\par
Despite this severe 'down-selection', we have found that at least three out of the ten stars harbor low-mass planetary companions. Around \object{HD\,192310} we could confirm the Neptune-mass planets recently announced by \citet{Howard:2010}, but identified a second companion of similar mass on a much wider orbit and lying at the outer edge of the habitable zone. In addition, a planet of only $m \sin i = 3.6\mathrm{[M_{\oplus}]}$ was found to orbit the star \object{HD\,85512}. Owing to the late spectral type of this star, the $P=58$\,day-orbit planet of probably rocky nature lies at the inner edge of its habitable zone. Finally, we reported the system around \object{HD\,20794} consisting of three super-Earths of only few Earth-masses each. While the two signals at an 18- and 90-day period leave little doubt on their planetary nature, we will need more data to definitively confirm the $m \sin i = 2.4\mathrm{[M_{\oplus}]}$ on the $P=40$\,days orbit. It must be noted, however, that the semi-amplitude induced by this planet on its parent star is only 0.56\ms, the lowest planetary signature ever detected. All these discoveries were made possible thanks to the ultra-precise radial velocities and the impressive measurements of the line bisector and the \rhk delivered by the HARPS instrument.
\par
Present and future discoveries of such low-mass planets will be of great importance per se, pushing the variety of known extrasolar planets one step further. They will furthermore offer, thanks to the proximity and magnitude of their parent star, an excellent opportunity for precise and in-depth photometric, polarimetric, spectroscopic and astrometric follow-up of the planet and its host star. A bunch of exciting and unprecedented observational data will follow, providing new input on the formation models and the understanding of the internal structure and the atmosphere of these planets. Finally, these candidates will be among the best targets to be considered by future space missions, which will need appropriate targets to look at and will have to take the best advantage of observational time.

\begin{acknowledgements}
We are grateful to all technical and scientific collaborators of the HARPS Consortium, ESO Head Quarter, and ESO La Silla, who have all contributed with their extra-ordinary passion and valuable work to the success of the HARPS project. This research made use of the SIMBAD database, operated at the CDS, Strasbourg, France. NCS acknowledges the support by the European Research Council/European Community under the FP7 through Starting Grant agreement number 239953. NCS also acknowledges the support from Funda\c{c}\~ao para a Ci\^encia e a Tecnologia (FCT) through program Ci\^encia\,2007 funded by FCT/MCTES (Portugal) and POPH/FSE (EC), and in the form of grants reference PTDC/CTE-AST/098528/2008 and PTDC/CTE-AST/098604/2008. FB wish to thank the French National Research Agency (ANR-08-JCJC-0102-01. Finally, we acknowledge the support by Swiss National Science Foundation for its continuous support in our research. 

\end{acknowledgements}

\bibliography{pepe}
\bibliographystyle{aa}

\end{document}